\documentclass[11pt]{article}

\usepackage{arxiv}

\usepackage[T1]{fontenc}
\usepackage{newtxtext}
\usepackage{amsmath}
\usepackage{amsthm}
\usepackage[varvw]{newtxmath}
\usepackage{microtype}
\usepackage{graphicx}
\usepackage{booktabs}
\usepackage{xcolor}
\usepackage{colortbl}
\usepackage{caption}
\usepackage{algorithm}
\usepackage{algpseudocode}
\usepackage{enumitem}
\usepackage{tikz}
\usetikzlibrary{arrows.meta}
\definecolor{cbBlue}{RGB}{0,114,178}
\definecolor{cbOrange}{RGB}{230,159,0}
\tikzset{
  >={Stealth[length=2mm,width=1.6mm]},
  every picture/.append style={font=\footnotesize},
  edge/.style={draw=black!65,line width=0.6pt},
  vtx/.style={circle,draw=black,line width=0.6pt,fill=white,minimum size=6mm,inner sep=0pt,font=\footnotesize},
  home/.style={vtx,line width=1pt},
  gh/.style={vtx,fill=black!70,text=white},
  cand/.style={vtx,fill=gray!25},
  far/.style={double,double distance=1.2pt,thick},
  omit/.style={edge,densely dotted},
  tgrid/.style={draw=gray!45},
  dot/.style={circle,fill=gray!55,inner sep=0pt,minimum size=2.4pt},
  ag/.style={circle,draw,thick,fill=white,minimum size=4.5mm,inner sep=0pt,font=\scriptsize},
  agA/.style={ag},
  agB/.style={ag,densely dashed},
  agC/.style={ag,double},
  move/.style={->,draw=black,line width=0.8pt,shorten >=1pt,shorten <=1pt},
  walk/.style={move,densely dashed},
  ruled/.style={move,densely dotted},
  dir/.style={move},
  outw/.style={move},
  retw/.style={move,dash dot},
  sbox/.style={draw=black!40,rounded corners=2pt,fill=black!3,align=center,font=\footnotesize},
  lab/.style={font=\footnotesize},
  ttl/.style={font=\footnotesize,anchor=west},
  note/.style={font=\footnotesize,align=center},
}
\usepackage{url}
\usepackage[numbers,sort&compress]{natbib}
\usepackage[colorlinks=true,linkcolor=blue!55!black,citecolor=green!45!black,urlcolor=blue!55!black]{hyperref}
\usepackage{doi}
\usepackage[capitalise,nameinlink,noabbrev]{cleveref}

\theoremstyle{plain}
\newtheorem{theorem}{Theorem}
\newtheorem{lemma}{Lemma}
\newtheorem{proposition}{Proposition}
\newtheorem{corollary}{Corollary}
\newtheorem{observation}{Observation}
\theoremstyle{definition}
\newtheorem{definition}{Definition}
\theoremstyle{remark}
\newtheorem{remark}{Remark}
\crefname{observation}{Observation}{Observations}
\Crefname{observation}{Observation}{Observations}
\crefname{algorithm}{Algorithm}{Algorithms}
\Crefname{algorithm}{Algorithm}{Algorithms}

\makeatletter\renewcommand{\theHALG@line}{\thealgorithm.\arabic{ALG@line}}\makeatother
\setlist[enumerate]{topsep=3pt,itemsep=1pt}
\setlist[itemize]{topsep=3pt,itemsep=1pt}

\newcommand{\PE}{\textsf{PE}}
\newcommand{\Home}{H}
\newcommand{\Hole}{u_g}
\newcommand{\cw}{\mathit{cw}}
\newcommand{\ccw}{\mathit{ccw}}

\title{A Match Made in Hell: \\Perpetual Exploration with a Gray Hole and Byzantine Agents}

\author{%
  \normalfont
  \begin{tabular}{@{}c@{\hspace{1cm}}c@{}}
    \textbf{Adri Bhattacharya}
      & \textbf{Prabhat Kumar Chand} \\
    LUISS Guido Carli University
      & Indian Institute of Technology Indore \\
    Rome, Italy
      & Indore, India \\
    \href{mailto:adri.bhattacharya@gmail.com}{\texttt{adri.bhattacharya@gmail.com}}
      & \href{mailto:pchand744@gmail.com}{\texttt{pchand744@gmail.com}} \\[1.5em]
    \textbf{Partha Sarathi Mandal}
      & \textbf{Debasish Pattanayak} \\
    Indian Institute of Technology Guwahati
      & Indian Institute of Technology Indore \\
    Guwahati, India
      & Indore, India \\
    \href{mailto:psm@iitg.ac.in}{\texttt{psm@iitg.ac.in}}
      & \href{mailto:debasish@iiti.ac.in}{\texttt{debasish@iiti.ac.in}}
  \end{tabular}%
}

\date{}
\renewcommand{\shorttitle}{Perpetual Exploration with a Gray Hole and Byzantine Agents}

\hypersetup{
  pdftitle={A Match Made in Hell: Perpetual Exploration with a Gray Hole and Byzantine Agents},
  pdfauthor={Adri Bhattacharya, Prabhat Kumar Chand, Partha Sarathi Mandal, Debasish Pattanayak},
  pdfkeywords={Mobile agents, Perpetual exploration, Gray hole, Byzantine agents, Ring networks, Whiteboard model, Face-to-face communication, Lower bounds},
}

\begin{document}
\maketitle

\begin{abstract}
Mobile agents can monitor a network by revisiting its nodes to assess changing local conditions. In this work, we study how a team of such agents can monitor a ring in the presence of at most $f$ \emph{Byzantine} agents, which are controlled by an adversary and may deviate arbitrarily from the algorithm. The remaining agents are honest and follow the algorithm. The agents act in common rounds, agree on a clockwise direction, and start together at a safe home node. The ring contains one \emph{gray hole}, a node that the same adversary may activate in any round to destroy every honest agent on it; all other nodes are safe. The goal of \emph{perpetual exploration} is to ensure that every safe node is visited infinitely often by at least one honest agent. A key challenge is that a Byzantine agent can deliberately fail to return, making its absence indistinguishable from the destruction of an honest agent. We consider the \emph{whiteboard} model, in which agents communicate through records stored on whiteboards at the nodes, and the \emph{face-to-face} model, in which agents communicate when they occupy the same node. In the whiteboard model, we show that $2f+3$ agents are necessary and sufficient when the ring size $n$ is known and $n\ge4$. On a triangle ($n=3$), $2f+2$ agents are necessary and sufficient. When $n$ is unknown, $3f+3$ or fewer agents do not suffice, whereas $3f+5$ agents suffice. In the face-to-face model, $2f+2$ or fewer agents do not suffice for $n\ge4$, whereas $3f+3$ agents suffice when $n$ is known. For $f=0$ and known $n\ge4$, three agents are necessary and sufficient, improving the previous upper bound of four agents due to Bhattacharya et al.\ [DISC~2025]. On rings of known size with four or five nodes, five agents tolerate one Byzantine agent, and this bound is optimal. When $n$ is unknown and all agents are honest, exactly five agents are necessary and sufficient---two more than for known $n\ge4$. Finally, with up to $f$ Byzantine agents and unknown $n$, we show that $3f+3$ or fewer agents do not suffice, whereas $5f+5$ agents suffice.
\end{abstract}


\section{Introduction}
\label{sec:intro}

Mobile agents are autonomous programs that move through a network and act on the nodes they visit. A team of such agents can monitor a network continuously, for instance to check the state of its nodes or to collect fresh data, as long as it keeps coming back to every node. Repeatedly revisiting the nodes is called \emph{perpetual exploration}. It becomes much harder in a hostile network, where some nodes may destroy the agents, and even more difficult when some of the agents themselves cannot be trusted. The agents must then keep exploring while they learn which nodes to avoid, even though a Byzantine agent may lie about what it has seen, or simply disappear, exactly as an agent destroyed by the network would.

The classical node fault is the \emph{black hole}, which destroys every agent that enters it and leaves no trace. Since its introduction by Dobrev et al.~\cite{DBLP:journals/algorithmica/DobrevFPS07}, black-hole search has become a large line of research on mobile agents in dangerous networks, and surveys may be found in~\cite{FlocchiniPS19,Pattanayak2026}. Later work studied arbitrary networks~\cite{DBLP:journals/dc/DobrevFPS06}, synchronous agents~\cite{DBLP:journals/fuin/CzyzowiczKMP06,DBLP:journals/cpc/CzyzowiczKMP07}, communication weaker than whiteboards, such as tokens and pebbles~\cite{DBLP:conf/ciac/DobrevKSS06,DBLP:journals/tcs/DobrevFKS13,DBLP:journals/algorithmica/FlocchiniIS12}, agents starting at different nodes~\cite{DBLP:journals/ijfcs/DobrevSS08,DBLP:journals/tcs/Chalopin0LM13}, the time needed to find the black hole~\cite{DBLP:journals/dmaa/BalamohanFMS11}, and networks whose links appear and disappear over time~\cite{DBLP:conf/icdcs/LunaFPS21,DBLP:conf/opodis/LunaFPS23,DBLP:conf/sand/BhattacharyaI024,DBLP:journals/jgaa/BhattacharyaIM25,DBLP:conf/icdcn/KaurS0025}. Two classical techniques for these faults are the \emph{cautious walk}~\cite{DBLP:journals/algorithmica/DobrevFPS07,DobrevFPS2004CautiousWalk}, in which one agent moves to the next node and returns while another waits at a safe node, so that a safe return proves that the next node is safe, and the \emph{pendulum walk}~\cite{DBLP:conf/icdcs/LunaFPS21} of synchronous agents, in which one agent stays at a safe base node while another swings out to the first unexplored node and back, one node farther each time it returns safely. Both techniques fail against a \emph{gray hole}. In each round, an adversary decides whether the gray hole is active, and an active gray hole destroys the agents on it, so it models a node that fails only from time to time. A node that an agent has visited safely may destroy it on a later visit, so a safe return does not prove that the node is safe. Until the gray hole is identified, an agent cannot assume that any node other than the home node is safe for waiting. In the pendulum walk, the moving agent passes nodes that were found safe earlier and may be destroyed on any of these passes, so the agents may suspect the wrong node. Faults of this kind first appeared in work on periodic data retrieval~\cite{DBLP:conf/sirocco/KralovicM10,DBLP:journals/tcs/BampasLMPP15} and later in work on perpetual exploration~\cite{DBLP:conf/opodis/GoswamiBD024,DBLP:conf/wdag/BhattacharyaGB025}.

Faulty agents are a second, independent concern. Byzantine agents, which may behave arbitrarily, have been studied for black-hole search, for example in~\cite{DBLP:conf/sss/TsuchidaOI19,DBLP:conf/icdcs/LunaFPPPS25}, where the black hole never spares an honest visitor. In this paper, both the network and some of the agents are faulty. The network is an oriented ring of $n$ nodes with one gray hole, and up to $f$ agents are \emph{Byzantine}. A single adversary controls both the gray hole and the Byzantine agents, so the two faults can act together against the honest agents. All agents start at a safe home node $\Home$ and move in synchronous rounds, and the goal is that every safe node be visited infinitely often by at least one honest agent. We study two communication models, the whiteboard model and the face-to-face model. An identifier is a distinct label assigned to an agent. Our Byzantine agents are weakly Byzantine, as in Di Luna et al.~\cite{DBLP:conf/icdcs/LunaFPPPS25}, so they cannot forge the identifier of another agent. The restriction on identifier forgery is essential because our algorithms count evidence from distinct agents. Without authenticated identifiers, one Byzantine agent could impersonate $f+1$ agents. Since one adversary controls both faults, the adversary can keep the gray hole inactive whenever a Byzantine agent visits the gray hole, so that the Byzantine agent passes unharmed and later claims that the node is safe. Moreover, an honest agent that misses its scheduled return may have been destroyed, but a Byzantine agent can also stay away deliberately. A single missed return therefore does not establish that an honest agent was destroyed.

In the \emph{whiteboard} model, the agents communicate only through records written on the nodes, and every activation of the gray hole also erases its whiteboard. The gray hole introduced by Kr\'alovi\v{c} and Mikl\'ik~\cite{DBLP:conf/sirocco/KralovicM10} and studied by Bampas et al.~\cite{DBLP:journals/tcs/BampasLMPP15} only destroys agents, while the Byzantine black hole of Goswami et al.~\cite{DBLP:conf/opodis/GoswamiBD024} may also erase its whiteboard, at the choice of the adversary. Our gray hole is the special case in which the gray hole erases its whiteboard whenever it is active. We first give an informal overview of the main ideas. The difficulty in the whiteboard model is to distinguish destruction from a deliberate absence. Our algorithms use two kinds of evidence. First, agents leave $\Home$ one at a time, on trips of growing length, each going outward from $\Home$ and returning along the same path by a fixed return deadline, and missed returns count only when $f+1$ distinct agents miss the same trip, since $f+1$ distinct agents include at least one honest agent. Such missed returns show that an honest agent was destroyed, but not where. Second, an honest agent that finds one of its own records missing identifies its node as the gray hole. To notice a missing record, however, the agent must return to the gray hole, where it may be destroyed. The repeated trips make sure that every honest agent still following the schedule has a record at the gray hole before such an agent makes a longer trip beyond it. Hence, once an honest agent is destroyed on a longer trip, every honest agent that returns to the gray hole either finds its record missing or is destroyed. At most $f+2$ honest agents are destroyed during the scheduled trips, so $2f+3$ agents suffice, which is optimal on rings with $n\ge4$. Without $n$, an agent that has found one neighbor of the gray hole cannot count the edges to the other neighbor. Hence, we run the known-size algorithm once in each direction. An agent that comes from the other side can learn $n$ from one of its own records from the first run. The two-run algorithm uses $3f+5$ agents. Our lower bounds build executions that look the same to the surviving agents but place the gray hole on different nodes.

In the \emph{face-to-face} model, there are no whiteboards, and agents exchange messages only when they stand on the same node. A destroyed agent leaves no trace, and the only sign of a destruction is a \emph{missed meeting}, which only the agent that expected the missing one notices. The adversary can exploit the local detection of a missed meeting by destroying a second agent before the agent that noticed the first missed meeting can warn it, so that the last survivor cannot tell two positions of the gray hole apart. Our three-agent algorithm allows the warning to arrive in time by slowing the agents down. Before a missed meeting, each step of the schedule lasts $n$ rounds, and each scheduled move occurs in its last round. The waiting time lets an agent that notices a missed meeting walk around the ring, avoid the nodes that may be the gray hole, and warn the third agent before the third agent moves. For $f=0$, three agents are optimal, and we show that $f+1$ groups, each containing three agents, can tolerate $f$ Byzantine agents, requiring a total team size of $3f+3$ agents. Without $n$, the agents know neither how long a step must last nor how long the walk around the ring is. We show that four honest agents are then not enough. The adversary repeats the attack above far away from $\Home$, with two possible ring sizes instead of two possible nodes, so that the last two agents have identical local observations in three possible ring instances. Five honest agents suffice. Our five-agent algorithm requires no warning walk around the ring, because its three moving agents form a \emph{tight chain}: one stays ahead, one stays behind, and one moves between them. Every forward departure is observed, at the departure node, by a partner, that is, an agent that is scheduled to meet the departing agent. A survivor that has passed the gray hole walks on to $\Home$ and learns $n$ on the way, while a survivor that has not passed it has identified the gray hole exactly and returns to $\Home$, where two agents wait to repeat the chain from the other side.

\subsection{Model and Problem Definition}
\label{sec:model}

\textbf{The Network and the Agents.} The network is an oriented\footnote{The orientation only simplifies the description. All agents start at the home node $\Home$ and can agree there on which port leads clockwise. An agent keeps its direction by leaving each node through the port it did not enter, and reverses by leaving through the port it entered. Hence, our algorithms also work on unoriented rings.} ring $R$ of $n\ge3$ nodes, which we denote by $u_0,u_1,\dots,u_{n-1}$ in clockwise order. The nodes have no identifiers that the agents can read, so they are anonymous except for the \emph{home node} $\Home=u_0$, which is safe. Each node $u_i$ has two ports $\{0,1\}$, port $0$ leading clockwise ($\cw$) to $u_{(i+1)\bmod n}$ and port $1$ leading counterclockwise ($\ccw$) to $u_{(i-1)\bmod n}$. In the \emph{known-size} setting, every agent knows $n$. In the \emph{unknown-size} setting, the agents know neither $n$ nor any upper bound on $n$. Initially, they cannot distinguish the triangle ($n=3$) from a larger ring. An algorithm for the unknown-size setting must therefore work on every ring with $n\ge3$. The known-size triangle algorithms apply only when the agents know that $n=3$; being placed on a triangle does not tell them that $n=3$. A team of $N$ agents $\mathcal{A}=\{a_1,\dots,a_N\}$ starts at $\Home$ in round~$1$. The agents have distinct identifiers from a totally ordered set, and we index them in increasing order of identifier. Every agent initially knows the team size $N$ and the complete ordered list of agent identifiers. The agents use the common ordered list to assign scheduled turns and form groups. Every agent knows $f$, the maximum number of Byzantine agents. The honest agents execute the same deterministic algorithm and have unbounded local memory, which holds their internal state. Any number of agents may occupy a node or traverse an edge in the same round, in either direction. An agent recognizes the ports through which it enters and leaves a node. An agent sees nothing beyond its current node.

\textbf{Communication Models.} In the \emph{whiteboard model}, every node has an initially empty whiteboard of unbounded capacity, and whiteboards are the only means of communication. An agent can read a whiteboard and append records to it only while the agent stands on the node of the whiteboard. Each record carries the authenticated identifier of its writer, so an agent writes only under its own identifier. An agent can neither erase nor alter an existing record, nor recreate a record under another identifier. A record stays until an activation of the gray hole erases it, so records on safe nodes persist forever. We call this storage append-only because an agent may add records but cannot erase or change existing records. An agent cannot sense other agents, even on its own node, and learns about another agent only through the records of that agent. In the \emph{face-to-face model}, there are no whiteboards. Agents on the same node in the same round see each other's identifiers and exchange messages. Every agent on a node hears every message that an agent sends there, so no agent can tell different things to different agents on the same node. Agents that cross the same edge in opposite directions do not meet and cannot sense each other. In both models, an agent that the gray hole destroys leaves no trace.

\textbf{Synchrony.} The agents share a common clock and act in synchronous rounds. In each round, every surviving agent first acts at its node and then stays or traverses one incident edge. With whiteboards, the action at the node is a \emph{Read--Compute--Write} cycle. In the face-to-face model, the action is an exchange of messages followed by local computation. An agent that reaches a node at the end of a round acts there in the next round, provided that the agent survives. Computation, whiteboard access, and messages take no extra time. Synchrony lets an agent check whether an expected event happened by its deadline. With whiteboards, the expected event is a recorded return, and its deadline is the \emph{return deadline}. In the face-to-face model, the expected event is the presence of an agent on the node where the schedule puts that agent. Neither check requires sensing an agent on another node.

\textbf{Gray Hole.} One node $\Hole\neq\Home$, at clockwise distance $g$ from $\Home$, is the \emph{gray hole}, and all other nodes are safe. The location of $\Hole$ never changes, and the agents initially do not know it. At the beginning of each round, the adversary decides whether $\Hole$ is active in the round. An inactive gray hole behaves like a safe node. An active gray hole destroys every honest agent already on it, before the agent can act or move, and every honest agent that enters it during the round. With whiteboards, an active gray hole also erases its whiteboard. Whether $\Hole$ is active is the same for all honest agents in a round, so the adversary cannot spare one honest agent and destroy another on $\Hole$ in the same round.

\textbf{Byzantine Agents and the Adversary.} At most $f$ of the $N$ agents are Byzantine, and the others are honest and follow the algorithm. Honest agents do not initially know which agents are Byzantine. A Byzantine agent may deviate arbitrarily from the algorithm. A Byzantine agent may stay idle, move in either direction, or miss its return deadline. With whiteboards, a Byzantine agent may omit records or write false ones under its own identifier. In the face-to-face model, a Byzantine agent may send false messages or stay silent. A Byzantine agent still traverses at most one edge per round and cannot forge the identifier of another agent. A Byzantine agent may also \emph{stop acting}: it never moves, writes, or sends another message. In the face-to-face model, such an agent remains visible to agents on its node unless the gray hole destroys it. Whether an active gray hole destroys a Byzantine agent on it is up to the adversary. Unless the gray hole destroys it, a Byzantine agent stays somewhere on the ring. A single adversary controls the Byzantine agents and the gray hole. An execution is the sequence of rounds determined by the algorithm and the adversary, including the agents' states, moves, messages, and whiteboard records. The adversary knows the algorithm and observes the entire execution. The adversary fixes the location of the gray hole and the set of Byzantine agents before the execution begins. Then, in every round, the adversary chooses adaptively the actions of the Byzantine agents and whether the gray hole is active. The adversary may activate the gray hole in any number of rounds.

\begin{definition}\label{def:pe}
An agent visits a node when it reaches or occupies that node. The agents achieve \emph{perpetual exploration} (\PE) in $R$ if every node of $R\setminus\{\Hole\}$ is visited infinitely often by at least one honest agent.
\end{definition}

All algorithms in this paper are deterministic: the same internal state and observations lead an honest agent to the same action. We call an agent that has not been destroyed a surviving agent. An algorithm \emph{solves} \PE{} if it achieves \PE{} on every ring that the setting allows, for every position of the gray hole, every choice of the Byzantine agents, and every behavior of the adversary. The agents need not agree on, or report, the location of the gray hole. We measure the cost of an algorithm by the total team size $N$, including the Byzantine agents.

\textbf{Exploration After Identification.} Removing $\Hole$ leaves the \emph{safe path} of all $n-1$ safe nodes. Once an honest agent can recognize both endpoints of the safe path, the agent achieves \PE{} by the \emph{omission tour} of $\Hole$. In the omission tour, the agent walks back and forth between the two endpoints and never enters $\Hole$. If the agents know $n$, it suffices to identify one safe neighbor of $\Hole$ and the port leading to $\Hole$. An agent that has identified the neighbor and the port can then walk $n-2$ edges away from $\Hole$, turn, and repeat. If the agents do not know $n$, an agent cannot count $n-2$ edges. The agent must then learn enough to recognize both endpoints of the safe path or the length of the safe path.

\subsection{Our Contributions}
\label{sec:results}

We study \PE{} in the whiteboard model and in the face-to-face model, for known and unknown ring size $n$, and establish lower and upper bounds on the team size needed to tolerate a gray hole and up to $f$ Byzantine agents. In several cases, the lower and upper bounds agree. A simple baseline for Byzantine agents comes from algorithms for honest agents. Any algorithm for $k$ honest agents tolerates $f$ Byzantine agents with $k(f+1)$ agents, because the team splits into $f+1$ \emph{independent groups} of $k$ agents, at least one of which is entirely honest. Each group runs the algorithm and listens only to its own members, and since no agent can forge the identifier of another agent, the other groups cannot disturb the entirely honest group. Our results improve on the independent-group baseline as follows.

\begin{itemize}
  \item \emph{Whiteboard model, known $n$.} Goswami et al.~\cite{DBLP:conf/opodis/GoswamiBD024} showed that three honest agents are necessary and sufficient, which gives a natural extension with $3f+3$ agents. We show that $2f+2$ agents do not suffice on any ring with $n\ge4$, and we give an algorithm with $2f+3$ agents, which is therefore optimal. On the triangle ($n=3$), $2f+2$ agents are necessary and sufficient.
  \item \emph{Whiteboard model, unknown $n$.} The algorithm of Goswami et al.~\cite{DBLP:conf/opodis/GoswamiBD024} for known $n$ extends naturally to five honest agents. Three of them run it clockwise, and once the gray hole is identified, the survivor returns to $\Home$ and runs it counterclockwise with the two agents waiting there, which reveals the other neighbor of the gray hole. This gives a baseline of $5f+5$ agents. We show that $3f+3$ agents do not suffice, and we give an algorithm with $3f+5$ agents. Thus, an algorithm that works for all unknown ring sizes needs at least $f+1$ more agents than the optimal known-size algorithm for $n\ge4$.
  \item \emph{Face-to-face model, known $n$.} Goswami et al.~\cite{DBLP:conf/opodis/GoswamiBD024} showed that three honest agents are necessary and five suffice, and Bhattacharya et al.~\cite[Remark~9]{DBLP:conf/wdag/BhattacharyaGB025} improved the upper bound to four, which gives an extension of $4f+4$ agents. Bhattacharya et al.\ also asked whether three agents suffice. We answer this question with an algorithm for three honest agents on every ring with $n\ge4$, which is optimal, and independent groups of three improve the bound to $3f+3$ agents. As with whiteboards, $2f+2$ agents do not suffice on any ring with $n\ge4$. On the triangle, $2f+2$ agents are necessary and sufficient, and on rings of four or five nodes, $\lceil 5(f+1)/2\rceil$ agents suffice, which matches the lower bound of $2f+3$ agents for $f=0$ and $f=1$.
  \item \emph{Face-to-face model, unknown $n$.} Running the four-agent algorithm of Bhattacharya et al.~\cite{DBLP:conf/wdag/BhattacharyaGB025} for paths once in each direction gives a baseline of eight honest agents, and of $8f+8$ agents with independent groups. We show that four honest agents do not suffice, although three suffice when $n$ is known, and we give an algorithm for five honest agents, which is therefore optimal. Independent groups of five then give $5f+5$ agents. Our whiteboard proof that $3f+3$ agents do not suffice also holds in this model, so at least $\max\{5,3f+4\}$ agents are necessary.
\end{itemize}

\Cref{tab:summary} lists all our bounds next to earlier results (to compare precisely, we adapt earlier results to our settings).

\begin{table}[t]
\centering
\footnotesize
\crefname{theorem}{Thm.}{Thms.}\crefname{corollary}{Cor.}{Cors.}\crefname{proposition}{Prop.}{Props.}
\renewcommand{\crefpairconjunction}{, }\renewcommand{\crefmiddleconjunction}{, }\renewcommand{\creflastconjunction}{, }
\renewcommand{\crefpairgroupconjunction}{, }\renewcommand{\crefmiddlegroupconjunction}{, }\renewcommand{\creflastgroupconjunction}{, }
\setlength{\tabcolsep}{5pt}
\renewcommand{\arraystretch}{1.12}
\begin{tabular}{@{}lcccccc@{}}
\toprule
\textbf{Work} & \textbf{Ring size} & \textbf{Faulty node} & \textbf{Byz.\ agents} & \textbf{Task} & \textbf{Necessary} & \textbf{Sufficient} \\
\midrule
\rowcolor{gray!25}\multicolumn{7}{@{}c@{}}{\textit{Whiteboard model}} \\
Goswami et al.~\cite{DBLP:conf/opodis/GoswamiBD024} & known & BBH & none & \PE{} & $3$ & $3$ \\
Di Luna et al.~\cite{DBLP:conf/icdcs/LunaFPPPS25} & known & black hole & $\boldsymbol{f}$ & search & $2f+2^{*}$ & $2f+2^{*}$ \\
\rowcolor{gray!8}\textbf{\cref{prop:wb-triangle}} & $n=3$ & gray hole & $\boldsymbol{f}$ & \PE{} & $\boldsymbol{2f+2}$ & $\boldsymbol{2f+2}$ \\
\rowcolor{gray!8}\textbf{\cref{thm:2f+2,thm:known_n}} & known, $n\ge4$ & gray hole & $\boldsymbol{f}$ & \PE{} & $\boldsymbol{2f+3}$ & $\boldsymbol{2f+3}$ \\
Di Luna et al.~\cite{DBLP:conf/icdcs/LunaFPPPS25} & unknown & black hole & $\boldsymbol{f}$ & search & $3f+3^{*}$ & $3f+3^{*}$ \\
\rowcolor{gray!8}\textbf{\cref{thm:3f+3-unknown,thm:unknown_n}} & unknown & gray hole & $\boldsymbol{f}$ & \PE{} & $\boldsymbol{3f+4}$ & $\boldsymbol{3f+5}$ \\
\midrule
\rowcolor{gray!25}\multicolumn{7}{@{}c@{}}{\textit{Face-to-face model}} \\
Goswami et al.~\cite{DBLP:conf/opodis/GoswamiBD024} & known & gray hole & none & \PE{} & $3$ & $5$ \\
Bhattacharya et al.~\cite{DBLP:conf/wdag/BhattacharyaGB025} & known & gray hole & none & \PE{} & -- & $4$ \\
Di Luna et al.~\cite{DBLP:conf/icdcs/LunaFPPPS25} & known & black hole & $\boldsymbol{f}$ & search & $2f+2^{*}$ & $2f+2^{*}$ \\
\rowcolor{gray!8}\textbf{\cref{prop:f2f-triangle}} & $n=3$ & gray hole & $\boldsymbol{f}$ & \PE{} & $\boldsymbol{2f+2}$ & $\boldsymbol{2f+2}$ \\
\rowcolor{gray!8}\textbf{\cref{thm:f2f-lower,thm:inch3}} & known, $n\ge4$ & gray hole & none & \PE{} & $\boldsymbol{3}$ & $\boldsymbol{3}$ \\
\rowcolor{gray!8}\textbf{\cref{thm:f2f-lower,cor:f2f-known}} & known, $n\ge4$ & gray hole & $\boldsymbol{f}$ & \PE{} & $\boldsymbol{2f+3}$ & $\boldsymbol{3f+3}$ \\
\rowcolor{gray!8}\textbf{\cref{thm:f2f-lower,cor:f2f-small}} & $n\in\{4,5\}$ & gray hole & $\boldsymbol{f}$ & \PE{} & $\boldsymbol{2f+3}$ & $\boldsymbol{\lceil 5(f+1)/2\rceil}$ \\
Bhattacharya et al.~\cite{DBLP:conf/wdag/BhattacharyaGB025} & unknown & gray hole & none & \PE{} & -- & $9^{*}$ \\
Di Luna et al.~\cite{DBLP:conf/icdcs/LunaFPPPS25} & unknown & black hole & $\boldsymbol{f}$ & search & $3f+3^{*}$ & $6f+4^{*}$ \\
\rowcolor{gray!8}\textbf{\cref{thm:f2f-unknown-lower,thm:f2f-unknown-five}} & unknown & gray hole & none & \PE{} & $\boldsymbol{5}$ & $\boldsymbol{5}$ \\
\rowcolor{gray!8}\textbf{\cref{thm:f2f-unknown-byz-lower,cor:f2f-unknown-five}} & unknown & gray hole & $\boldsymbol{f}$ & \PE{} & $\boldsymbol{\max\{5,3f+4\}}$ & $\boldsymbol{5f+5}$ \\
\bottomrule
\end{tabular}
\caption{Our results and the closely related results from the literature; all results concern a ring of size $n$ with one faulty node and at most $f$ Byzantine agents. For unknown ring size, an algorithm must work on every ring with $n\ge3$. \PE{} denotes perpetual exploration; ``search'' means only locating the faulty node. Our bounds are in bold, next to the results that prove them. ``--'' means that the work states no lower bound for rings. BBH is the Byzantine black hole of Goswami et al., a gray hole that may also erase its whiteboard. ${}^{*}$We derive the bound from a bound for general graphs, with maximum degree $\Delta=2$. For Bhattacharya et al.\ with unknown $n$, running their four-agent path algorithm once in each direction gives eight agents instead of nine (\cref{sec:f2f-unknown}). In~\cite{DBLP:conf/icdcs/LunaFPPPS25}, the Byzantine agents survive the black hole.}
\label{tab:summary}
\end{table}

\subsection{Related Work}
\label{sec:related}

Dobrev et al.~\cite{DBLP:journals/algorithmica/DobrevFPS07} introduced the black-hole search problem and showed, with the cautious walk, that two agents that start together are necessary and sufficient on a ring with whiteboards. In 2-connected networks~\cite{DBLP:journals/dc/DobrevFPS06}, $\Delta+1$ agents are necessary in the worst case and suffice without a map, where $\Delta$ is the maximum degree, while two agents suffice with a map. With weaker communication, two agents still suffice with three tokens in total on a ring~\cite{DBLP:conf/ciac/DobrevKSS06} and with one pebble each in any network of known topology~\cite{DBLP:journals/algorithmica/FlocchiniIS12}. Our agents are synchronous, as in the work of Czyzowicz et al.~\cite{DBLP:journals/fuin/CzyzowiczKMP06,DBLP:journals/cpc/CzyzowiczKMP07}. There, an agent that misses its return deadline has been destroyed by the black hole, and the question becomes how fast two agents can find the black hole. All of these results rest on the fact that the black hole never spares a visitor. Hence, a single safe return proves that a node is safe.

Kr\'alovi\v{c} and Mikl\'ik~\cite{DBLP:conf/sirocco/KralovicM10} introduced the gray hole, a node that may destroy any agent at any time but otherwise behaves as a safe node. For periodic data retrieval by asynchronous agents on a ring with whiteboards, a task close to perpetual exploration, they showed that two agents are not enough, while nine agents suffice. Bampas et al.~\cite{DBLP:journals/tcs/BampasLMPP15} later showed that four agents are necessary and sufficient. Goswami et al.~\cite{DBLP:conf/opodis/GoswamiBD024} introduced the \emph{Byzantine black hole}, a gray hole that may also erase its whiteboard when it acts as a black hole, that is, when it is active. They studied \PE{} of rings of known size by synchronous agents. For agents that start together, they showed that three agents are necessary and sufficient with whiteboards, while in the face-to-face model at least three agents are necessary and five agents suffice. Bhattacharya et al.~\cite{DBLP:conf/wdag/BhattacharyaGB025} studied the face-to-face model on arbitrary networks of unknown size and topology. They consider two versions of the task: the agents must perpetually explore at least one of the components that remain after the gray hole is removed, or they must perpetually explore the component that contains the home node. In trees, four agents are necessary and sufficient to explore one component, and six agents are necessary and sufficient to explore the component of the home node~\cite{DBLP:conf/wdag/BhattacharyaGB025}. In general graphs, $3\Delta+3$ agents suffice for the component of the home node. When the agents do not know the graph, $2\Delta-1$ agents are necessary for some graphs with $\Delta\ge 4$. Their four-agent algorithm also improves the face-to-face bound of Goswami et al.~\cite{DBLP:conf/opodis/GoswamiBD024} for rings of known size to four agents. Bhattacharya et al.~\cite{DBLP:conf/wdag/BhattacharyaGB025} asked whether three agents suffice.

All of these works assume that every agent is honest. Earlier work studied Byzantine agents for tasks in safe networks, such as gathering~\cite{DBLP:journals/talg/DieudonnePP14}. For gathering, Dieudonn\'e et al.~\cite{DBLP:journals/talg/DieudonnePP14} distinguish weakly Byzantine agents, which cannot change their identifier, from strongly Byzantine agents, which can. Tsuchida et al.~\cite{DBLP:conf/sss/TsuchidaOI19} studied Byzantine agents in black-hole search. Their synchronous agents start together, know a map of the network, and communicate when they meet. In that work, the Byzantine agents may survive the black hole or disappear anywhere on their own. They showed that $2f+1$ agents are not enough, even when the Byzantine agents only crash, and that $2f+2$ agents suffice. Di Luna et al.~\cite{DBLP:conf/icdcs/LunaFPPPS25} let weakly Byzantine agents collude with the black hole and be immune to it. With a map, $2f+2$ agents are necessary and sufficient, in the whiteboard model and in the face-to-face model alike. Without a map, $(f+1)(\Delta+1)$ agents are necessary in both models. These agents suffice with whiteboards, and $(f+1)(\Delta+1)+3f+1$ agents suffice in the face-to-face model. See a recent survey~\cite{Pattanayak2026} for other variants. To the best of our knowledge, no earlier work lets Byzantine agents face a gray hole.

\section{Perpetual Exploration in the Whiteboard Model}
\label{sec:whiteboard}

In the whiteboard model, an activation of the gray hole destroys the honest agents on the gray hole and erases the whiteboard of the gray hole. The records that a destroyed agent wrote on safe nodes remain. In this section, we first show the limits of the evidence that these records give, with lower bounds for known and unknown $n$. We then give an algorithm for known $n$ that matches the lower bound and therefore uses the minimum possible number of agents. For unknown $n$, our algorithm uses $3f+5$ agents, one more than the $3f+4$ agents that the lower bound requires.

\subsection{Limits of Whiteboard Evidence}
\label{sec:wb-impossibility}

Both lower bounds compare executions that place the gray hole on different nodes. The gray hole destroys every agent that could tell these executions apart. In an execution in which the node where such an agent would be destroyed is safe, the adversary instead chooses this agent to be Byzantine, and the agent stops acting at that node.

\subsubsection{Lower Bounds for Known Ring Size}
\label{sec:impossibility}
The lower bound uses two neighboring possible locations of the gray hole. An honest agent that could distinguish the locations is destroyed, while Byzantine agents reproduce the same absence in the other execution. A ring with at least four nodes provides the distinct nodes needed for the construction.
\begin{theorem}\label{thm:2f+2}
Suppose that $n\ge4$, that the agents know $n$, and that at most $f$ agents are Byzantine. Then no algorithm solves \PE{} in the whiteboard model with $2f+2$ or fewer agents.
\end{theorem}

\begin{figure}[t]
\centering
\begin{tikzpicture}[x=1.15cm,y=0.9cm]
  \foreach \row/\name/\g in {0/1/1,-1.72/2/2}{
    \node[anchor=east] at (-0.75,\row) {$I_{\name}$};
    \foreach \i/\lab in {0/\Home,1/v,2/w,3/x}{
      \ifnum\i=\g
        \node[gh] (b\name-\i) at (\i,\row) {$\lab$};
      \else\ifnum\i=0
        \node[home] (b\name-\i) at (\i,\row) {$\lab$};
      \else
        \node[vtx] (b\name-\i) at (\i,\row) {$\lab$};
      \fi\fi}
    \draw[edge] (b\name-0)--(b\name-1)--(b\name-2)--(b\name-3);
    \draw[omit] (b\name-0)--(-0.5,\row);
    \draw[omit] (b\name-3)--(3.5,\row);
    \begin{scope}[shift={(0,\row)}]
      \draw[move] (0.13,-0.32)--(0.87,-0.32);
      \draw[move] (1.13,0.32)--node[above] {$M$} (1.87,0.32);
      \draw[move] (2.87,-0.32)--node[below] {$X$} (2.13,-0.32);
    \end{scope}}
\end{tikzpicture}
\caption{The two executions of \cref{thm:2f+2}, one in each row. The home node is $u=\Home$.
The dark-gray node is the gray hole, $v$ in $I_1$ and $w$ in $I_2$.
The agents of $M$ are scheduled to enter the unvisited node
$w$ from $v$. They are destroyed at $v$ before moving in
$I_1$, and upon entering $w$ in $I_2$.
The left arrow shows later visits to $v$.
The agents of $X$ are scheduled to enter $w$ from $x$
in the same round as $M$.
Each agent shown on the two outer arrows is destroyed
in one execution. In the other, the adversary chooses it to be Byzantine, and it stops acting.
Thus, all whiteboards except that of $v$ have the same
records in both executions.}
\label{fig:uvwx-chain}
\end{figure}
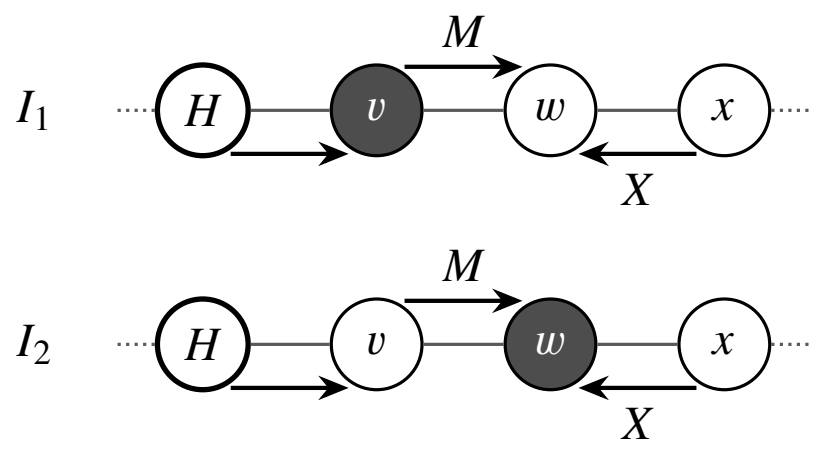

Fix a deterministic algorithm, and let $E_0$ be its execution in which all agents are honest and the gray hole is never active. Since the algorithm solves \PE{} wherever the gray hole is, every node is visited in $E_0$. Let $r$ be the first round of $E_0$ in which an agent moves from a node $v\neq\Home$ to a node $w$ that no agent has visited before. Round $r$ exists because $n\ge4$, so some node is not a neighbor of $\Home$, and the first agent that enters it comes from a node other than $\Home$. By symmetry, we assume that the move from $v$ to $w$ is clockwise. Let $u$ be the counterclockwise neighbor of $v$, and $x$ the clockwise neighbor of $w$. Every node visited before round $r$ was first entered from $\Home$, so $v$ is a neighbor of $\Home$ and $u=\Home$. Since $n\ge4$, the four nodes $u,v,w,x$ of \cref{fig:uvwx-chain} are distinct. Let $M\neq\emptyset$ be the agents that move from $v$ to $w$ in round $r$, and $X$ the agents that move from $x$ to $w$ in round $r$.

We build the two executions $I_1$ and $I_2$ shown in the two rows of \cref{fig:uvwx-chain}. Both copy $E_0$ before round $r$. We choose different gray-hole locations, activation schedules, and Byzantine agents so that the continuing honest agents see the same records. In $I_1$, the gray hole is $v$, inactive before round $r$ and active from round $r$ on. In $I_2$, the gray hole is $w$, active in round $r$ and inactive in every other round. The agents of $M$, which take the middle arrow of the figure, are destroyed in both executions, at $v$ before they can move in $I_1$, and on entering $w$ in $I_2$. Every other agent that stands on $v$ in round $r$, or enters $v$ in round $r$ or later, from $u$ or from $w$, is destroyed at $v$ in $I_1$. In $I_2$, the node $v$ is safe, so the adversary chooses such an agent to be Byzantine, and the agent stops acting at $v$. For the agents on the other nodes, such an agent disappears exactly as in $I_1$. Symmetrically, the agents of $X$, which take the right arrow, are destroyed on entering $w$ in $I_2$, while in $I_1$ the adversary chooses them to be Byzantine, and they stop acting at $w$. All other agents follow the algorithm in both executions. The adversary may fix these Byzantine agents in advance, because a Byzantine agent may behave like an honest one until it stops acting, and, as the next lemma shows, the agents that enter $v$ are the same in both executions.

\begin{lemma}\label{lem:two-executions}
In the executions $I_1$ and $I_2$ constructed above, the \emph{view} of an agent consists of its own state and everything it has read on whiteboards. Consider a round $t\ge r$ and an agent $a$ that follows the algorithm in both executions through round $t$, is not in $X$, and neither stands on $v$ at the start of round $r$ nor enters $v$ in any of the rounds $r,\dots,t$. Then, up to round $t$, agent $a$ has the same view and makes the same moves in $I_1$ and $I_2$.
\end{lemma}

\begin{proof}
The two executions treat only the nodes $v$ and $w$ differently, so it suffices to compare the whiteboards of these two nodes. At the end of round $r$, the whiteboard of $w$ is empty in both executions. No agent visited $w$ before round $r$, and in round $r$ every agent that enters $w$ is destroyed or stops acting there. After round $r$, the node $w$ is safe in $I_1$ and is an inactive gray hole in $I_2$, so it behaves as a safe node in both executions. Moreover, destroyed agents and agents that have stopped acting write nothing. Hence, by induction on the rounds, every whiteboard other than that of $v$ carries the same records in $I_1$ and $I_2$. An agent that does not enter $v$ reads only these whiteboards, so it has the same view in both executions, and since the algorithm is deterministic, it also makes the same moves.
\end{proof}

\begin{proof}[Proof of \cref{thm:2f+2}]
Let $D$ be the set of agents that visit $v$ from round $r$ on, that is, the agents of $M$ together with every agent that enters $v$ from $u$ in round $r$ or later (the left arrow of \cref{fig:uvwx-chain}), or stands on $v$ in round $r$, or enters $v$ from $w$ afterwards. By \cref{lem:two-executions}, the same agents enter $v$, at the same rounds, in both executions. In $I_1$, every agent of $D$ is destroyed at $v$. In $I_2$, the agents of $M$ are destroyed at $w$, and the other agents of $D$ are exactly the agents that the adversary chooses to be Byzantine at $v$. Suppose first that $|D|\ge f+2$. Then at most $f$ agents lie outside $D$, and the adversary chooses all of them to be Byzantine in $I_1$. These agents include the agents of $X$. Every honest agent of $I_1$ then belongs to $D$ and is destroyed, so \PE{} fails in $I_1$. Suppose next that $|D|\le f+1$. Since $M$ is not empty, at most $f$ agents are chosen to be Byzantine in $I_2$. In $I_2$, the node $v$ is safe, but after round $r$ every agent that reaches $v$ is Byzantine, so no honest agent visits $v$ again, and \PE{} fails in $I_2$. In both cases, the failing execution has at most $f$ Byzantine agents. If it has fewer, the adversary chooses further agents to be Byzantine and lets them follow the algorithm, which changes nothing.
\end{proof}

For $f=0$, \cref{thm:2f+2} gives the lower bound of three agents proved by Goswami et al.~\cite{DBLP:conf/opodis/GoswamiBD024} for the Byzantine black hole. The proof also shows why one more agent helps. With $2f+3$ agents and $|D|=f+2$, the $f+1$ agents outside $D$ include at least one honest agent. Our algorithm (\cref{sec:algorithms_known_n}) uses the survival of an honest agent beyond the destruction bound. On the triangle ($n=3$), both nodes other than $\Home$ are neighbors of $\Home$, so an algorithm can first visit each node from $\Home$, and the four distinct nodes $u,v,w,x$ of \cref{fig:uvwx-chain} do not exist. \Cref{prop:wb-triangle} shows that $2f+2$ agents then suffice.
On a known triangle, we partition the identifiers into two groups of $f+1$ agents. The first group repeatedly walks from $\Home$ to $u_1$ and back, and the second group repeatedly walks from $\Home$ to $u_2$ and back. Each agent stores its assigned neighbor $v$ and alternates between $\Home$ and $v$; no whiteboard records are needed. The schedule is given in \cref{alg:triangle}, and the following proposition proves that one honest agent survives in the group assigned to the safe neighbor, while fewer agents cannot suffice.

\begin{algorithm}[tbp]
\caption{Exploration of a known triangle by an honest agent $a_i$.}
\label{alg:triangle}
\begin{algorithmic}[1]
\State set $v\gets u_1$ if $i\le f+1$, and $v\gets u_2$ otherwise
\Loop
  \State move from node $\Home$ to node $v$ in the current round
  \State return to $\Home$ in the next round
\EndLoop
\end{algorithmic}
\end{algorithm}
\begin{proposition}\label{prop:wb-triangle}
Suppose that the agents know that $n=3$ and that at most $f$ agents are Byzantine. Then, in the whiteboard model, $2f+2$ agents are necessary and sufficient for \PE{}.
\end{proposition}

\begin{proof}
For sufficiency, use the two-group schedule described above. The group whose neighbor is safe is never destroyed, and it contains an honest agent, which visits both safe nodes infinitely often. For necessity, fix a correct algorithm with $N$ agents. Let all agents be honest, and let the gray hole be $u_1$, active in every round. Let $D$ be the set of agents that ever enter $u_1$ in this execution, and $S$ the others. If $|S|\le f$, the adversary chooses the agents of $S$ to be Byzantine and lets them follow the algorithm, which changes nothing, and every honest agent then enters $u_1$ and is destroyed. If $|D|\le f$, we consider instead the execution in which the gray hole is $u_2$ and is never active, and in which the adversary chooses the agents of $D$ to be Byzantine. Each of them follows the algorithm until it enters $u_1$, and then stops acting at $u_1$. In both executions, the agents of $D$ write nothing at $u_1$ after entering it, and the agents of $S$ never read the whiteboard of $u_1$. Hence, by induction on the rounds, the whiteboards of $\Home$ and $u_2$ carry the same records in both executions, and the agents of $S$ act in the same way. So, in the second execution, no honest agent ever visits the safe node $u_1$. In both cases, \PE{} fails, so $|D|\ge f+1$, $|S|\ge f+1$, and $N\ge 2f+2$.
\end{proof}

\subsubsection{Lower Bounds for Unknown Ring Size}
\label{sec:impossibility_unknown}
\begin{figure}[t]
\centering
\begin{tikzpicture}
\def\R{1.15}
\foreach \xoffset in {0,4.2,8.4}{
  \begin{scope}[xshift=\xoffset cm]
    \draw[edge] (210:\R) arc (210:450:\R);
    \draw[omit] (90:\R) arc (90:210:\R);
    \node[home] at (90:\R) {$\Home$};
    \node[lab,align=center] at (-0.42,0.2) {$d-1$\\edges};
  \end{scope}
}
\begin{scope}[xshift=0cm]
  \node[vtx]    at (210:\R) {$w$};
  \node[gh,far] at (270:\R) {$v$};
  \node[lab] at (0,-1.85) {$I_1$: $d+1$ nodes, $z=v$};
\end{scope}
\begin{scope}[xshift=4.2cm]
  \node[vtx]    at (210:\R) {};
  \node[vtx]    at (0:\R) {$v$};
  \node[gh,far] at (270:\R) {$w$};
  \node[lab] at (0,-1.85) {$I_2$: $d+2$ nodes, $z=w$};
\end{scope}
\begin{scope}[xshift=8.4cm]
  \node[vtx]     at (210:\R) {};
  \node[gh]      at (30:\R) {$v$};
  \node[vtx]     at (330:\R) {$w$};
  \node[vtx,far] at (270:\R) {$z$};
  \node[lab] at (0,-1.85) {$I_3$: $d+3$ nodes, $z$ safe};
\end{scope}
\end{tikzpicture}
\caption{The three executions of \cref{thm:3f+3-unknown} for general $d$, with clockwise to the right of $\Home$. Dark-gray nodes mark the gray hole, and double borders mark $z$. Each dotted path has $d-1$ edges and $d-2$ omitted internal nodes. One more solid edge leads to $z$, giving counterclockwise distance $d$ from $\Home$ in every ring.}
\label{fig:three-rings}
\end{figure}
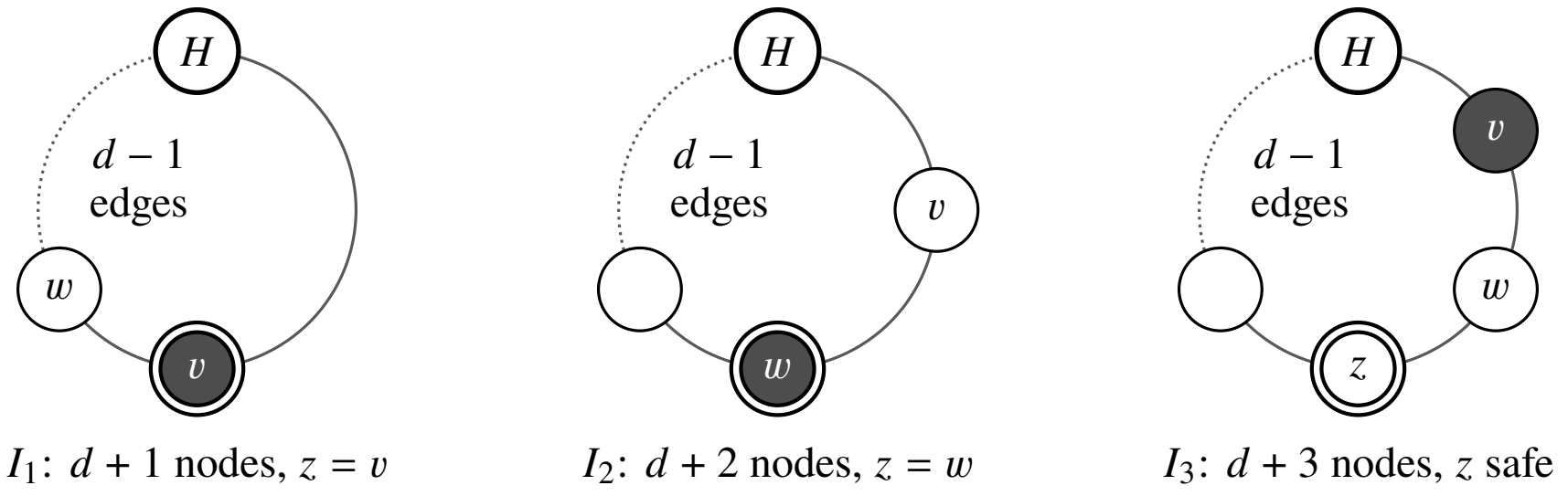

When $n$ is known, an agent that locates the gray hole from one side can count along the ring and locate it from the other side as well. Without $n$, the agent cannot determine the distance to the other side by counting. We use the uncertainty about that distance to obtain a stronger lower bound. We compare three rings of different sizes (\cref{fig:three-rings}). In each of them, some agents must enter the clockwise neighbor $v$ of $\Home$ from $\Home$, and some agents must enter the node $z$, at a fixed counterclockwise distance from $\Home$, while they walk counterclockwise, since either node may be safe. In the smallest ring, however, $v$ and $z$ are the same node, the gray hole, so the gray hole destroys the agents that make either move.

\begin{theorem}\label{thm:3f+3-unknown}
Suppose that the agents do not know $n$ and that at most $f$ agents are Byzantine. Then no algorithm solves \PE{} in the whiteboard model with $3f+3$ or fewer agents. This remains true even if the agents know that $n$ is at least some fixed number $n_0$.
\end{theorem}

\begin{proof}
Fix a deterministic algorithm for at most $3f+3$ agents and an integer $d\ge\max\{3,n_0\}$. Define $E_0$, $r$, $v$, $w$, and $M$ as in the proof of \cref{thm:2f+2}. Thus $r$ is the first round in which an agent moves from a node $v\neq\Home$ to a node $w$ that no agent has visited before, we assume that this move is clockwise, and $M$ is the nonempty set of agents that make it. Before round $r$, agents move only between $\Home$ and its two neighbors, so up to round $r$ the execution is the same on every ring with at least four nodes. Let $z$ be the node at counterclockwise distance $d$ from $\Home$.

We consider the three executions of \cref{fig:three-rings}, which agree with $E_0$ until round $r$. In $I_1$, the ring has $d+1$ nodes and the gray hole is $v$, which is also $z$. In $I_2$, the ring has $d+2$ nodes and the gray hole is $w$, which is also $z$. In $I_3$, the ring has $d+3$ nodes, the gray hole is $v$, and $z$ is safe. In $I_1$ and $I_3$, the gray hole is active in round $r$ and in every later round. In $I_2$, it is active in round $r$ and in every later round in which an agent enters it. The agents of $M$ are destroyed in all three executions, at $v$ in $I_1$ and $I_3$, and on entering $w$ in $I_2$. Let $P$ be the set of the other agents that are at $v$ in round $r$ or move from $\Home$ to $v$ in round $r$ or later. They are destroyed at $v$ in $I_1$ and $I_3$. In $I_2$, the node $v$ is safe. In the construction of $I_2$, the adversary chooses every agent of $P$ to be Byzantine, and each such agent stops acting at node $v$ as soon as it joins $P$. This construction is a valid execution with at most $f$ Byzantine agents whenever $|P|\le f$; the counting argument below selects a valid failing execution. Let $F$ be the set of agents that move into $z$ from the node at counterclockwise distance $d-1$. They are destroyed in $I_1$ and $I_2$, where $z$ is the gray hole. In $I_3$, where $z$ is safe, the adversary chooses them to be Byzantine, and they stop acting at $z$. All other agents follow the algorithm in all three executions.

These other agents cannot tell the three executions apart. From round $r$ on, such an agent stays on $\Home$ and on the nodes at counterclockwise distance $1,\dots,d-1$. In each of the three rings, these are $d$ safe nodes, and the only edges that leave them lead to $v$ and to $z$, where every arriving agent is destroyed or stops acting. Destroyed agents and agents that have stopped acting write nothing. Hence, by induction on the rounds, these $d$ nodes carry the same records in $I_1$, $I_2$, and $I_3$, and every agent outside $M\cup P\cup F$ has the same view, and makes the same moves, in all three executions. Call these $d$ safe nodes the common safe path. The equal views and moves on the common safe path determine the same sets $M$, $P$, and $F$ in all three constructions. The sets are disjoint: an agent never returns to the common safe path after it joins one of them. Determinism therefore fixes all three sets before the adversary selects the failing instance; agents are not reassigned as Byzantine during an execution.

We now count. If $|F|\le f$, then $I_3$ has at most $f$ Byzantine agents. The safe node $z$ is entered only by them, so no honest agent ever visits $z$, and \PE{} fails in $I_3$. If $|P|\le f$, then $I_2$ has at most $f$ Byzantine agents. The node $v$ is safe in $I_2$, but after round $r$ every agent that reaches it from $\Home$ belongs to $P$ and is Byzantine. An agent approaching from the other side is destroyed at $w$ before reaching $v$. Hence honest agents visit $v$ only finitely often, and \PE{} fails in $I_2$. Otherwise, $|F|\ge f+1$ and $|P|\ge f+1$, and since $M$ is not empty, at least $2f+3$ agents are destroyed in $I_1$. The adversary chooses the at most $f$ remaining agents to be Byzantine, so every honest agent of $I_1$ is destroyed, and \PE{} fails in $I_1$. In each case, the failing execution has at most $f$ Byzantine agents. If it has fewer, the adversary chooses further agents to be Byzantine and lets them follow the algorithm, which changes nothing.
\end{proof}

For $f=0$, \cref{thm:3f+3-unknown} says that three honest agents cannot explore a ring of unknown size, although three suffice when $n$ is known~\cite{DBLP:conf/opodis/GoswamiBD024}. The proof uses only the three sizes $d+1$, $d+2$, and $d+3$, for any large $d$, so knowing a lower bound on $n$ does not help. With $3f+4$ agents, the last case of the counting may leave $f+1$ agents alive in $I_1$, at least one of which is honest, so the three-ring argument gives no bound above $3f+3$. The same lower bound holds in the face-to-face model, with the same three rings (\cref{thm:f2f-unknown-byz-lower}).

\subsection{Exploration from Whiteboard Records}
\label{sec:wb-algorithms}

\subsubsection{Optimal Exploration with Known Ring Size}
\label{sec:algorithms_known_n}
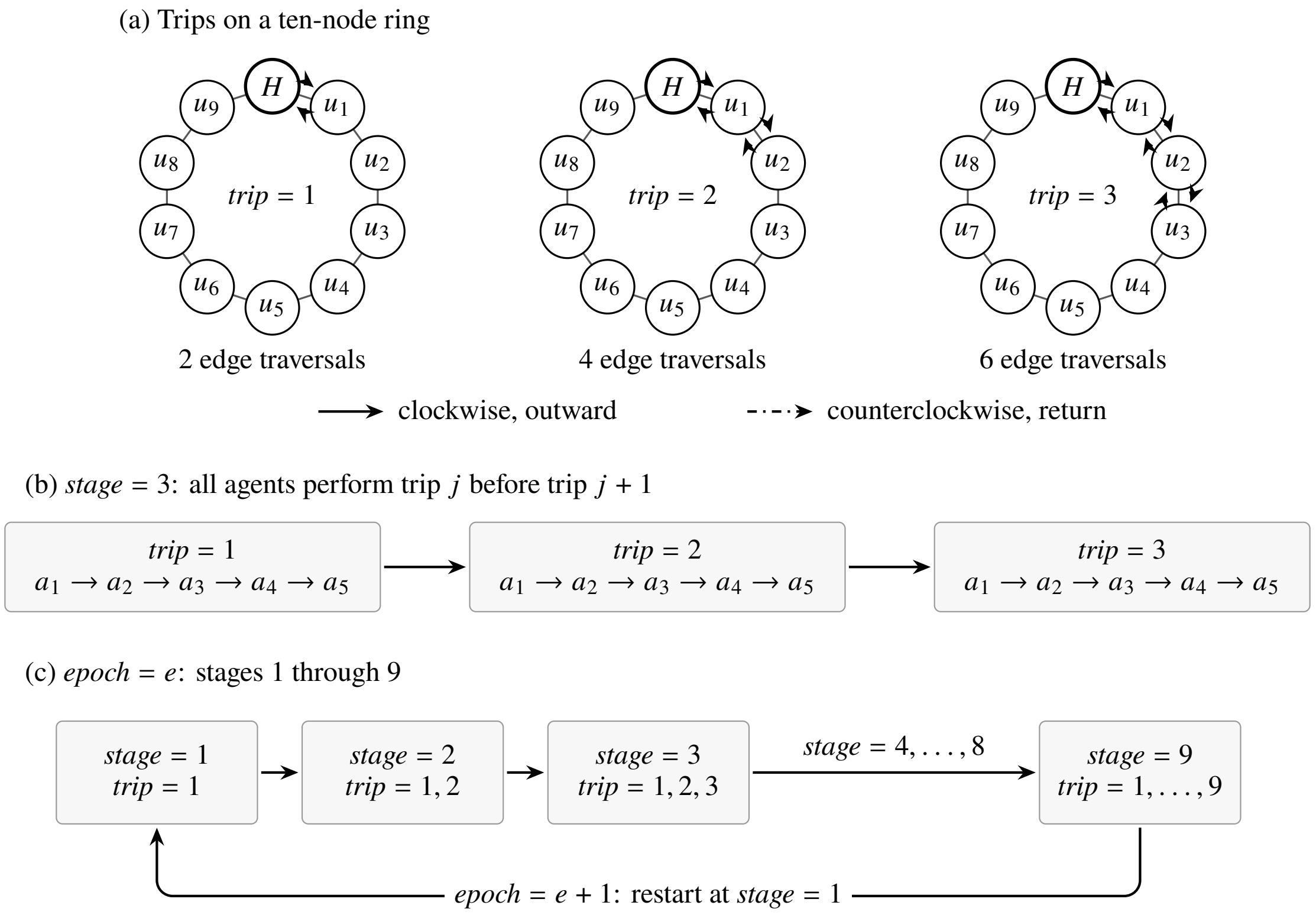
\begin{figure}[t]
\centering
\begin{tikzpicture}[x=0.8cm,y=0.8cm]
\node[ttl] at (-2.3,2.45) {(a) Trips on a ten-node ring};
\foreach \trip/\xx in {1/0,2/5.6,3/11.2}{
  \begin{scope}[shift={(\xx,0)}]
    \foreach \v in {0,...,9}{\coordinate (p\v) at ({90-36*\v}:1.55);}
    \foreach \v in {0,...,9}{
      \pgfmathtruncatemacro{\next}{mod(\v+1,10)}
      \draw[edge] (p\v) -- (p\next);
    }
    \foreach \v in {1,...,9}{\node[vtx] (v\v) at (p\v) {$u_{\v}$};}
    \node[home] (v0) at (p0) {$\Home$};
    \foreach \v in {1,...,\trip}{
      \pgfmathtruncatemacro{\prev}{\v-1}
      \draw[outw] (v\prev) to[bend left=22] (v\v);
      \draw[retw] (v\v) to[bend left=22] (v\prev);
    }
    \node[lab] at (0,0) {$\mathit{trip}=\trip$};
    \pgfmathtruncatemacro{\steps}{2*\trip}
    \node[lab] at (0,-2.3) {$\steps$ edge traversals};
  \end{scope}
}
\draw[outw] (0.6,-3.0) -- (1.6,-3.0) node[right,lab,text=black] {clockwise, outward};
\draw[retw] (6.6,-3.0) -- (7.6,-3.0) node[right,lab,text=black] {counterclockwise, return};
\end{tikzpicture}\\[3mm]
\begin{tikzpicture}
\node[ttl] at (-2,0) {(b) $\mathit{stage}=3$: all agents perform trip $j$ before trip $j+1$};
\foreach \j/\xx in {1/0,2/5.2,3/10.4}{
  \node[sbox,minimum width=4.2cm,minimum height=1cm] (t\j) at (\xx,-0.9) {$\mathit{trip}=\j$\\$a_1\to a_2\to a_3\to a_4\to a_5$};
}
\draw[move] (t1.east) -- (t2.west);
\draw[move] (t2.east) -- (t3.west);
\node[ttl] at (-2,-2.1) {(c) $\mathit{epoch}=e$: stages $1$ through $9$};
\foreach \k/\xx/\trips in {1/-.4/1, 2/2.35/{1,2}, 3/5.1/{1,2,3}, 9/10.6/{1,\ldots,9}}{
  \node[sbox,minimum width=2.25cm,minimum height=1.15cm] (s\k) at (\xx,-3.2) {$\mathit{stage}=\k$\\$\mathit{trip}=\trips$};
}
\draw[move] (s1.east) -- (s2.west);
\draw[move] (s2.east) -- (s3.west);
\draw[move] (s3.east) -- node[above,lab] {$\mathit{stage}=4,\ldots,8$} (s9.west);
\draw[move,rounded corners=5pt] (s9.south) -- ++(0,-0.8) -| (s1.south);
\node[lab,fill=white,inner sep=3pt] at (5.1,-4.58) {$\mathit{epoch}=e+1$: restart at $\mathit{stage}=1$};
\end{tikzpicture}
\caption{The schedule on a ten-node ring with five agents, without missed returns. (a) On trip $j$, an agent walks $j$ edges clockwise (solid arrows) and back (dash-dotted arrows). (b) In a stage, all agents perform trip $1$ in identifier order, then trip $2$, and so on. (c) Stage $k$ consists of trips $1,\ldots,k$, and after stage $n-1=9$, a new epoch starts.}
\label{fig:movement-pattern}
\end{figure}

As noted in~\cref{sec:results}, an immediate way to solve \PE{} with $3f+3$ agents when $n$ is known is to create $f+1$ independent groups, each executing the three-agent algorithm of Goswami et al.~\cite{DBLP:conf/opodis/GoswamiBD024}. However, in such an algorithm, no group uses what the other groups learn. In our new algorithm, all agents work together. Still, no honest agent relies on the records of a single other agent. An absent return record by the scheduled deadline or a missing record counts as evidence only if it involves $f+1$ distinct agents, or if the missing record is the honest agent's own. The agents explore clockwise from $\Home$ in trips of growing length, one agent at a time, and every agent writes records on the nodes it visits. The repeated trips expose the gray hole as follows. Recall that the gray hole is $\Hole=u_g$, so $g$ is its clockwise distance from $\Home$. When the gray hole $u_g$ destroys an honest agent on a trip $j>g$, every honest agent that is still in the schedule has already left a record on $u_g$. The activation erases every record on $u_g$. From then on, an honest agent that returns to $u_g$ and survives finds its own record missing, and thus identifies $u_g$ as the gray hole. To keep the honest agents from identifying the gray hole, the adversary must activate the gray hole whenever an honest agent returns to $u_g$, so the gray hole destroys the agent. Each destruction prevents the honest agent from returning by its deadline. After $f+1$ missed returns on the same trip, the surviving agents identify $u_g$ without entering it. Byzantine agents can miss returns on purpose. However, they miss at most $f$ returns in total, so they cannot reach the threshold of $f+1$ on their own. We first describe the movement pattern of the agents.
On \emph{trip} $j$, an agent walks clockwise from $\Home$ to $u_j$ and returns to $\Home$ along the same path. \emph{Stage} $k$ consists of trips $1,\dots,k$ in this order. The \emph{participants}, that is, the agents that remain in the schedule, perform each trip one after another, in increasing order of identifier, so only one agent is scheduled to be away from $\Home$ at any time. An \emph{epoch} consists of stages $1,\dots,n-1$, after which a new epoch starts (\cref{fig:movement-pattern}). The rounds that the schedule reserves for one trip of one agent form the \emph{turn} of that agent. A turn has a fixed length, even if the agent returns early. Hence every agent knows the schedule and the \emph{return deadline} of every turn, that is, the round by which the agent of the turn must be back at $\Home$.

On every node it visits, an agent writes an arrival record and a departure record. A record written by agent $a_i$ in round $t$ has the form
$$
\rho^{a_i}_t := \langle e,k,j \mid \mathit{event},\mathit{dir} \mid \mathit{state} \rangle .
$$
The \emph{schedule descriptor} $(e,k,j)$ gives the epoch, the stage, and the trip. The fields $\mathit{event}\in\{\mathsf{arr},\mathsf{dep}\}$ and $\mathit{dir}\in\{\mathsf{cw},\mathsf{ccw}\}$ tell whether the record marks an arrival or a departure, and in which direction its writer travels. The $\mathit{state}$ is $\mathsf{exploring}$ during the trip and becomes $\mathsf{returned}$ in the \emph{return record}, that is, the arrival record that the agent writes at $\Home$ at the end of the trip. Every record carries the identifier of its writer, which no agent can alter. A Byzantine agent may omit records or fill the other fields of its own records with false values. The round $t$ is not part of the record, and we use it only in the analysis. We write $\mathcal{W}_u^t$ for the set of records present on the whiteboard of node $u$ when it is read in round $t$, that is, the records written at $u$ before round $t$ and not erased since.

Consider agent $a_i$ on trip $j$ in stage $k$ of epoch $e$. When it reaches $u_\ell$ on its way out and survives, it makes two checks before it writes its arrival record there. The first check uses its own records. Agent $a_i$ identifies $u_\ell$ as the gray hole if some record $\rho^{a_i}_{t'}$ that it wrote at $u_\ell$ earlier is missing, that is, $\rho^{a_i}_{t'}\notin\mathcal{W}_{u_\ell}^{t}$ for the current round $t$. The second check uses the records of other agents, which $a_i$ read at $u_{\ell-1}$ in round $\tau=t-1$, just before it moved from $u_{\ell-1}$ to $u_\ell$. Let $S$ be the set of agents $a_h$ such that $\mathcal{W}_{u_{\ell-1}}^{\tau}$ contains a record $\rho^{a_h}_s=\langle e',k',j'\mid\mathsf{arr},\mathsf{ccw}\mid\sigma\rangle$ with $\sigma\in\{\mathsf{exploring},\mathsf{returned}\}$ and $j'\ge \ell$, but $\mathcal{W}_{u_\ell}^{t}$ contains no record $\langle e',k',j'\mid\mathsf{arr},\mathsf{cw}\mid\mathsf{exploring}\rangle$ of $a_h$. In words, $S$ contains the agents that, according to these records, came back from $u_\ell$ or from beyond it on some trip, but whose arrival record from that trip is missing on $u_\ell$. If $|S|\ge f+1$, agent $a_i$ identifies $u_\ell$ as the gray hole. If neither check identifies the gray hole, $a_i$ writes $\langle e,k,j\mid\mathsf{arr},\mathsf{cw}\mid\mathsf{exploring}\rangle$ at $u_\ell$ and continues. When it is back at $\Home$, it writes its return record $\langle e,k,j\mid\mathsf{arr},\mathsf{ccw}\mid\mathsf{returned}\rangle$.

A turn for trip $j$ that starts in round $s$ consists of the $2j+2$ rounds $s,\dots,s+2j+1$. The agent leaves $\Home$ in round $s$. If it completes the trip, it is back at $\Home$ at the end of round $s+2j-1$ and writes its return record in round $s+2j$, which is the return deadline, and the agents waiting at $\Home$ read the return record in round $s+2j+1$.
A \emph{withdrawal notification} is a record at $\Home$ containing the writer's schedule descriptor $(e,k,j)$ and its stated clockwise distance of the gray hole. It replaces the return record when the agent identifies or accepts the gray-hole location; the writing and withdrawal rules are specified below.

In the last round of each turn, the agents at $\Home$ look for the return record or the withdrawal notification of the scheduled agent, with its descriptor $(e,k,j)$, written by the return deadline. If there is none, the agent has \emph{missed its return} on trip $j$. The algorithm removes the agent from the schedule permanently, so each agent misses at most one return. The participants then start a new epoch at stage~$1$ and keep all records, so that a participant never repeats a trip with the same schedule descriptor $(e,k,j)$. Once at least $f+1$ distinct agents have missed their returns on the same trip, let $d$ be the smallest such trip. At least one of the agents that missed their returns on trip $d$ is honest. The gray hole destroyed an honest agent among them on trip $d$, so the gray hole is one of $u_1,\dots,u_d$. From then on, the agents make only trips shorter than $d$, in the usual stage order, so that stage $k$ consists of trips $1,\dots,\min\{k,d-1\}$. On the shorter trips, they check their reference records: the records that they wrote on $u_1,\dots,u_{d-1}$ in the last stage $d-1$ that was completed before trip $d$ was scheduled for the first time. An agent that finds one of these records missing identifies the node of the missing record as the gray hole. An agent that completes a trip $d-1$ after $d$ is defined has checked all of $u_1,\dots,u_{d-1}$. If all of these records are still present, none of these nodes is the gray hole, since the activation that destroyed the honest agent came after these records were written and would have erased the record on its node. Hence the gray hole is $u_d$, and the agent identifies $u_d$ as the gray hole without entering $u_d$. For $d=1$, the agents identify $u_1$ and write withdrawal notifications at once, without waiting for a new turn. The immediate notification for $d=1$ is the only notification that need not be written during the writer's own turn. We call the identification rule the \emph{threshold rule}.

An agent that identifies the gray hole returns to $\Home$ at once. The way back is safe, because the agent could read the whiteboard of the gray hole only while the gray hole was inactive, and the agent leaves the gray hole in that same round. At $\Home$, during its own turn and in place of its return record, the agent writes its withdrawal notification. A withdrawal notification written by the return deadline counts as a return. An agent also accepts a location when, before a trip, it finds matching withdrawal notifications of $f+1$ distinct agents at $\Home$. It then writes its own withdrawal notification during its current turn. Any agent that has written a withdrawal notification is removed from future turns. From then on, the agent walks the omission tour of the gray hole (\cref{sec:model}), back and forth between the two neighbors of the gray hole through $\Home$. The omission tour visits every safe node.

The schedule descriptor $(e,k,j)$, and the agent whose turn it is, follow from the round number and from the missed returns and withdrawal notifications at $\Home$. Every honest participant at $\Home$ reads the missed-return information and withdrawal notifications in every round, so all honest participants agree on the schedule. For each trip $j$, an agent keeps the set of agents that have missed their returns on trip $j$. From these sets, it computes the threshold index $d$, which is undefined as long as no trip has $f+1$ missed returns. An agent also keeps the list of records it has written, each with its node and round, which the first check and the threshold rule need. Finally, $g$ holds the clockwise distance of the gray hole once the agent has identified or accepted it, and is undefined before. \Cref{alg:known_n} summarizes the behavior of an honest agent.

\begin{algorithm}[tbp]
\caption{Perpetual exploration with known ring size, for an honest agent $a_i$.}
\label{alg:known_n}
\begin{algorithmic}[1]
\State $g\gets\bot$
\Repeat
  \State wait at $\Home$ until the next turn of $a_i$ or until $d=1$; in every round, update the missed returns and $d$
  \If{$d=1$}
    \State $g\gets 1$ \Comment{identify immediately; no new turn is required}
  \ElsIf{$\Home$ holds matching withdrawal notifications of $f+1$ distinct agents}
    \State $g\gets$ the stated distance \Comment{acceptance}
  \Else
    \State perform trip $j$ of the turn (once $d$ is defined, only trips shorter than $d$, in stage order), applying the two record checks at every node reached on the outward route
    \If{a check identifies $u_\ell$ as the gray hole}
      \State $g\gets\ell$;\quad cancel the rest of the trip and return to $\Home$
    \ElsIf{$d$ is defined, the trip just made is trip $d-1$, and all reference records from the last completed stage $d-1$ before the first trip $d$ are still present}
      \State $g\gets d$ \Comment{threshold rule}
    \EndIf
  \EndIf
\Until{$g\neq\bot$}
\State write a withdrawal notification stating $g$ by the current turn's return deadline, or immediately if $d=1$; remove $a_i$ from future turns
\Loop
  \State walk from $u_{g-1}$ to $u_{g+1}$ through $\Home$ and back, never entering $u_g$ \Comment{omission tour}
\EndLoop
\end{algorithmic}
\end{algorithm}

The correctness proof relies on the persistence and authentication of records on safe nodes.

\begin{observation}\label{obs:record-preservation}
A record of an honest agent never disappears from a safe node, and no other agent can write a copy of it. Hence, if an honest agent $a_i$ wrote a record $\rho^{a_i}_{s}$ at node $u_\ell$ and later finds it missing, that is, $\rho^{a_i}_{s}\notin\mathcal{W}_{u_\ell}^{t}$ for some round $t>s$, then $u_\ell$ is the gray hole.
\end{observation}

\begin{lemma}\label{lem:sound-identification}
In \cref{alg:known_n}, every node that an honest agent identifies as the gray hole, or accepts as the gray hole, is the gray hole.
\end{lemma}

\begin{proof}
We argue by induction on the round in which an honest agent identifies or accepts a location. An identification by the first check is correct by \cref{obs:record-preservation}. In the second check, at least one agent $a_h\in S$ is honest. An honest agent writes its arrival record on every node that it reaches on its way out, unless a check identifies the node as the gray hole. Hence either the clockwise arrival record of $a_h$ at $u_\ell$ was erased, and then $u_\ell=u_g$ by \cref{obs:record-preservation}, or $a_h$ identified $u_\ell$ as the gray hole in an earlier round, and this identification is correct by induction. Now consider an honest agent $a_i$ that applies the threshold rule with index $d$. At least one of the $f+1$ agents that missed their returns on trip $d$ is honest, and we call it $a_h$. The gray hole destroyed $a_h$ on trip $d$, at some node $u_m$ with $1\le m\le d$. Agent $a_i$ wrote its records at $u_1,\dots,u_{d-1}$ in stage $d-1$, before trip $d$ was first scheduled, and hence before the destruction of $a_h$. The index $d$ becomes defined only after $a_h$ has missed its return, and $a_i$ applies the rule only after a trip $d-1$ that it makes after this missed return, so $a_i$ checked each of these records after the destruction of $a_h$. If one of these records of $a_i$ is missing at some $u_\ell$, then $u_\ell=u_g$ by \cref{obs:record-preservation}. If none of them is missing, then $m<d$ is impossible, because the activation that destroyed $a_h$ at $u_m$ would have erased the record of $a_i$ at $u_m$. Hence $m=d$, and the gray hole is $u_d$. For $d=1$, the gray hole is $u_1$ directly. Finally, an accepted location is stated in the withdrawal notifications of $f+1$ distinct agents. At least one of these agents, say $a_h$, is honest, and the identification by $a_h$ is correct by induction.
\end{proof}

\begin{lemma}\label{lem:long-trip}
In \cref{alg:known_n}, let $u_g$ be the gray hole. Consider the first honest agent that the gray hole destroys on a trip $j$ with $j>g$. After the destruction of that first honest agent, every honest participant that reaches $u_g$ again is either destroyed when it arrives or identifies $u_g$ as the gray hole. Hence the gray hole destroys at most one honest agent on trips $j$ with $j>g$.
\end{lemma}

\begin{proof}
The gray hole can destroy an honest agent only at $u_g$. Let the first honest agent destroyed on a trip with index $j>g$ be destroyed in stage $k\ge j$. Every honest participant made trip $g$ earlier in stage $k$, so each of them has a record at $u_g$, and the activation that destroys this honest agent erases these records. During this activation, every other honest participant waits at $\Home$, since at most one honest participant is away from $\Home$ on its scheduled trip. The missed return starts a new epoch at stage~$1$, and the stages keep their order also after the threshold index is defined, so each honest participant makes trip $g$ before any longer trip. When it next reaches $u_g$, it is either destroyed on arrival or finds its record missing and identifies $u_g$ as the gray hole by the first check. In the second case, it returns safely and leaves the schedule. Hence no honest participant makes a trip longer than $g$ afterwards.
\end{proof}

\begin{lemma}\label{lem:loss-bound}
Suppose that at most $f$ agents are Byzantine. In \cref{alg:known_n}, the gray hole destroys at most $f+2$ honest agents.
\end{lemma}

\begin{proof}
Trips with index less than $g$ never reach $u_g$, and by \cref{lem:long-trip}, the gray hole destroys at most one honest agent on longer trips. Every other honest agent that the gray hole destroys is destroyed on trip $g$, and each such agent misses its return on trip $g$. Byzantine agents cause at most $f$ missed returns on any trip, so the index $d$ of the threshold rule is never smaller than $g$. Once $f+1$ distinct agents have missed their returns on trip $g$, the threshold rule applies with $d=g$, and no honest participant makes trip $g$ again. Hence the gray hole destroys at most $f+1$ honest agents on trip $g$, and at most $f+2$ in total.
\end{proof}

\begin{theorem}\label{thm:known_n}
Suppose that the agents know $n$ and that at most $f$ agents are Byzantine. Then \cref{alg:known_n} solves \PE{} in the whiteboard model with $2f+3$ agents.
\end{theorem}

\begin{proof}
At least $f+3$ agents are honest, and by \cref{lem:loss-bound}, the gray hole destroys at most $f+2$ of them, so at least one honest agent survives. If a surviving honest agent identifies or accepts a location, the location is the gray hole by \cref{lem:sound-identification}, and the agent walks the omission tour of the gray hole forever. Otherwise, the surviving honest agents follow the schedule forever. Then no trip ever collects $f+1$ missed returns, since the threshold rule would make the survivors identify the gray hole. Each agent misses at most one return, so missed returns interrupt only finitely many epochs, and the survivors complete infinitely many epochs, in each of which every honest participant visits every safe node. Byzantine agents cannot stop the schedule, because every turn has a fixed return deadline.
\end{proof}

\subsubsection{Exploration Without Known Ring Size}
\label{sec:algorithms_unknown_n}
Without the ring size, identifying the gray hole from one side does not reveal the length of the safe path. We therefore use the same team in two runs. The first run explores clockwise and obtains the clockwise distance $g$ of the gray hole. The second run explores counterclockwise and determines $n$, either by identifying the gray hole from the other side or by recognizing one of the agent's own first-run records. A \emph{circuit} is one complete traversal around the ring, from $\Home$ back to $\Home$. Since a scheduled trip may contain several circuits, the extension must distinguish distance traveled from position on the ring.

We use the same trips and return deadlines, but an epoch has no last stage; a missed return starts the next epoch. A trip of length $j$ consists of $j$ outward edge traversals followed by the reverse route, unless the agent withdraws earlier. Thus a trip may pass through $\Home$ before its scheduled end. Each agent distinguishes the number of outward edge traversals from its position on the ring. Before completing a circuit, its position is its distance from $\Home$ in the direction of the run. When the agent completes a circuit and recognizes $\Home$, it learns $n$ and thereafter maintains its position modulo $n$. On the return route, the position changes in the opposite direction. An honest agent's position therefore lies in $\{0,\ldots,n-1\}$, even on a trip longer than $n$. An identification reports the position of the gray hole, rather than the total distance traveled.

Every record includes the run number, either $1$ or $2$, and the writer's position. A visit sequence number distinguishes repeated visits within a trip; the agent starts the counter at $0$ at $\Home$ and increases it on each arrival at a node, using the same number for the arrival and departure of that visit. The first check compares complete records, including their sequence numbers. The second check asks whether a record has the required run, schedule descriptor, event, direction, and state, and ignores the visit sequence number. The second check uses only records from the current run. Earlier records are used only for the first check or for the size calculation described below, and an agent uses a recorded position to compute $n$ only if it wrote the record itself. Recognizing $\Home$ during a trip does not count as a scheduled return unless the agent ends its trip or withdraws. The run number also tags schedule descriptors and withdrawal notifications. At the start of each run, the epoch, stage, and trip counters start at $1$, the missed-return sets are empty, and the threshold index is undefined. Threshold decisions and acceptance use only the current run's records and notifications. The second run retains the first-run clockwise distance $g$ separately and uses $h$ for its own counterclockwise gray-hole distance.

In the first run, agents explore clockwise. If an agent identifies the gray hole while standing on it, it leaves in that same round and moves counterclockwise to the next occurrence of $\Home$, using $g$ safe edges rather than retracing any completed circuits. If it identifies or accepts the location at $\Home$, it stays there. The agent writes its withdrawal notification and waits at $\Home$, even if it has already learned $n$. The first run ends when $f+1$ distinct agents have written notifications with the same clockwise distance $g$. The common distance $g$ is correct because at least one writer is honest. The second run starts with the next turn, with directions reversed. The participants of the second run are the agents with initial identifiers that were not removed for a missed return in the first run. Agents that withdrew normally rejoin. Resetting the schedule and missed-return sets does not restore a removed identifier, so every surviving honest agent participates and no destroyed honest agent is scheduled again.

In the second run, let $q$ be the number of outward edge traversals from the starting occurrence of $\Home$. The agent stops its outward route as soon as one of the following events occurs: (a) it identifies the gray hole at counterclockwise distance $h$, and computes $n=g+h$; (b) before returning to $\Home$, it reaches a non-home node containing one of its own first-run records with clockwise position $\ell$, and computes $n=\ell+q$; or (c) it recognizes $\Home$ after a positive number $q$ of outward edge traversals, and computes $n=q$. In (b), $1\le q<n$ and $1\le\ell<n$, so the sum is the ring size, not the length of several circuits. An agent that already learned $n$ in the first run may instead withdraw at the start of its second-run turn.

After learning $n$, an agent already at $\Home$ needs no return travel. Otherwise, it reverses the outward segment already taken and attempts to return to $\Home$. The reverse-route return finishes by the original return deadline if the agent survives. In case (b), its return may cross the gray hole, so learning $n$ does not itself guarantee survival. If the agent is destroyed on the reverse route, its absence is counted as a missed return on the original trip $j$. A survivor writes a second-run withdrawal notification giving $h=n-g$ by its return deadline and begins the omission tour of the gray hole, using $g$ and $n$. An agent that accepts $h$ from $f+1$ matching second-run notifications also obtains $n=g+h$ and withdraws. The immediate $d=1$ rule writes notifications without waiting for another turn; in the first run the notifying agents remain at $\Home$ for the transition to the second run.

The two runs reuse the same team and the same Byzantine fault bound. The number of honest agents required is therefore determined by the total number that the gray hole can destroy across the two runs. The proof below shows that an honest agent survives with $3f+5$ agents.

In addition to the variables of the known-size algorithm, each agent stores its run number, position, visit sequence number, and the ring size once learned. During a trip, $q$ counts outward edge traversals from the starting occurrence of $\Home$; position is maintained separately. The clockwise distance $g$ is retained after the first run, while $h$ is the distance reported in the second run. The agent also retains its own earlier records and the identifiers permanently removed for missed returns. The run number determines which records contribute to the current schedule and threshold. The wrapper in \cref{alg:whiteboard-unknown} summarizes how the existing turns are used in the two runs.

\begin{algorithm}[tbp]
\caption{Two-run whiteboard algorithm for an honest agent $a_i$ when $n$ is unknown.}
\label{alg:whiteboard-unknown}
\begin{algorithmic}[1]
\State initialize run $1$ with the clockwise schedule, no last stage, and the record conventions above
\While{fewer than $f+1$ distinct agents have matching first-run withdrawal notifications at $\Home$}
  \If{$a_i$ has not withdrawn}
    \State while waiting for the turn of $a_i$, update the common schedule; act immediately if $d=1$
    \State perform that turn with the record checks and threshold rule of \cref{alg:known_n}; if $g$ is identified at the gray hole during the turn, leave in the same round and return counterclockwise along the $g$ safe edges to $\Home$
    \State after identifying or accepting $g$, write the withdrawal notification by the return deadline, or immediately for $d=1$, and wait at $\Home$
  \Else
    \State wait at $\Home$ and read the first-run withdrawal notifications
  \EndIf
\EndWhile
\State retain the common clockwise distance $g$ and every learned value of $n$
\State start run $2$ at the next turn, reverse direction, reset the schedule and missed-return sets, and restore agents that withdrew but missed no return
\Loop
  \State update the common schedule while waiting for the next turn of $a_i$; act immediately if the threshold gives $d=1$
  \If{$d=1$}
    \State set $h\gets1$ and $n\gets g+h$
  \ElsIf{$n$ is already known}
    \State set $h\gets n-g$
  \ElsIf{$f+1$ matching second-run notifications give $h$}
    \State set $n\gets g+h$
  \Else
    \State during the second-run turn, apply the record checks and threshold rule, stopping the outward route as soon as rule (a), (b), or (c) gives $n$; if away from $\Home$, attempt the reverse route by the original deadline
  \EndIf
  \If{$n$ is known and $a_i$ is at $\Home$}
    \State write a second-run withdrawal notification with $h=n-g$ by the turn's deadline, or immediately for $d=1$
    \State leave the schedule and walk the omission tour of the gray hole forever
  \EndIf
\EndLoop
\end{algorithmic}
\end{algorithm}

\begin{corollary}\label{thm:unknown_n}
Suppose that the agents do not know $n$ and that at most $f$ agents are Byzantine. Then \cref{alg:whiteboard-unknown} solves \PE{} in the whiteboard model with $3f+5$ agents.
\end{corollary}

\begin{proof}
To reuse \cref{lem:sound-identification,lem:long-trip,lem:loss-bound}, we check that the extended trips preserve the record evidence and the order of shorter trips before longer trips. None of those proofs requires a last stage in an epoch. We use $g$ for the distance in the first run and $h=n-g$ for the distance in the second run. The loss bound concerns honest agents that still participate in the schedule. Honest agents that have already started an omission tour avoid the gray hole and cause no additional losses. Allowing a trip to traverse several circuits does not affect the proof: before the first trip with index larger than the distance of the gray hole, every remaining honest participant has completed the shorter trip to that node and left its record there. Likewise, ending an outward route early when $n$ is learned cannot create a new loss; a destruction on the reverse route is charged to the original trip. Thus the gray hole still destroys at most $f+2$ honest agents in each run. In the first run, a threshold index $d>g$, including $d\ge n$, cannot cause a false identification: the activation that destroyed an honest agent erased its reference record at $u_g$, and the restarted stage order brings each remaining honest participant to $u_g$ before it completes trip $d-1$. It then either identifies $u_g$ or is destroyed. The same reasoning applies in the second run with the directions exchanged and $h$ in place of $g$. In the second run, an agent that learns $n$ attempts the reverse route of its trip. If it survives, it returns by its deadline; otherwise, its destruction contributes to the same loss count. Hence every location that an honest agent identifies is the gray hole, and rules~(a) and~(c) give the correct value of $n$. So does rule~(b), because the agent wrote the record itself, on the node at counterclockwise distance $q$, when it stood at clockwise distance $\ell$ from $\Home$. There are at least $2f+5$ honest agents initially, and the gray hole destroys at most $2(f+2)$ honest agents in the two runs. Hence at least one honest agent survives both runs. If the first run never ends, at most $f$ honest agents write withdrawal notifications, so at least three honest agents follow its schedule forever. As in the proof of \cref{thm:known_n}, only finitely many missed returns can restart an epoch. The stages therefore grow without bound, and every trip with index at least $n-1$ visits every node. If an honest agent survives its second-run withdrawal, it starts the correct omission tour. Otherwise, consider the surviving honest participants after the last destruction and the last Byzantine missed return; both kinds of event occur only finitely often. At least one such participant exists by the loss bound. If a threshold is defined, the threshold rule makes a surviving participant identify the gray hole at counterclockwise distance $h$ and compute $n=g+h$. If no threshold is defined, the stages grow, and a surviving participant eventually completes an outward circuit and learns $n$ upon recognizing $\Home$, unless it learns $n$ earlier. After the last destruction it survives its return and withdraws, contradicting the assumption. Thus some honest agent walks the omission tour of the gray hole and visits every safe node infinitely often.
\end{proof}

\begin{remark}\label{rem:whiteboard}
The algorithms analyzed here use unbounded, append-only whiteboards. Replacing all records of an agent by those of its latest visit does not preserve the second record check: a return visit could overwrite the outward-arrival record at a safe node. A bounded-storage variant therefore requires a separate record-retention rule and correctness proof; no such bound is claimed here. Our lower bounds are proved even with unbounded whiteboards and consequently also apply when storage is restricted.
\end{remark}

\section{Perpetual Exploration in the Face-to-Face Model}
\label{sec:inch3}
In the face-to-face model, an agent notices a destruction through a \emph{missed meeting}: it expects a partner on a node in a specified round, but the partner is missing. If the partner is honest and was required to follow the schedule until the meeting, its absence shows that it was destroyed along its route since the partners last met. The detecting agent must restrict the possible locations of the gray hole and inform the other survivors. Exact identification is not always needed: two survivors can avoid two different \emph{candidates}, that is, nodes that have not been ruled out as the gray hole, so that at least one of them walks the omission tour of the gray hole. The algorithms below specify when an agent leaves the schedule and how the survivors continue. Our main result in the face-to-face model concerns honest agents on rings of known size with a gray hole\footnote{In the face-to-face model, the Byzantine black hole of~\cite{DBLP:conf/opodis/GoswamiBD024,DBLP:conf/wdag/BhattacharyaGB025} coincides with our gray hole.}. We show that three honest agents suffice, which answers the question of Bhattacharya et al.~\cite{DBLP:conf/wdag/BhattacharyaGB025} whether three agents suffice. The three-agent algorithm also serves as our building block for Byzantine agents, through $f+1$ independent groups of three. We first show that, as with whiteboards, $2f+2$ agents are not enough. We then give the three-agent algorithm and an optimal five-agent patrol algorithm that tolerates one Byzantine agent on rings of four or five nodes. In \cref{sec:f2f-unknown}, we consider rings of unknown size.

\subsection{Agent Lower Bounds in the Face-to-Face Model}
\label{sec:f2f-impossibility}\label{sec:f2f-lower}

With whiteboards, $2f+2$ agents are not enough on rings with $n\ge4$ (\cref{thm:2f+2}). The same bound holds in the face-to-face model. Its proof needs a refined construction, in which a missed meeting takes the place of a missing record.

\begin{theorem}\label{thm:f2f-lower}
Suppose that $n\ge4$, that the agents know $n$, and that at most $f$ agents are Byzantine. Then no algorithm solves \PE{} in the face-to-face model with $2f+2$ or fewer agents.
\end{theorem}

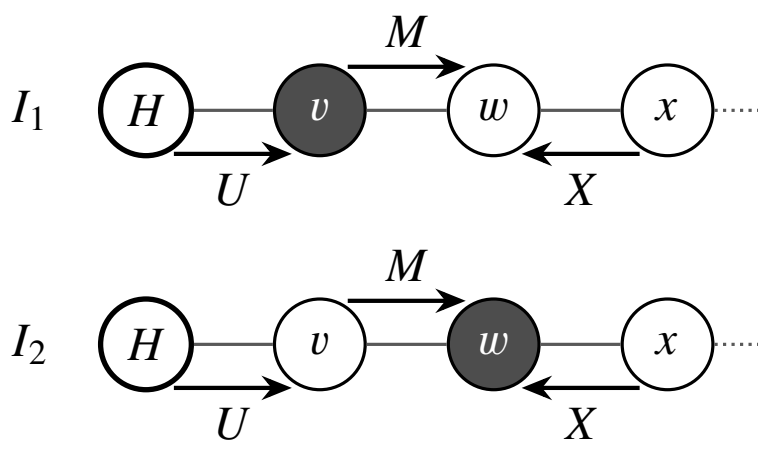
\begin{figure}[t]
  \centering
  \begin{tikzpicture}[x=1.15cm,y=0.9cm]
    \foreach \row/\name/\g in {0/1/1,-1.72/2/2}{
      \node[anchor=east] at (-0.45,\row) {$I_{\name}$};
      \foreach \i/\lab in {0/\Home,1/v,2/w,3/x}{
        \ifnum\i=0
          \node[home] (a\name-\i) at (\i,\row) {$\lab$};
        \else\ifnum\i=\g
          \node[gh] (a\name-\i) at (\i,\row) {$\lab$};
        \else
          \node[vtx] (a\name-\i) at (\i,\row) {$\lab$};
        \fi\fi}
      \draw[edge] (a\name-0)--(a\name-1)--(a\name-2)--(a\name-3);
      \draw[omit] (a\name-3)--(3.52,\row);
      \begin{scope}[shift={(0,\row)}]
        \draw[move] (0.13,-0.32)--node[below] {$U$} (0.87,-0.32);
        \draw[move] (1.13, 0.32)--node[above] {$M$} (1.87, 0.32);
        \draw[move] (2.87,-0.32)--node[below] {$X$} (2.13,-0.32);
      \end{scope}}
  \end{tikzpicture}
  \caption{The two executions of \cref{lem:f2f-loss}, one in each row. The dark-gray node is the gray hole, $v$ in $I_1$ and $w$ in $I_2$. The agents of $M$ are scheduled to make the first move to distance two from $\Home$ and are destroyed at $v$ before moving in $I_1$, and on entering $w$ in $I_2$. Arrows $U$ and $X$ label the entry routes $\Home\to v$ and $x\to w$, respectively. For the agents outside $\{v,w\}$, an agent that takes one of the two entry arrows disappears in both executions.}
  \label{fig:f2f-lower}
\end{figure}

In the face-to-face model, agents on the same node see each other. We therefore refine the two-execution construction used for whiteboards in \cref{thm:2f+2}. Let $v$ be a neighbor of $\Home$, and let $w$ be the next node in the same direction. Placing the gray hole on $v$ in one execution and on $w$ in the other is not enough: agents outside $\{v,w\}$ must also see the same agents and hear the same messages. The following lemma builds two such executions, illustrated in \cref{fig:f2f-lower}, with identical observations outside $\{v,w\}$. Here and below, distance without a specified direction means the minimum number of edges on a path between two nodes.

\begin{lemma}\label{lem:f2f-loss}
Suppose that $n\ge4$, and consider an algorithm that solves \PE{} in the face-to-face model when at most $f$ agents are Byzantine. Then the algorithm has an execution in which the gray hole destroys at least $f+2$ honest agents and every Byzantine agent follows the algorithm until it stops acting.
\end{lemma}

\begin{proof}
Let $E_0$ be the execution in which all agents are honest and the gray hole is never active. The execution $E_0$ is the same wherever the gray hole is, so the agents visit every node in $E_0$. Let $r$ be the first round of $E_0$ in which an agent moves to a node at distance two from $\Home$. It exists because $n\ge4$. Name the nodes $\Home,v,w,x$ in the direction of the first move to distance two, so that the move goes from $v$ to $w$. Let $P$ be the agents on $v$ at the start of round $r$, let $M\subseteq P$ be those that move to $w$ in round $r$, and let $Q=P\setminus M$. Then $|M|\ge 1$.

The adversary must fix the Byzantine agents before the execution starts, so it first predicts which agents will enter $v$ or $w$. To do so, it runs the algorithm outside $\{v,w\}$. It follows $E_0$ up to the messages of round $r$, and then removes $P$ and every agent that enters $v$ or $w$, in round $r$ or later. The remaining agents see only each other. Let $U_t$ be the agents removed on a move from $\Home$ to $v$ in rounds $r,\dots,t$, and $X_t$ the agents removed on a move from $x$ to $w$ in rounds $r,\dots,t$. The moves $\Home\to v$ and $x\to w$ are the only two ways into $\{v,w\}$ from outside. When $n=4$, the node $x$ is a neighbor of $\Home$, and an agent may enter $w$ from $x$ already in round $r$. Set $U_{r-1}=X_{r-1}=\emptyset$, and let $t^*\ge r$ be the first round with $|Q|+|U_{t^*}|\ge f+1$ or $|X_{t^*}|\ge f+1$, or $t^*=\infty$ if there is no such round. The simulation outside $\{v,w\}$ is deterministic, so $t^*$ and the sets $U_t$ and $X_t$ are known in advance. Let $\widehat U$ and $\widehat X$ contain the agents removed on the respective entry routes strictly before $t^*$. Thus, for finite $t^*$, $\widehat U=U_{t^*-1}$ and $\widehat X=X_{t^*-1}$; if $t^*=\infty$, they are $\bigcup_{t\ge r}U_t$ and $\bigcup_{t\ge r}X_t$.

We now build two executions in which the agents outside $\{v,w\}$ act as in the simulation outside $\{v,w\}$. In both, the gray hole is inactive before round $r$ and active from round $r$ on. In $I_1$, the gray hole is $v$. It destroys $P$ at the start of round $r$ and every later visitor of $v$. The adversary chooses the agents of $\widehat X$ to be Byzantine, and each of them follows the algorithm until it enters $w$, and then stops acting at $w$. In $I_2$, the gray hole is $w$, and the adversary chooses the agents of $Q\cup \widehat U$ to be Byzantine. An agent of $Q$ sends its messages of round $r$ and then stops acting at $v$. The agents of $M$ therefore hear what they hear in $E_0$, move to $w$, and are destroyed there. An agent of $\widehat U$ follows the algorithm until it enters $v$, and then stops acting at $v$, and the gray hole destroys every agent that enters $w$ from $x$. In both executions, every removed agent is destroyed or stops acting inside $\{v,w\}$, and agents never meet on an edge. Hence, by induction on the rounds, up to round $t^*$, or in every round if $t^*=\infty$, the agents outside $\{v,w\}$ see exactly what they see in the simulation above, and they act in the same way.

Suppose first that $t^*$ is finite. If $|Q|+|U_{t^*}|\ge f+1$, we take $I_1$. It has at most $f$ Byzantine agents by the choice of $t^*$, and by round $t^*$ the gray hole destroys the honest agents of $P\cup U_{t^*}$ at $v$, and there are $|M|+|Q|+|U_{t^*}|\ge f+2$ of them. Otherwise, $|X_{t^*}|\ge f+1$, and we take $I_2$. It has $|Q|+|\widehat U|\le f$ Byzantine agents, and the gray hole destroys the honest agents of $M\cup X_{t^*}$ at $w$, again at least $f+2$ of them. Finally, $t^*$ cannot be infinite. Otherwise, $I_2$ with the Byzantine agents $Q\cup\bigcup_{t\ge r}U_t$, at most $f$ of them, is a valid execution in which only Byzantine agents enter the safe node $v$ after round $r$, so \PE{} fails.
\end{proof}

\begin{proof}[Proof of \cref{thm:f2f-lower}]
Suppose that an algorithm solves \PE{} with $N\le 2f+2$ agents. \Cref{lem:f2f-loss} gives an execution with $b\le f$ Byzantine agents in which the gray hole destroys at least $f+2$ honest agents. At most $N-b-(f+2)\le f-b$ honest agents survive. The adversary chooses them to be Byzantine from the start and lets them follow the algorithm, which changes nothing. The execution then has at most $f$ Byzantine agents, and no honest agent survives, so \PE{} fails.
\end{proof}

As with whiteboards, one agent fewer suffices on the triangle, where no node lies at distance $2$ from $\Home$.

\begin{proposition}\label{prop:f2f-triangle}
Suppose that the agents know that $n=3$ and that at most $f$ agents are Byzantine. Then $2f+2$ agents are necessary and sufficient for \PE{} in the face-to-face model.
\end{proposition}

\begin{proof}
For sufficiency, split the team into two groups of $f+1$ agents. One group walks back and forth between $\Home$ and $u_1$, and the other between $\Home$ and $u_2$. The group assigned to the safe neighbor contains an honest agent and never enters the gray hole, so an honest agent visits both safe nodes infinitely often. For necessity, consider an execution with all agents honest and $u_1$ active in every round. Let $D$ contain the agents that ever enter $u_1$, and let $S$ contain the remaining agents. If $|S|\le f$, the adversary chooses the agents of $S$ to be Byzantine and lets them follow the algorithm, while every honest agent is destroyed at $u_1$. If $|D|\le f$, consider instead the execution in which the gray hole is $u_2$ and is never active. The adversary chooses the agents of $D$ to be Byzantine; each follows the algorithm until it enters $u_1$ and then stops acting at node $u_1$. The agents of $S$ never enter $u_1$, so they see the same agents and hear the same messages in both executions. By induction on the rounds, they make the same moves and never visit $u_1$, which is safe in this execution. Thus both $|D|$ and $|S|$ must be at least $f+1$, giving $N\ge2f+2$. The construction is the same as the one in \cref{prop:wb-triangle}, with meetings replacing whiteboard observations.
\end{proof}

\subsection{Perpetual Exploration with Three Honest Agents}
\label{sec:f2f-algorithms}

We now give the three-agent algorithm. Its final behavior uses omission tours, either after exact identification or after the survivors reduce the possible gray-hole locations to two candidates. For any node $x\neq\Home$, the \emph{omission tour} $O(x)$ walks back and forth along the other $n-1$ nodes and never enters $x$. If $x$ is the gray hole, an honest agent that walks $O(x)$ visits every safe node infinitely often. Consequently, two survivors with two candidates can each avoid one candidate, provided that their candidates contain the gray hole.

\label{sec:inch3-protocol}
All three agents are honest and know $n$. We call them $A$, $B$, $C$ in increasing order of identifier. For $n=4$, agents $A$, $B$, and $C$ walk the omission tours of $u_1$, $u_2$, and $u_3$, respectively. One of the three tours omits the gray hole. The known triangle case is covered by \cref{prop:f2f-triangle}. We therefore describe the remaining case $n\ge5$. Write node $u_i$ as $i$, with indices modulo $n$, so $\Home=0$ and the gray hole is $g$.
A missed meeting tells an agent that the gray hole has destroyed its \emph{partner}, the agent that it was scheduled to meet. The last meeting with the partner tells the agent where to look for the gray hole. The difficulty is to warn the third agent, which has seen nothing and may be about to move onto the node that destroyed the partner. We therefore separate consecutive scheduled moves by $n$ rounds. An agent that misses a meeting leaves at once, takes the route around the other side of the ring if needed, and reaches the third agent before the third agent moves again. We choose the meeting places so that at most two adjacent nodes, that is, nodes joined by an edge, are ever candidates, and the third agent never waits on a candidate.

\begin{figure}[t]
\centering
\begin{tikzpicture}[x=1.3cm,y=0.74cm]
  \foreach \c/\l in {0/{p-1},1/p,2/{p+1},3/{p+2},4/{p+3}} \node at (\c,0.95) {$\l$};
  \foreach \j in {0,...,6}{
    \node[anchor=east] at (-0.6,-\j) {$j=\j$};
    \draw[tgrid] (-0.4,-\j)--(4.4,-\j);
    \foreach \c in {0,...,4} \node[dot] at (\c,-\j) {};}
  \node[agA] at (0.80,0) {$A$}; \node[agB] at (1.20,0) {$B$}; \node[agC] at (3,0) {$C$};
  \node[agB] at (1,-1) {$B$}; \node[agA] at (2,-1) {$A$}; \node[agC] at (3,-1) {$C$};
  \node[agB] at (0,-2) {$B$}; \node[agA] at (2,-2) {$A$}; \node[agC] at (3,-2) {$C$};
  \node[agB] at (0,-3) {$B$}; \node[agA] at (2.80,-3) {$A$}; \node[agC] at (3.20,-3) {$C$};
  \node[agB] at (1,-4) {$B$}; \node[agA] at (2.80,-4) {$A$}; \node[agC] at (3.20,-4) {$C$};
  \node[agB] at (1,-5) {$B$}; \node[agA] at (3,-5) {$A$}; \node[agC] at (4,-5) {$C$};
  \node[agB] at (1,-6) {$B$}; \node[agA] at (2,-6) {$A$}; \node[agC] at (4,-6) {$C$};
  \draw[move] (0.9,-0.3)--(1.9,-0.7);
  \draw[move] (0.92,-1.3)--(0.08,-1.7);
  \draw[move] (2.08,-2.3)--(2.78,-2.7);
  \draw[move] (0.08,-3.3)--(0.92,-3.7);
  \draw[move] (3.22,-4.3)--(3.92,-4.7);
  \draw[move] (2.92,-5.3)--(2.08,-5.7);
  \draw[move] (1.08,-6.3)--(1.85,-6.85) node[right,lab] {$B$ joins $A$ on $p+1$: row $0$ of period $p+1$};
\end{tikzpicture}
\caption{Period $p$ of the three-agent chain, on the nodes $p-1,\ldots,p+3$. Row $j$ shows where the agents stand during step $4+7p+j$, and each arrow is the move at the end of step $4+7p+j$. Two agents on the same node meet. Agent $A$ has a solid outline, $B$ a dashed outline, and $C$ a double outline. The last arrow starts period $p+1$.}
  \label{fig:f2f-inchworm}
\end{figure}
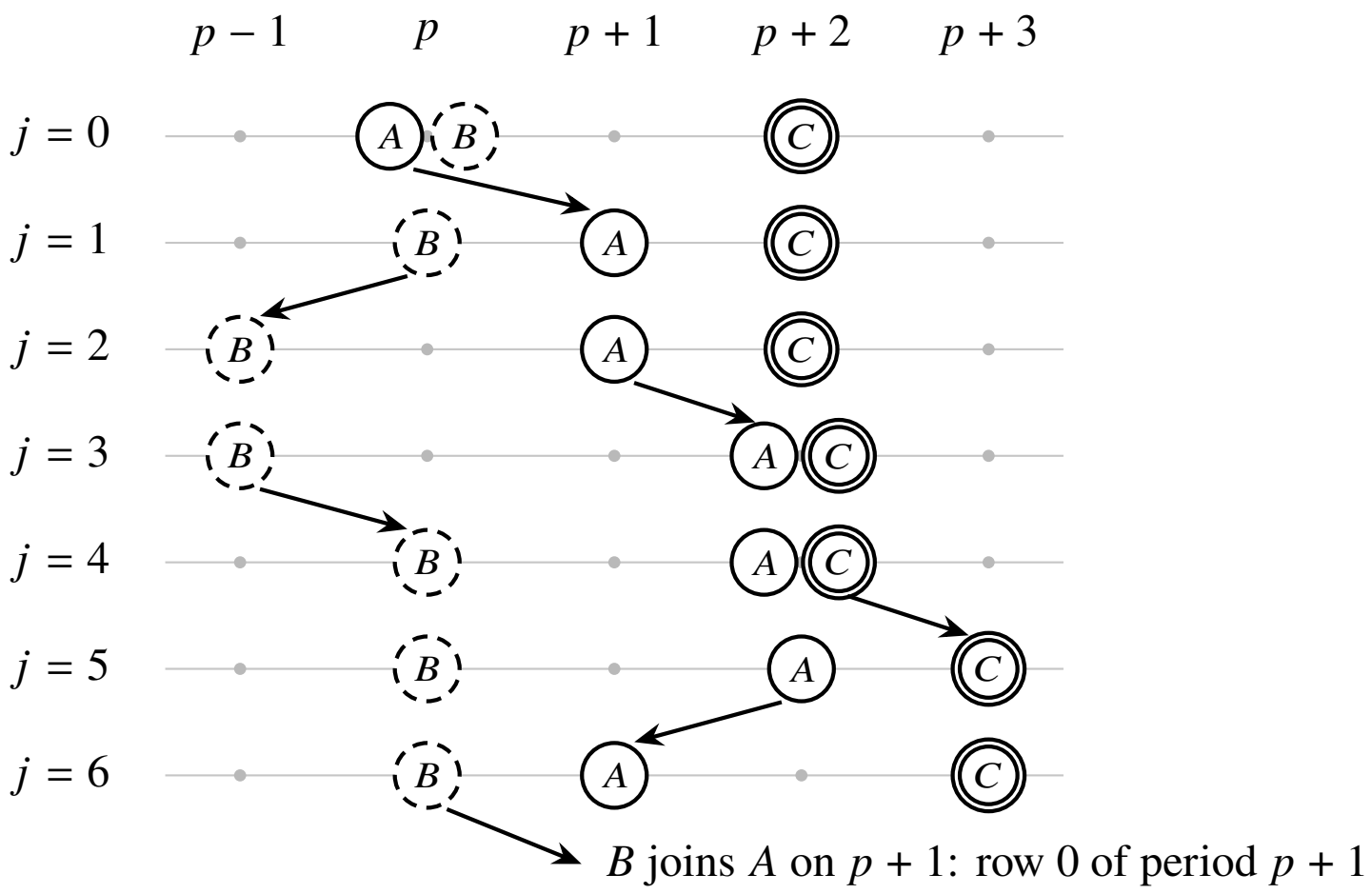

All three agents start at $\Home=0$. Set $T=n$. Each \emph{step} lasts $T$ rounds, with one agent scheduled to move in the final round. At the start, $C$ first moves from node $0$ to node $1$. In the second step, $A$ joins $C$ there. In the third step, $C$ moves to node $2$. In the fourth step, $A$ returns to $B$ at node $0$. Thus, after $4$ steps, or $4n$ rounds, $A$ and $B$ stand on node $0$ and $C$ on node $2$. From the configuration with $A$ and $B$ on node $0$ and $C$ on node $2$, the agents move around the ring as a short \emph{chain}. Each \emph{period} consists of seven successive steps that advance the configuration one node in a clockwise direction (\cref{fig:f2f-inchworm}). Period $p$ starts with $A$ and $B$ on node $p$ and $C$ on node $p+2$. In the first step, $A$ moves to $p+1$. In the second, $B$ moves to $p-1$. In the third, $A$ moves to $p+2$ and meets $C$. In the fourth, $B$ returns to $p$. In the fifth, $C$ moves to $p+3$. In the sixth, $A$ returns to $p+1$. Finally, in the seventh step, $B$ joins $A$ on $p+1$. The agents have now reached the same configuration shifted one node clockwise, with $A$ and $B$ on $p+1$ and $C$ on $p+3$. Since each step lasts $n$ rounds, each period takes $7n$ rounds.

As long as no agent misses a meeting, every agent follows a fixed \emph{quiet schedule} $Q$. During step $s$, agent $a$ stands on node $Q_a(s)$, and in the last round of the step, the \emph{move round}, it moves to $Q_a(s+1)$. Steps $0$ to $3$ form the \emph{start-up} described above, and step $s=4+7p+j$, with $p\ge0$ and $0\le j\le6$, is step $j$ of period $p$, shown in row $j$ of \cref{fig:f2f-inchworm}.

When an agent first misses a partner, it becomes the \emph{detector}. It computes the possible gray-hole locations from the agents' records of past meetings and immediately starts a \emph{warning walk} to the node where the third agent waits in the current step. The detector takes a shortest path, measured by the number of edges, that avoids the candidates after its starting node; it chooses the clockwise path if both paths have the same length. In the first round after reaching the destination, it exchanges messages with the third agent if the third agent is present. Otherwise, it marks the third agent as missing. The survivors combine their histories and recompute the candidates. If only $x$ remains, every survivor walks the omission tour $O(x)$. If two candidates $x$ and $x+1$ remain, two agents survive: the survivor with the smaller identifier walks the omission tour $O(x)$, and the other walks the omission tour $O(x+1)$.

To find the gray hole after a destruction, each agent keeps a history of the others. For every agent $c$, agent $X$ keeps the last round $\mathrm{seen}_X(c)$ in which $X$ knows that $c$ was alive (initially $1$), the round $\mathrm{dead}_X(c)$ in which $X$ marked $c$ as missing (initially unset), and the list $\mathrm{route}_X(c)$ of the moves that $c$ made after it left the quiet schedule, with their rounds (initially empty). In every round, $X$ sets $\mathrm{seen}_X(c)$ to the current round for itself and for every agent $c$ on its node. Agents on the same node exchange their histories and keep, for each agent, the latest known round of survival, the earliest recorded round of absence, and the longest recorded route. Together with the quiet schedule, the recorded routes specify the positions each agent would occupy if it had survived. The candidate calculation uses the scheduled or reported routes between the last confirmed round of survival and the missed meeting. We say that an agent \emph{uses} a node in a round if it stands on the node at the start of the round or enters it during the round, so an active gray hole destroys every honest agent that uses it. A node $x\neq\Home$ is a \emph{candidate} for a missing agent $y$ if, in some round $r$ with $\mathrm{seen}_X(y)\le r<\mathrm{dead}_X(y)$, agent $y$ used $x$ and no agent $z$ with $\mathrm{seen}_X(z)\ge r+1$ used $x$ in round $r$. The second condition is needed because such an agent $z$ shows that $x$ was not active in round $r$. The same gray hole destroyed every agent marked missing in the execution under consideration, so $X$ intersects the candidate sets of all agents that it has marked missing, and we call the result $K_X$. The detector uses the name $Z$ for the third agent, $t$ for the detection round, and $s$ for its current step. After the warning walk, $K$ denotes the final candidate set obtained from the survivors' combined histories, or from the detector's history if it is alone. The variables $s$, $p$, and $j$ follow the schedule above, and $T=n$ remains fixed. \Cref{alg:inch3} gives the three procedures: \textsc{Quiet} follows the schedule, \textsc{Warn} performs the warning walk, and \textsc{Final} assigns the omission tours.

\begin{algorithm}[tbp]
\caption{The three-agent algorithm, for agent $X$ ($n\ge5$, $T=n$).}
\label{alg:inch3}
\begin{algorithmic}[1]
\Statex \textbf{Initially:} $\mathrm{seen}_X(c)\gets 1$, $\mathrm{dead}_X(c)\gets\bot$, $\mathrm{route}_X(c)\gets\emptyset$ for every agent $c$.
\Statex \textbf{In every round $t$:} set $\mathrm{seen}_X(c)\gets t$ for $X$ and every agent $c$ on its node; exchange histories and retain the latest survival round, earliest absence round, and longest route for each agent.
\Statex
\Procedure{Quiet}{} \Comment{the quiet schedule of \cref{fig:f2f-inchworm}}
  \For{every step $s=0,1,2,\dots$ and every round $t$ of step $s$}
    \If{some agent $y$ with $Q_y(s)=Q_X(s)$ is missing}
      \State $\mathrm{dead}_X(y)\gets t$; \Call{Warn}{$t,s$} \Comment{$X$ becomes the detector}
    \ElsIf{an agent not scheduled on the current node of $X$ arrives with a warning}
      \State compute $K$ from the merged histories; \Call{Final}{$K$}
    \ElsIf{$t=(s+1)T$}
      \State move to $Q_X(s+1)$
    \EndIf
  \EndFor
\EndProcedure
\Statex
\Procedure{Warn}{$t,s$} \Comment{$Z$ is the third agent}
  \State compute $K_X$, and a shortest path to $Q_Z(s)$ that avoids $K_X$ after its first node (clockwise on a tie)
  \State walk the path, one edge per round from round $t$ on, and record the moves in $\mathrm{route}_X(X)$
  \State let $t'$ be the first round after arrival at $Q_Z(s)$
  \If{$Z$ is on $Q_Z(s)$ in round $t'$}
    \State merge histories with $Z$, and compute $K$
  \Else
    \State $\mathrm{dead}_X(Z)\gets t'$; recompute $K_X$ using both missing agents; $K\gets K_X$
  \EndIf
  \State \Call{Final}{$K$}
\EndProcedure
\Statex
\Procedure{Final}{$K$}
  \If{$K=\{x\}$}
    \State walk the omission tour $O(x)$ forever
  \Else \Comment{$K=\{x,x+1\}$, two survivors}
    \State the survivor with the smaller identifier walks the omission tour $O(x)$ forever, and the other walks the omission tour $O(x+1)$ forever
  \EndIf
\EndProcedure
\end{algorithmic}
\end{algorithm}

\begin{theorem}\label{thm:inch3}
Suppose that $n\ge4$, that the agents know $n$, and that all agents are honest. Then three agents solve \PE{} in the face-to-face model.
\end{theorem}

\begin{proof}
For $n=4$, one of the three assigned omission tours avoids the gray hole and visits every safe node infinitely often. Now let $n\ge5$. If no agent is ever destroyed, all meetings take place, and agent $B$, for example, visits every node infinitely often, and \PE{} is obtained. Otherwise, a survivor notices a missing partner, and becomes the detector, before all three agents can be destroyed. Indeed, under the quiet schedule, no node other than the safe node $\Home$ holds all three agents (\cref{fig:f2f-inchworm}), so the first activation that destroys an agent leaves a survivor, and the four cases below and the two cases of \cref{fig:inch3-double} cover every way in which a second agent can be destroyed. Two agents that stand on the same node in the same round are destroyed together or not at all, because an active gray hole destroys every honest agent on it. Hence an agent can find a partner missing only when the schedule brings the two agents together again after a move. Every move of the quiet schedule takes place in the last round of a step, so the first missed meeting occurs in the first round of a step.

One period of \cref{fig:f2f-inchworm} uses the five nodes $p-1,\dots,p+3$, and the whole argument takes place on $p-1,\dots,p+3$ and on node $p-2$, which $B$ uses in the previous period. After the start-up, a move brings two agents together in only two rows of \cref{fig:f2f-inchworm}. In row $0$ of period $p+1$, agent $B$ joins $A$ on node $p+1$ (the last arrow of the figure), and in row $3$ of period $p$, agent $A$ joins $C$ on node $p+2$. Each of the two partners can be the detector, so there are four cases, shown in \cref{fig:inch3-cases}. In each case, the detector looks at the scheduled route of its partner since their last meeting, and the candidates are the nodes of the missing partner's scheduled route on which the partner may have been destroyed.

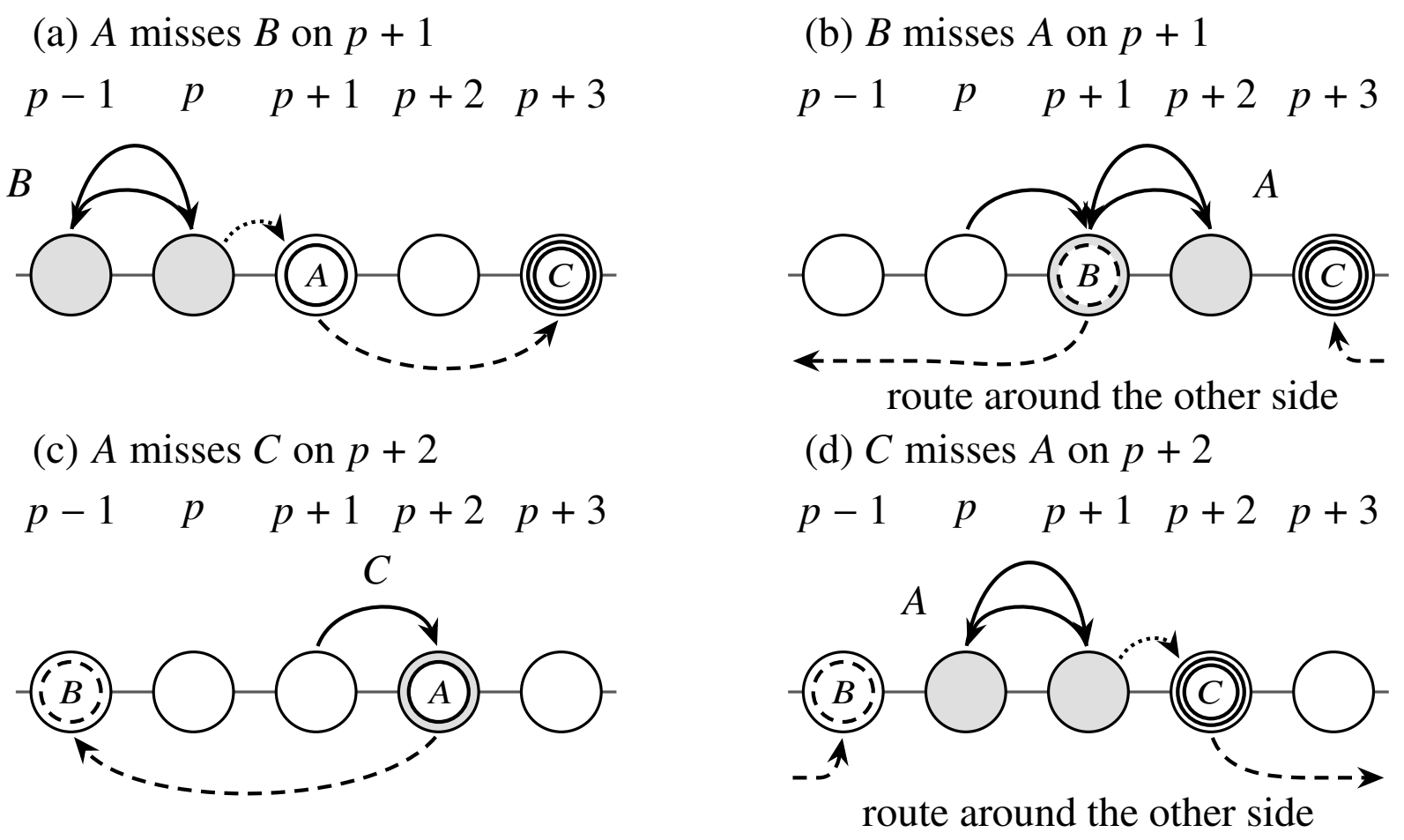
\begin{figure}[t]
\centering
\begin{tikzpicture}[x=0.92cm,y=0.92cm]
  \begin{scope}[shift={(0,0)}]
    \node[ttl] at (-0.45,1.95) {(a) $A$ misses $B$ on $p+1$};
    \foreach \i/\l in {0/{p-1},1/p,2/{p+1},3/{p+2},4/{p+3}} \node[lab] at (\i,1.45) {$\l$};
    \draw[edge] (-0.45,0)--(4.45,0);
    \node[cand] (a0) at (0,0) {};
    \node[cand] (a1) at (1,0) {};
    \node[vtx] (a2) at (2,0) {};
    \node[agA] at (2,0) {$A$};
    \node[vtx] (a3) at (3,0) {};
    \node[vtx] (a4) at (4,0) {};
    \node[agC] at (4,0) {$C$};
    \draw[move]  (a1.north) to[out=110,in=70,looseness=1.2] (a0.north);
    \draw[move]  (a0.north) to[out=80,in=100,looseness=2.4] (a1.north);
    \draw[ruled] (a1.north east) to[out=60,in=120,looseness=1.3] (a2.north west);
    \node[lab,fill=white,inner sep=1pt] at (-0.42,0.75) {$B$};
    \draw[walk] (a2.south) to[out=-50,in=-130,looseness=0.9] (a4.south);
  \end{scope}
  \begin{scope}[shift={(6.3,0)}]
    \node[ttl] at (-0.45,1.95) {(b) $B$ misses $A$ on $p+1$};
    \foreach \i/\l in {0/{p-1},1/p,2/{p+1},3/{p+2},4/{p+3}} \node[lab] at (\i,1.45) {$\l$};
    \draw[edge] (-0.45,0)--(4.45,0);
    \node[vtx] (b0) at (0,0) {};
    \node[vtx] (b1) at (1,0) {};
    \node[cand] (b2) at (2,0) {};
    \node[agB] at (2,0) {$B$};
    \node[cand] (b3) at (3,0) {};
    \node[vtx] (b4) at (4,0) {};
    \node[agC] at (4,0) {$C$};
    \draw[move] (b1.north) to[out=70,in=110,looseness=1.2] (b2.north);
    \draw[move] (b2.north) to[out=70,in=110,looseness=1.2] (b3.north);
    \draw[move] (b3.north) to[out=100,in=80,looseness=2.4] (b2.north);
    \node[lab,fill=white,inner sep=1pt] at (3.45,0.75) {$A$};
    \draw[walk] (b2.south) to[out=-110,in=0] (0.6,-0.7) -- (-0.45,-0.7);
    \draw[walk] (4.45,-0.7) -- (4.2,-0.7) to[out=180,in=-80] (b4.south);
    \node[lab] at (2.2,-0.98) {route around the other side};
  \end{scope}
  \begin{scope}[shift={(0,-3.4)}]
    \node[ttl] at (-0.45,1.95) {(c) $A$ misses $C$ on $p+2$};
    \foreach \i/\l in {0/{p-1},1/p,2/{p+1},3/{p+2},4/{p+3}} \node[lab] at (\i,1.45) {$\l$};
    \draw[edge] (-0.45,0)--(4.45,0);
    \node[vtx] (c0) at (0,0) {};
    \node[agB] at (0,0) {$B$};
    \node[vtx] (c1) at (1,0) {};
    \node[vtx] (c2) at (2,0) {};
    \node[cand] (c3) at (3,0) {};
    \node[agA] at (3,0) {$A$};
    \node[vtx] (c4) at (4,0) {};
    \draw[move] (c2.north) to[out=70,in=110,looseness=1.2] (c3.north);
    \node[lab,fill=white,inner sep=1pt] at (2.5,1.0) {$C$};
    \draw[walk] (c3.south) to[out=-130,in=-50,looseness=0.7] (c0.south);
  \end{scope}
  \begin{scope}[shift={(6.3,-3.4)}]
    \node[ttl] at (-0.45,1.95) {(d) $C$ misses $A$ on $p+2$};
    \foreach \i/\l in {0/{p-1},1/p,2/{p+1},3/{p+2},4/{p+3}} \node[lab] at (\i,1.45) {$\l$};
    \draw[edge] (-0.45,0)--(4.45,0);
    \node[vtx] (d0) at (0,0) {};
    \node[agB] at (0,0) {$B$};
    \node[cand] (d1) at (1,0) {};
    \node[cand] (d2) at (2,0) {};
    \node[vtx] (d3) at (3,0) {};
    \node[agC] at (3,0) {$C$};
    \node[vtx] (d4) at (4,0) {};
    \draw[move]  (d2.north) to[out=110,in=70,looseness=1.2] (d1.north);
    \draw[move]  (d1.north) to[out=80,in=100,looseness=2.4] (d2.north);
    \draw[ruled] (d2.north east) to[out=60,in=120,looseness=1.3] (d3.north west);
    \node[lab,fill=white,inner sep=1pt] at (0.58,0.75) {$A$};
    \draw[walk] (d3.south) to[out=-70,in=180] (3.8,-0.7) -- (4.45,-0.7);
    \draw[walk] (-0.45,-0.7) -- (-0.2,-0.7) to[out=0,in=-100] (d0.south);
    \node[lab] at (2.0,-0.98) {route around the other side};
  \end{scope}
\end{tikzpicture}
\caption{The four ways in which a meeting can fail, on the five nodes $p-1,\dots,p+3$ of \cref{fig:f2f-inchworm}. The small circles show the detector and the third agent ($A$ with a solid, $B$ with a dashed, and $C$ with a double outline). Solid arrows show the scheduled route of the missing partner since the last meeting, and a dotted arrow shows a move that the survival of the detector rules out. Light-gray nodes show candidate nodes; $\Home$ is excluded whenever one of the displayed node indices denotes it. The dashed arrow is the warning walk of the detector to the third agent, which avoids every candidate after its starting node, and in (b) and (d) it takes the route around the other side of the ring.}
\label{fig:inch3-cases}
\end{figure}

\begin{figure}[t]
\centering
\begin{tikzpicture}[x=0.92cm,y=0.92cm]
  \begin{scope}[shift={(0,0)}]
    \node[ttl] at (-0.45,1.95) {(e) $C$ misses $A$, then $B$};
    \foreach \i/\l in {0/{p-2},1/{p-1},2/p,3/{p+1},4/{p+2},5/{p+3}} \node[lab] at (\i,1.45) {$\l$};
    \draw[edge] (-0.45,0)--(5.45,0);
    \node[vtx] at (0,0) {};
    \node[vtx] (e1) at (1,0) {};
    \node[gh]  at (2,0) {};
    \node[vtx] at (3,0) {};
    \node[vtx] (e4) at (4,0) {}; \node[agC] at (4,0) {$C$};
    \node[vtx] at (5,0) {};
    \draw[thick] (-0.35,-0.42) -- (-0.35,-0.52) -- (2.3,-0.52) -- (2.3,-0.42);
    \node[lab] at (0.5,-0.75) {for $B$};
    \draw[thick] (1.7,-0.85) -- (1.7,-0.95) -- (3.35,-0.95) -- (3.35,-0.85);
    \node[lab] at (2.5,-1.2) {for $A$};
    \draw[walk] (e4.north) to[out=60,in=180] (4.8,0.8) -- (5.45,0.8);
    \draw[walk] (-0.45,0.8) -- (0.3,0.8) to[out=0,in=120] (e1.north);
  \end{scope}
  \begin{scope}[shift={(7.2,0)}]
    \node[ttl] at (-0.45,1.95) {(f) $B$ misses $A$, then $C$};
    \foreach \i/\l in {0/{p-2},1/{p-1},2/p,3/{p+1},4/{p+2},5/{p+3}} \node[lab] at (\i,1.45) {$\l$};
    \draw[edge] (-0.45,0)--(5.45,0);
    \node[vtx] at (0,0) {};
    \node[vtx] at (1,0) {};
    \node[vtx] at (2,0) {};
    \node[vtx] (f3) at (3,0) {}; \node[agB] at (3,0) {$B$};
    \node[gh]  at (4,0) {};
    \node[vtx] (f5) at (5,0) {};
    \draw[thick] (2.65,-0.42) -- (2.65,-0.52) -- (4.3,-0.52) -- (4.3,-0.42);
    \node[lab] at (3.5,-0.75) {for $A$};
    \draw[thick] (3.7,-0.85) -- (3.7,-0.95) -- (5.35,-0.95) -- (5.35,-0.85);
    \node[lab] at (4.5,-1.2) {for $C$};
    \draw[walk] (f3.north) to[out=120,in=0] (2.2,0.8) -- (-0.45,0.8);
    \draw[walk] (5.45,0.8) -- (5.3,0.8) to[out=180,in=90] (f5.north);
  \end{scope}
\end{tikzpicture}
\caption{When the third agent is missing too, the candidates of the two missing agents have exactly one node in common, shown in dark gray. (e) After case~(d) of \cref{fig:inch3-cases}, $C$ finds $B$ missing on $p-1$, and only $p$ is a candidate for both $A$ and $B$. (f) After case~(b), $B$ finds $C$ missing on $p+3$, and only $p+2$ is a candidate for both $A$ and $C$. The drawing is unrolled: node indices are interpreted modulo $n$, so $p-2$ and $p+3$ coincide when $n=5$. Brackets show the candidates of each missing agent, and dashed arrows show the warning walk around the other side of the ring.}
\label{fig:inch3-double}
\end{figure}
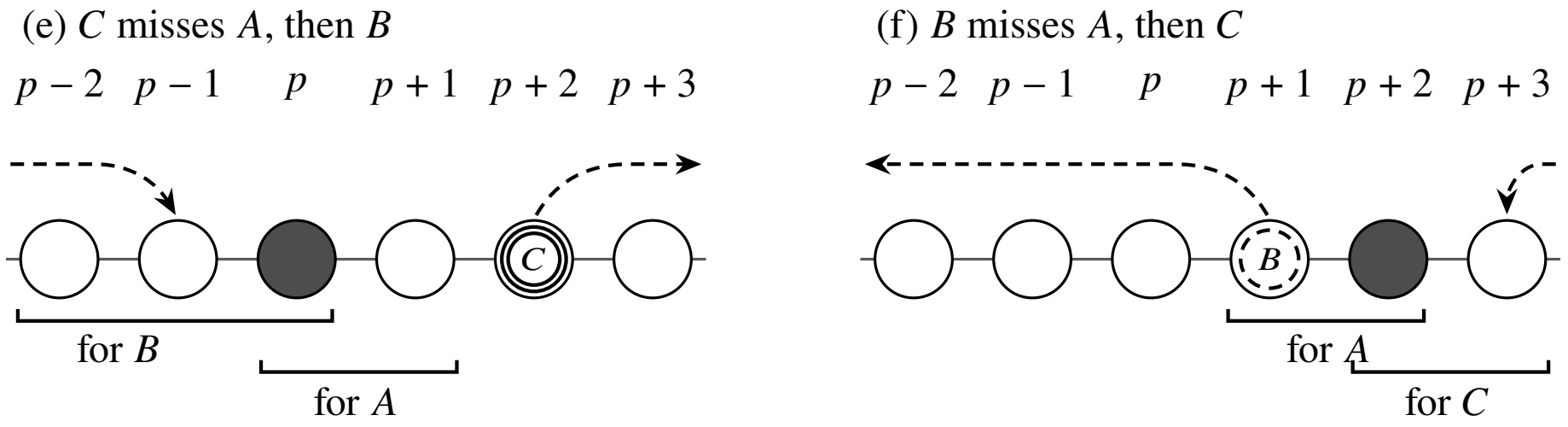

\begin{itemize}
  \item \emph{Case (a): $A$ misses $B$ on node $p+1$.} The two agents last met on node $p$, in row $0$ of period $p$. Since then, $B$ stood on node $p-1$ and on node $p$. Its last move, into node $p+1$, is ruled out, because $A$ survived on node $p+1$ in the round of the scheduled move into $p+1$. Hence the candidates lie in $\{p-1,p\}$. Agent $C$ waits on node $p+3$, and $A$ reaches it in two moves, through node $p+2$.
  \item \emph{Case (b): $B$ misses $A$ on node $p+1$.} Since their last meeting on node $p$, agent $A$ stood on nodes $p+1$ and $p+2$, so the candidates lie in $\{p+1,p+2\}$. Agent $C$ waits on node $p+3$. The direct route to it passes node $p+2$, so $B$ takes the route around the other side of the ring, counterclockwise through node $p$, with $n-2$ edges.
  \item \emph{Case (c): $A$ misses $C$ on node $p+2$.} The two agents last met on node $p+1$ in period $p-1$, just before $C$ moved on to node $p+2$, and since then $C$ has stood only on node $p+2$. Hence the only candidate is $p+2$. Agent $B$ waits on node $p-1$, and $A$ leaves node $p+2$ at once and reaches $B$ by a shortest path that avoids $p+2$, in at most three moves.
  \item \emph{Case (d): $C$ misses $A$ on node $p+2$.} The two agents last met on node $p+1$ in period $p-1$. Since then, $A$ stood on nodes $p$ and $p+1$. Its last move, into node $p+2$, is ruled out, because $C$ survived on node $p+2$ in the round of the scheduled move into $p+2$. Hence the candidates lie in $\{p,p+1\}$. Agent $B$ waits on node $p-1$. The direct route to it passes node $p+1$, so $C$ takes the route around the other side of the ring, clockwise through node $p+3$, with $n-3$ edges.
\end{itemize}

In all four cases, the candidates are one node or two adjacent nodes other than the safe node $\Home$, and they contain the gray hole. The third agent waits outside them and has no meeting of its own in the step of the missed meeting. The nodes outside the candidates form a path, so the warning walk has at most $n-2$ edges and avoids every candidate after its first node. Even if the detector starts on the gray hole, its survival means that the gray hole is inactive in the round of the missed meeting, so immediate departure is safe. The detector leaves in the first round of the step, so the warning arrives before the final round of the step, while the third agent, if alive, is still waiting.

After cases~(a) and~(c), the third agent cannot be missing as well. Since it was last seen, it used only nodes $p+2$ and $p+3$ in case~(a), and only nodes among $p-2$, $p-1$, and $p$ in case~(c). For $n\ge5$, the nodes just listed are disjoint from the candidates, which contain the gray hole. If the third agent is also missing after case~(b) or~(d), the candidates of the two missing agents have exactly one node in common (\cref{fig:inch3-double}). Suppose first that $C$ misses $A$ on node $p+2$, as in case~(d), and then misses $B$ on node $p-1$. The candidates for $A$ lie in $\{p,p+1\}$. Agent $C$ last heard of $B$ through $A$, whom $B$ had met on node $p-1$ at the start of period $p-1$, and since then $B$ used only nodes $p-2$, $p-1$, and $p$. Hence only node $p$ is a candidate for both $A$ and $B$. Suppose next that $B$ misses $A$ on node $p+1$, as in case~(b), and then misses $C$ on node $p+3$. The candidates for $A$ lie in $\{p+1,p+2\}$. Agent $B$ last heard of $C$ through $A$, when $A$ and $B$ met on node $p$, and $C$ had then already moved from node $p+1$ to node $p+2$, after which it used only nodes $p+2$ and $p+3$. Hence only node $p+2$ is a candidate for both $A$ and $C$.

The start-up needs a short separate check, because the chain is not yet formed. During its four steps, agent $B$ waits on the safe node $\Home=0$, and the other agents use only nodes $1$ and $2$. Suppose first that the gray hole is node $1$. If it destroys $C$ before $A$ arrives, and $A$ survives its arrival, then $A$ misses $C$ on node $1$, identifies node $1$ as the gray hole, since $C$ used no other node, and returns to $B$. If it destroys $A$ at node $1$, then $B$ misses $A$ when $A$ should return to $\Home$. Since $A$ used only $\Home$ and node $1$, agent $B$ identifies node $1$ as the gray hole and walks counterclockwise to node $2$ to warn $C$, if $C$ is still alive, without entering node $1$. Suppose next that the gray hole is node $2$. Agent $C$ waits there alone until $A$ joins it in row $3$ of period $0$ in \cref{fig:f2f-inchworm}, so a destruction on node $2$ is case~(c), or case~(b) followed by \cref{fig:inch3-double}(f), with $p=0$. Either $A$ misses $C$ on node $2$, or $A$ is destroyed there as well, and $B$ then misses $A$ on node $1$ and $C$ on node $3$, as in \cref{fig:inch3-double}(f) with $p=0$, and identifies node $2$. The start-up uses no other node.

Thus, either a survivor identifies $g$ as the gray hole and walks the omission tour $O(g)$, or two survivors share the candidate set $\{g_1,g_2\}$, which contains $g$. One survivor walks the omission tour $O(g_1)$, and the other walks the omission tour $O(g_2)$. One of the two assigned tours is $O(g)$, which avoids the gray hole, so its agent survives and visits every safe node infinitely often. Hence the algorithm achieves \PE{}.
\end{proof}

\subsection{Tolerating \texorpdfstring{$f$}{f} Byzantine Agents}
\label{sec:f2f-byzantine}

The three-agent algorithm extends to Byzantine agents in a simple way. No agent can forge the identifier of another agent, so independent groups of three cannot disturb each other. Hence the team splits into $f+1$ independent groups, at least one of which is entirely honest.

\begin{corollary}\label{cor:f2f-known}
Suppose that $n\ge4$, that the agents know $n$, and that at most $f$ agents are Byzantine. Then, in the face-to-face model, $3f+3$ agents suffice for \PE{}, and at least $2f+3$ agents are necessary. For $f=0$, three agents are necessary and sufficient.
\end{corollary}

\begin{proof}
At $\Home$, the agents form $f+1$ groups of three in the order of their identifiers. Each group runs the algorithm of \cref{thm:inch3} and uses only the messages of its own members. The entirely honest group runs the three-agent algorithm undisturbed, so \cref{thm:inch3} applies to it, and one of its agents visits every safe node infinitely often. The lower bound comes from \cref{thm:f2f-lower}.
\end{proof}

\subsection{Five Agents on Rings of Four or Five Nodes}
\label{sec:f2f-small}
\begin{figure}[t]
\centering
\begin{tikzpicture}[x=1.3cm,y=0.74cm]
  \foreach \c/\l in {0/\Home,1/w,2/z} \node at (\c,0.95) {$\l$};
  \foreach \i in {0,...,4}{
    \draw[tgrid] (-0.4,-\i)--(2.4,-\i);
    \foreach \c in {0,1,2} \node[dot] at (\c,-\i) {};}
  \node[ag] at (-0.20,0) {$P$};  \node[ag] at (0.20,0) {$E$};
  \node[ag] at (0.80,-1) {$P$};  \node[ag] at (1.20,-1) {$E$};
  \node[ag] at (1,-2) {$E$};     \node[ag] at (2,-2) {$P$};
  \node[ag] at (0.80,-3) {$P$};  \node[ag] at (1.20,-3) {$E$};
  \node[ag] at (-0.20,-4) {$P$}; \node[ag] at (0.20,-4) {$E$};
  \draw[move] (-0.1,-0.3)--(0.78,-0.72);
  \draw[move] (0.22,-0.3)--(1.1,-0.72);
  \draw[move] (0.9,-1.3)--(1.92,-1.72);
  \draw[move] (1.92,-2.3)--(0.9,-2.72);
  \draw[move] (0.78,-3.3)--(-0.1,-3.72);
  \draw[move] (1.1,-3.3)--(0.22,-3.72);
  \foreach \i/\t in {0/{Leave $\Home$ together},1/{Check for partner},2/{Prober returns to $w$},3/{Check again; return to $\Home$},4/{Report at the census}}
    \node[anchor=west,lab] at (2.65,-\i) {\t};
\end{tikzpicture}
\caption{An escorted probe; time runs from top to bottom, and each row shows where the prober $P$ and the escort $E$ stand. The prober visits $z$ alone while the escort waits on $w$. Both agents move from $\Home$ to $w$ together and return to $\Home$ together.}
\label{fig:f2f-probe}
\end{figure}
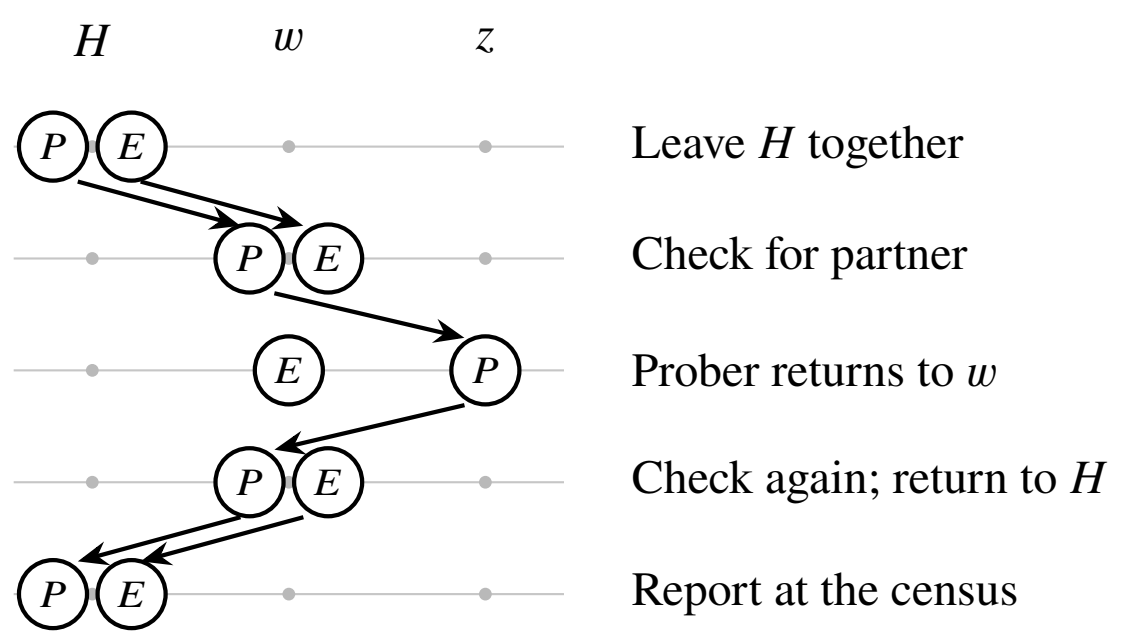

For $f=1$, at least five agents are necessary on every ring with $n\ge4$ (\cref{thm:f2f-lower}). On rings of four or five nodes, every node is at distance at most two from $\Home$, so five agents can also suffice. We use two agents, called the \emph{patrol}, to visit the nodes while three agents wait at $\Home$. In an \emph{escorted probe}, a \emph{prober} and an \emph{escort} move together from $\Home$ to a neighbor $w$; the prober visits the next node $z$ alone while the escort waits on $w$, and both return to $\Home$. Each partner reports whether it saw the other partner on $w$ both before the prober visited $z$ and after the prober returned. A \emph{positive report} confirms both meetings; a \emph{negative report} means that at least one was missed. At $\Home$, the agents check the present identifiers and the two reports. We call the check the \emph{census}; it succeeds exactly when all five agents are present and both reports are positive. The timed probe is shown in \cref{fig:f2f-probe}.
The two agents with the smallest identifiers form the patrol, and the other three agents, $q_1,q_2,q_3$ in increasing order of identifier, wait at $\Home$. In each \emph{cycle}, the patrol makes two escorted probes on the clockwise side, with $(w,z)=(u_1,u_2)$, first with the smaller-identifier partner as the prober and then with the other partner as the prober. It repeats the two probes on the counterclockwise side, with $(w,z)=(u_{n-1},u_{n-2})$. When $n=4$, the two choices of $z$ coincide. At each census, a partner's response counts as positive only if it is a single positive report; any missing, conflicting, or malformed response counts as negative. The patrol continues after a successful census. At the first census that fails, the agents decide from the identifiers they see and the reports they hear. We write $\mathcal{P}_3$ for the three-agent algorithm of \cref{sec:f2f-algorithms}, started by three agents at $\Home$ that ignore all other agents. At a new start of $\mathcal{P}_3$, the three agents number their local rounds from $1$, so the schedule uses their common restart round as its origin.
An escorted probe with nodes $(w,z)$ that starts in round $t$ takes four rounds, and row $i$ of \cref{fig:f2f-probe} shows where the two partners stand at the start of round $t+i$. In round $t$, both partners move from $\Home$ to $w$. In round $t+1$, each partner checks whether the other partner is on $w$, and the prober moves from $w$ to $z$. In round $t+2$, the prober moves from $z$ back to $w$. In round $t+3$, each partner checks again whether the other partner is on $w$, and both partners move from $w$ to $\Home$. In round $t+4$, the census takes place on $\Home$, and the next escorted probe starts in the same round. Thus a cycle of four escorted probes takes $16$ rounds. An honest partner follows the four-round probe even if it misses the other partner on $w$. Its report at the census is positive if it saw the other partner on $w$ in both rounds $t+1$ and $t+3$, and negative otherwise. Every agent on a node hears every message sent there, so all honest agents on $\Home$ see the same agents, hear the same reports, and apply the same rule.

The first census that fails falls into one of the following cases, each with its own rule, where $(w,z)$ are the nodes of the last escorted probe.

\begin{enumerate}[label=\textit{Case \arabic*.},wide,labelindent=0pt,topsep=3pt,itemsep=2pt]
  \item \emph{A waiting agent is missing from $\Home$.}
  \begin{enumerate}[label=(\alph*),leftmargin=*,topsep=2pt,itemsep=1pt]
    \item \emph{The escort is missing.} Every agent walks the omission tour $O(w)$.
    \item \emph{Only the prober is missing.} Every agent walks the omission tour $O(z)$.
    \item \emph{Both partners are back.} The three agents at $\Home$ with the smallest identifiers start $\mathcal{P}_3$.
  \end{enumerate}
  \item \emph{All three waiting agents are present.}
  \begin{enumerate}[label=(\alph*),leftmargin=*,topsep=2pt,itemsep=1pt]
    \item \emph{Both partners are back but a report is negative, or the escort is missing while the prober is back.} Agents $q_1,q_2,q_3$ start $\mathcal{P}_3$.
    \item \emph{Only the escort is back.} The escort walks the omission tour $O(z)$, while $q_1,q_2,q_3$ start $\mathcal{P}_3$.
    \item \emph{Neither partner is back.} Agents $q_1$ and $q_2$ walk the omission tour $O(w)$, and $q_3$ walks the omission tour $O(z)$.
  \end{enumerate}
\end{enumerate}
The decisions preserve an honest explorer, as we prove in \cref{lem:patrol-cases}. To implement the probe schedule, each agent stores a probe index $c\in\{0,1,2,3\}$, initially $0$, and its starting round $t$, initially $1$. The index determines $(w,z)$ and the prober: indices $0,1$ use $(u_1,u_2)$, indices $2,3$ use $(u_{n-1},u_{n-2})$, and the smaller-identifier patrol agent is the prober exactly when $c$ is even. Each partner records two yes/no values, initially no, indicating whether it sees the other partner on $w$ in rounds $t+1$ and $t+3$. The report is positive exactly when both values are yes. Every agent stores the identifiers present and the two reports at the census, together with the last pair $(w,z)$ needed by \textsc{Decide}. After a successful census, it sets $c\gets(c+1)\bmod4$ and $t\gets t+4$ and resets the two meeting values for the next probe. The waiting identifiers $q_1,q_2,q_3$ remain fixed. \Cref{alg:patrol} summarizes the probe loop and the decision procedure.

\begin{algorithm}[tbp]
\caption{The five-agent patrol for $n\in\{4,5\}$ and $f=1$, for an honest agent $X$.}
\label{alg:patrol}
\begin{algorithmic}[1]
\Statex \textbf{Initially:} $c\gets0$, $t\gets1$; set both meeting values to no.
\Statex \textbf{Cycle:} $(u_1,u_2)$, $(u_1,u_2)$, $(u_{n-1},u_{n-2})$, $(u_{n-1},u_{n-2})$; the smaller-identifier patrol agent is the prober when $c$ is even.
\Repeat \Comment{one escorted probe, rounds $t,\dots,t+4$}
  \If{$X$ belongs to the patrol}
    \State walk the escorted probe as prober or escort, check for the other partner on $w$ in rounds $t+1$ and $t+3$, and send the report at the census
  \Else
    \State wait on $\Home$
  \EndIf
  \If{the census succeeds}
    \State $c\gets(c+1)\bmod4$; $t\gets t+4$; reset both meeting values to no
  \EndIf
\Until{the census fails}
\State \Call{Decide}{} \Comment{$(w,z)$ are the nodes of the last escorted probe}
\Statex
\Procedure{Decide}{}
  \If{a waiting agent is missing} \Comment{Case~1: the patrol is honest}
    \If{the escort is missing}
      \State every agent on $\Home$ walks the omission tour $O(w)$ \Comment{Case~1(a)}
    \ElsIf{the prober is missing}
      \State every agent on $\Home$ walks the omission tour $O(z)$ \Comment{Case~1(b)}
    \Else
      \State the three agents on $\Home$ with the smallest identifiers start $\mathcal{P}_3$ \Comment{Case~1(c)}
    \EndIf
  \ElsIf{both partners are back and a report is negative, or only the prober is back}
    \State $q_1$, $q_2$, and $q_3$ start $\mathcal{P}_3$ \Comment{Case~2(a)}
  \ElsIf{only the escort is back}
    \State the escort walks the omission tour $O(z)$, and $q_1$, $q_2$, and $q_3$ start $\mathcal{P}_3$ \Comment{Case~2(b)}
  \Else \Comment{Case~2(c): neither partner is back}
    \State $q_1$ and $q_2$ walk the omission tour $O(w)$, and $q_3$ walks the omission tour $O(z)$
  \EndIf
\EndProcedure
\end{algorithmic}
\end{algorithm}

The analysis uses the fact that two honest partners use $w$ in exactly the same rounds. In round $t$, both partners enter $w$. In round $t+1$, both stand on $w$ at the start of the round. In round $t+2$, the escort stands on $w$ and the prober enters $w$ from $z$. In round $t+3$, both stand on $w$ again. The escort uses no node other than $\Home$ and $w$, and the prober uses $z$ only in rounds $t+1$ and $t+2$.

\begin{lemma}\label{lem:honest-pair}
Consider a four-round escorted probe with nodes $(w,z)$, starting from the safe home node in round $t$, in the face-to-face model. Suppose that both the prober and the escort are honest. Then the following statements hold.
\begin{enumerate}[label=(\roman*)]
  \item If the escort is missing at the census, then the prober is missing too, and $w$ is the gray hole.
  \item If only the prober is missing at the census, then $z$ is the gray hole.
  \item If both partners are back, then both reports are positive.
\end{enumerate}
\end{lemma}

\begin{proof}
Whether the gray hole is active is the same for all honest agents in a round. (i) The escort uses only $\Home$ and $w$, and $\Home$ is safe, so the gray hole is $w$ and destroyed the escort in some round among $t,\dots,t+3$. The prober uses $w$ in the same destruction round, so the gray hole also destroys the prober. (ii) The escort survives every round $t,\dots,t+3$ on $w$, so $w$ is not active in any round from $t$ to $t+3$, which are the only rounds in which the prober uses $w$. Hence the prober was destroyed at node $z$. (iii) Neither partner was destroyed, so both stood on $w$ at the start of rounds $t+1$ and $t+3$, and each partner saw the other one in both rounds.
\end{proof}

\begin{lemma}\label{lem:patrol-cases}
Suppose that five agents execute the patrol algorithm on a ring of known size $n\in\{4,5\}$ and that at most one agent is Byzantine. At the first census that fails, the rule applied by \textsc{Decide} leaves an honest agent that visits every safe node infinitely often.
\end{lemma}

\begin{proof}
An honest waiting agent stays on the safe node $\Home$, so it is never missing.

In Case~1, a waiting agent is missing. The missing agent is then the Byzantine agent, so the other four agents are honest, and in particular the patrol is honest. If the escort is missing, $w$ is the gray hole by \cref{lem:honest-pair}(i), and every surviving agent walks the omission tour of the gray hole. If only the prober is missing, $z$ is the gray hole by \cref{lem:honest-pair}(ii), and again every surviving agent walks the omission tour of the gray hole. If both partners are back, the four agents on $\Home$ are honest, and three of them start $\mathcal{P}_3$ together on $\Home$. They ignore all other agents, so by \cref{thm:inch3} one of them visits every safe node infinitely often.

In Case~2, all three waiting agents are present. By \cref{lem:honest-pair}(i) and (iii), if the patrol is honest, a census never finds both partners back with a negative report, and never finds the prober back without the escort. Hence, in Case~2(a), the Byzantine agent belongs to the patrol, the agents $q_1$, $q_2$, and $q_3$ are honest, and $\mathcal{P}_3$ achieves \PE{} by \cref{thm:inch3}. In Case~2(b), only the escort is back. If the patrol is honest, $z$ is the gray hole by \cref{lem:honest-pair}(ii), and the honest escort walks the omission tour of the gray hole. Otherwise, $q_1$, $q_2$, and $q_3$ are honest, and $\mathcal{P}_3$ achieves \PE{}. In Case~2(c), neither partner is back. If $w$ is the gray hole, at most one of $q_1$ and $q_2$ is Byzantine, so the other one walks the omission tour of the gray hole. Otherwise, $w$ is safe. An honest escort uses only the safe nodes $\Home$ and $w$, so it would be back, and hence the escort is the Byzantine agent. The prober is then honest, and it follows its route even without the escort, so it uses only $\Home$, $w$, and $z$, and the prober was destroyed at node $z$. The honest agent $q_3$ then walks the omission tour of the gray hole.

The cases are exhaustive: a missing waiting agent gives Case~1, and otherwise the partners' presence and reports determine Case~2(a), 2(b), or 2(c). The remaining combination, all five agents present with two positive reports, is a successful census.
\end{proof}

The patrol algorithm also improves the bound for general $f$. A five-agent group tolerating one Byzantine agent can fail only if it contains at least two Byzantine agents, whereas an entirely honest group of three succeeds by the three-agent algorithm. We therefore use as many five-agent groups as possible. Write $f+1=2s+r$, where $s=\lfloor(f+1)/2\rfloor$ and $r\in\{0,1\}$. At $\Home$, the agents form $s$ groups of five in identifier order and, when $r=1$, one group of three. Each five-agent group runs the patrol algorithm, the group of three runs $\mathcal{P}_3$, and each group uses only the messages of its own members. The construction uses $5s+3r=\lceil5(f+1)/2\rceil$ agents. The following corollary proves both the five-agent guarantee and the resulting bound for general $f$.

\begin{corollary}\label{cor:f2f-small}
Suppose that $n\in\{4,5\}$ and that the agents know $n$. If at most one agent is Byzantine, five agents are necessary and sufficient for \PE{} in the face-to-face model. For every $f\ge0$, if at most $f$ agents are Byzantine, $\lceil 5(f+1)/2\rceil$ agents suffice.
\end{corollary}

\begin{proof}
Let $f=1$. If every census succeeds, no honest agent is ever destroyed, since a destroyed agent would be missing at the next census. At least one partner is honest, and the partners take turns as the prober. Hence, in every cycle, an honest partner walks the escorted probes $(u_1,u_2)$ and $(u_{n-1},u_{n-2})$ as the prober, and visits $\Home,u_1,u_2,u_{n-1},u_{n-2}$, which are all the nodes of the ring. Otherwise, a census fails, and \cref{lem:patrol-cases} gives an honest agent that visits every safe node infinitely often. Thus five agents suffice for $f=1$, and four agents do not suffice by \cref{thm:f2f-lower}. For general $f$, consider the grouping described above. If every group failed, each of the $s$ five-agent groups would contain at least two Byzantine agents, and the group of three, when present, would contain at least one. There would then be at least $2s+r=f+1$ Byzantine agents, a contradiction. Therefore, at least one group succeeds and achieves \PE{}. The number of agents is $5s+3r$, equal to $5(f+1)/2$ when $f+1$ is even and $5f/2+3$ when $f+1$ is odd, hence $\lceil5(f+1)/2\rceil$ in both cases. For example, $f=0$ gives one group of three, $f=1$ one group of five, and $f=2$ one group of five and one group of three, using eight agents instead of nine.
\end{proof}

For known $n$ and one Byzantine agent, our general bounds leave open whether five or six agents are needed, since $2f+3=5$ agents are necessary and $3f+3=6$ agents suffice. The patrol shows that five agents suffice for $n=4$ and $n=5$. For $n\ge6$, escorted probes that reach two edges from $\Home$ no longer visit every node, and whether five agents suffice for one Byzantine agent remains open.

\section{Perpetual Exploration of Rings of Unknown Size in the Face-to-Face Model}
\label{sec:f2f-unknown}

In the three-agent algorithm of \cref{sec:f2f-algorithms}, each step lasts $n$ rounds so that an agent detecting a missed meeting can warn the third agent before its next scheduled move. The warning walk may go around the other side of the ring to avoid the possible locations of the gray hole. Without an upper bound on $n$, the agents cannot choose a waiting time long enough for every possible ring.

For comparison, we can adapt the path algorithm of Bhattacharya et al.~\cite{DBLP:conf/wdag/BhattacharyaGB025} using two independent groups of four agents, one in each direction. Each group treats a complete walk around the ring as a path with a separate copy of $\Home$ at each end. To track its position, an agent starts with count zero, adds one for a move in the group's direction, and subtracts one for a move against it. Agents within a group use a meeting only when their counts are equal, keeping the two copies of $\Home$ separate. After a destruction, the cited algorithm guarantees a survivor that identifies the gray hole and explores the remaining path on its side of the gray hole. If the survivor is beyond the gray hole in the chosen direction, it reaches the returning copy of $\Home$, learns $n$ from its count, and walks the omission tour of the gray hole. Otherwise, it repeatedly explores the safe nodes between the starting copy of $\Home$ and the gray hole. Hence either a survivor walks the omission tour of the gray hole or the two groups repeatedly explore the two parts of the safe path. If no agent is destroyed, the path exploration already visits every safe node infinitely often. The construction uses eight honest agents, or $8f+8$ agents with independent groups when at most $f$ agents are Byzantine.

In this section, we show that five honest agents are necessary and sufficient when $n$ is unknown. Our five-agent algorithm works on every ring with $n\ge3$, including the triangle. The lower bound uses larger rings; it does not assert that five agents are necessary on a triangle when its size is known. Finally, when at most $f$ agents are Byzantine, we show that $3f+3$ agents do not suffice, whereas $5f+5$ agents suffice. Throughout the section, $v=u_1$ and $w=u_2$ are the first two nodes clockwise from $\Home$, and $c_i=u_{(n-i)\bmod n}$ is the node at counterclockwise distance $i$ from $\Home$. Thus $c_0=\Home$, $c_{n-1}=v$, and $c_{n-2}=w$.

\subsection{Four Honest Agents Are Not Enough}
\label{sec:f2f-unknown-lower}

In this subsection, all agents are honest, do not know $n$, and run a fixed deterministic algorithm in the face-to-face model. The adversary chooses the ring size, the gray hole, and the rounds in which the gray hole is active. The known-size lower bound in \cref{thm:f2f-lower} compares two possible locations of the gray hole when an agent first moves two nodes away from $\Home$, from $v$ to $w$. The moving agent can be destroyed at $v$ or at $w$, while the agents outside $\{v,w\}$ meet the same agents and hear the same messages in both cases. As in the whiteboard model with unknown $n$ (\cref{sec:impossibility_unknown}), we also arrange a destruction far from $\Home$ on the counterclockwise side. There, changing the ring size changes which node is the gray hole without changing what the surviving agents have seen. The construction leaves at most two agents unable to tell three possible rings apart. Every move that could distinguish the rings enters the gray hole of at least one of them. The proof shows that, whichever moves the last agents choose, there is a ring on which perpetual exploration fails.

An agent's \emph{view} includes its stored information, including its next instruction. In the face-to-face model, the view also includes the agents it has seen and the messages it has heard in each round so far. When comparing rings with more than $i$ nodes, nodes with the same name $c_i$ are at the same counterclockwise distance from $\Home$. We say that two executions \emph{agree} up to a round $t$ if, at the end of every round up to $t$, the same agents survive, each of them stands on the same node $c_i$ in both executions, and each of them has the same view in both executions. Since the algorithm is deterministic, these agents also make the same moves. Recall also that an agent \emph{uses} a node in a round if it stands on the node at the start of the round or enters it during the round; an active gray hole destroys exactly the agents that use it. The following observation states when additional unvisited nodes leave the surviving agents with the same views.

\begin{observation}\label{obs:same-rings}
Consider two executions of the same algorithm, on two rings that both have more than $m$ nodes. Suppose that at the end of round $s$, the same agents are left in both executions, and each of them stands on the same node $c_i$ and has the same view in both executions. Suppose also that in every round from $s+1$ to $t$, the same agents are destroyed in both executions, and every agent that is left at the end of the round either stayed where it was or moved between two neighboring nodes $c_i$ and $c_{i+1}$ with $0\le i<m$. Then at the end of every round from $s+1$ to $t$, every agent that is left stands on the same node $c_i$ and has the same view in both executions. In particular, the two executions agree up to round $t$.
\end{observation}

\begin{proof}
We argue by induction on the rounds from $s+1$ to $t$. Consider such a round, and suppose that the claim holds at the end of the previous round. The nodes $c_0,\dots,c_m$ are distinct in both rings, and every agent that is left stands on the same node $c_i$ in both rings. Hence every agent sees the same agents on its node and hears the same messages in both executions. Its view was the same, and the algorithm is deterministic, so it makes the same move in both executions, through the same port. The rings are oriented, so in both rings the counterclockwise port of $c_i$ leads to $c_{i+1}$ and the clockwise port leads to $c_{i-1}$. Hence an agent that moves between $c_i$ and $c_{i+1}$ reaches the same node in both rings. The same agents are destroyed in both executions, so the claim holds at the end of the round.
\end{proof}

The next lemma compares two possible ring sizes after at most two agents remain. For a sufficiently large natural number $\delta$, the remaining agents cannot distinguish whether the gray hole $v$ is the node $c_\delta$, on a ring of $\delta+1$ nodes, or the node $c_{\delta+1}$, on a ring of $\delta+2$ nodes. Hence the last agent can neither enter $c_\delta$ safely nor skip it. \Cref{fig:two-left} shows the two rings.

\begin{figure}[t]
\centering
\begin{tikzpicture}[scale=0.9]
\begin{scope}[xshift=0cm]
  \draw[edge] (90:1.25) arc (90:135:1.25);
  \draw[omit] (135:1.25) arc (135:240:1.25);
  \draw[edge] (240:1.25) arc (240:450:1.25);
  \node[home] at (90:1.25) {$\Home$};
  \node[vtx] at (135:1.25) {}; \node[lab,anchor=south east] at (135:1.45) {$c_1$};
  \node[vtx] at (240:1.25) {}; \node[lab,anchor=north east] at (240:1.45) {$c_{\delta-1}$};
  \node[gh] at (330:1.25) {}; \node[lab,anchor=west] at (330:1.55) {$c_\delta=v$};
  \node[lab] at (0,-2.05) {$\rho_1$: $\delta+1$ nodes};
\end{scope}
\begin{scope}[xshift=6cm]
  \draw[edge] (90:1.25) arc (90:135:1.25);
  \draw[omit] (135:1.25) arc (135:240:1.25);
  \draw[edge] (240:1.25) arc (240:450:1.25);
  \node[home] at (90:1.25) {$\Home$};
  \node[vtx] at (135:1.25) {}; \node[lab,anchor=south east] at (135:1.45) {$c_1$};
  \node[vtx] at (240:1.25) {}; \node[lab,anchor=north east] at (240:1.45) {$c_{\delta-1}$};
  \node[vtx] at (300:1.25) {}; \node[lab,anchor=north west] at (300:1.45) {$c_\delta$};
  \node[gh] at (0:1.25) {}; \node[lab,anchor=west] at (0:1.55) {$c_{\delta+1}=v$};
  \node[lab] at (0,-2.05) {$\rho_2$: $\delta+2$ nodes};
\end{scope}
\end{tikzpicture}
\caption{The two rings of \cref{lem:two-left}, with clockwise to the right of $\Home$. The dark-gray node is the gray hole $v$, and the dotted arc stands for the safe nodes $c_2,\dots,c_{\delta-2}$. The agents explore counterclockwise, and an agent $Z$ is the first scheduled to move from $c_\delta$ to $c_{\delta+1}$. In $\rho_1$, the node $c_\delta$ is the gray hole, and $Z$ is destroyed at $c_\delta$ before it moves. In $\rho_2$, the node $c_\delta$ is safe, and $Z$ is destroyed when it enters $c_{\delta+1}=v$. The last agent cannot tell the two rings apart: if it enters $c_\delta$, it is destroyed in $\rho_1$, and if it never does, a safe node of $\rho_2$ is never visited again.}
\label{fig:two-left}
\end{figure}

\begin{lemma}\label{lem:two-left}
Suppose that all agents are honest, do not know $n$, and run a fixed deterministic algorithm in the face-to-face model. Consider an execution on a ring with more than $D+2$ nodes, in which the gray hole is $v$. Suppose that up to some round $s$, the agents have used only $v$, $\Home$, and $c_1,\dots,c_D$, and that at the end of round $s$, at most two agents are left and none of them stands on $v$. Then, for every natural number $n_0$, there is a ring with at least $n_0$ nodes, whose gray hole is again $v$, on which the first $s$ rounds are the same as in the given execution and on which the adversary makes \PE{} fail.
\end{lemma}

\begin{proof}
On every ring with more than $D+2$ nodes, the visited nodes $v,\Home,c_1,\dots,c_D$ form the same oriented path, with at least one unvisited node between its endpoints. Reproducing the activations from the first $s$ rounds of the given execution therefore makes the agents meet the same agents and hear the same messages on each of these rings. From round $s+1$ on, the adversary activates the gray hole in every round in which an agent moves from $\Home$ to $v$, so an agent is always destroyed when it moves from $\Home$ to $v$. If no agent is left, \PE{} fails.

Otherwise, the agents must still visit the nodes far away on the counterclockwise side. Fix $\delta\ge\max\{D+2,n_0\}$, and look first at the ring with $\delta+3$ nodes. There, $c_{\delta+1}$ is a safe node between $c_\delta$ and $c_{\delta+2}=v$. At the end of round $s$, no agent stands on $v$, and an agent can enter $v$ only from $\Home$, where it is destroyed, or from $c_{\delta+1}$. Hence an agent can reach $c_{\delta+1}$ only from $c_\delta$. If no agent ever moves from $c_\delta$ to $c_{\delta+1}$, \PE{} fails on this ring. Otherwise, let $\tau$ be the first round in which an agent $Z$ moves from $c_\delta$ to $c_{\delta+1}$ on this ring. Before round $\tau$, the agents that are not destroyed stand only on $c_0,\dots,c_\delta$.

Now the adversary uses the two rings of \cref{fig:two-left}, both with gray hole $v$: the ring $\rho_1$ of $\delta+1$ nodes, where $v=c_\delta$, and the ring $\rho_2$ of $\delta+2$ nodes, where $v=c_{\delta+1}$ and $c_\delta$ is safe. In both rings, the adversary activates the gray hole in round $\tau$ and in every round in which an agent moves from $\Home$ to $v$. Before round $\tau$, these activations destroy the same agents on $\rho_1$, $\rho_2$ and the ring of $\delta+3$ nodes, with one exception: in some round, one agent uses $c_\delta$ and another agent moves from $\Home$ to $v$. In $\rho_1$, where $c_\delta=v$, the activation in that round destroys both agents, and \PE{} fails in $\rho_1$. If no move from $\Home$ to $v$ coincides with another agent using $c_\delta$, the executions on the three rings agree up to round $\tau-1$, by \cref{obs:same-rings} with $m=\delta$.

In round $\tau$, agent $Z$ is destroyed in both $\rho_1$ and $\rho_2$: in $\rho_1$ at $c_\delta=v$ before it moves, and in $\rho_2$ when it enters $c_{\delta+1}=v$. If no agent is left in one of the two rings, \PE{} fails in that ring. Otherwise, one agent $S$ is left, the same agent in both rings, and $S$ has the same view in both rings. From round $\tau+1$ on, the adversary activates the gray hole in both rings whenever $S$ moves from $\Home$ to $v$, and in $\rho_1$ also whenever $S$ uses $c_\delta$. By \cref{obs:same-rings} with $m=\delta-1$, agent $S$ makes the same moves in both rings until it moves from $\Home$ to $v$ or uses $c_\delta$. Each choice of $S$ fails:
\begin{itemize}
  \item if $S$ moves from $\Home$ to $v$, it is destroyed in both rings;
  \item if $S$ uses $c_\delta$, in round $\tau$ or later, it is destroyed in $\rho_1$;
  \item if $S$ does neither, the safe node $c_\delta$ of $\rho_2$ is never visited again.
\end{itemize}
Hence \PE{} fails in $\rho_1$ or in $\rho_2$, and both rings have at least $n_0$ nodes.
\end{proof}

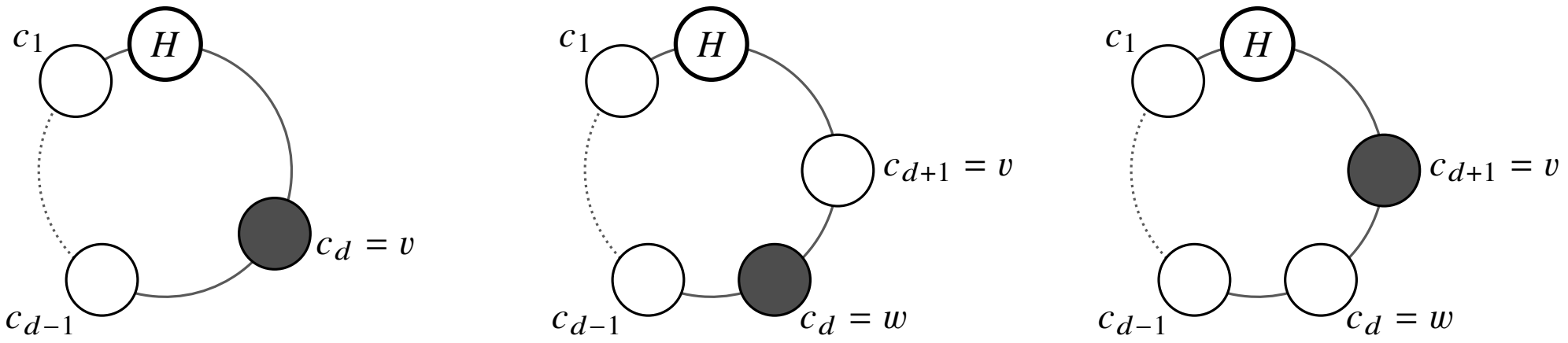
\begin{figure}[t]
\centering
\begin{tikzpicture}[scale=0.85]
\foreach \xx in {0,5.4,10.8}{
  \begin{scope}[xshift=\xx cm]
    \draw[edge] (90:1.25) arc (90:135:1.25);
    \draw[omit] (135:1.25) arc (135:240:1.25);
    \draw[edge] (240:1.25) arc (240:450:1.25);
    \node[home] at (90:1.25) {$\Home$};
    \node[vtx] at (135:1.25) {}; \node[lab,anchor=south east] at (135:1.45) {$c_1$};
    \node[vtx] at (240:1.25) {}; \node[lab,anchor=north east] at (240:1.45) {$c_{d-1}$};
  \end{scope}}
\begin{scope}[xshift=0cm]
  \node[gh] at (330:1.25) {}; \node[lab,anchor=west] at (330:1.55) {$c_d=v$};
  \node[lab] at (0,-2.05) {$R_1$: $d+1$ nodes};
\end{scope}
\begin{scope}[xshift=5.4cm]
  \node[gh] at (300:1.25) {}; \node[lab,anchor=north west] at (300:1.45) {$c_d=w$};
  \node[vtx] at (0:1.25) {}; \node[lab,anchor=west] at (0:1.55) {$c_{d+1}=v$};
  \node[lab] at (0,-2.05) {$R_2$: $d+2$ nodes};
\end{scope}
\begin{scope}[xshift=10.8cm]
  \node[vtx] at (300:1.25) {}; \node[lab,anchor=north west] at (300:1.45) {$c_d=w$};
  \node[gh] at (0:1.25) {}; \node[lab,anchor=west] at (0:1.55) {$c_{d+1}=v$};
  \node[lab] at (0,-2.05) {$R_3$: $d+2$ nodes};
\end{scope}
\end{tikzpicture}
\caption{The three rings of Case~2b in the proof of \cref{thm:f2f-unknown-lower}, with clockwise to the right of $\Home$. The dark-gray node is the gray hole, and the dotted arc stands for the safe nodes $c_2,\dots,c_{d-2}$, which are the same in all three rings. In each ring, the gray hole destroys the first agent scheduled to move two nodes clockwise from $\Home$ (round $r$) and the first agent scheduled to move from $c_d$ to $c_{d+1}$ (round $\tau$). In the subcase $Y=\{Z\}$, the at most two remaining agents have identical local views in all three ring instances. A move from $\Home$ to $v$ is safe only in $R_2$, and entering $c_d$ is safe only in $R_3$. In $R_1$, both moves lead into the gray hole.}
\label{fig:three-rings-f2f}
\end{figure}

We prove the next theorem by arranging a destruction near $\Home$ and, when needed, another far away on the counterclockwise side. The remaining agents then face either the two-ring obstruction of \cref{lem:two-left} or two possible boundary moves, each dangerous in one of the three rings of \cref{fig:three-rings-f2f}.

\begin{theorem}\label{thm:f2f-unknown-lower}
Suppose that the agents do not know $n$ and that all agents are honest. Then no algorithm solves \PE{} in the face-to-face model with four or fewer agents. This remains true even if the agents know that $n$ is at least some fixed number $n_0$.
\end{theorem}

\begin{proof}
Fix a deterministic algorithm for four agents and an integer $n_0$. We show that the algorithm fails on some ring with at least $n_0$ nodes. With fewer agents, every case below leaves even fewer agents, so the same argument applies.

We start from the execution $E_0$ in which the gray hole is never active. In this execution, the agents must visit the nodes at distance two from $\Home$. Let $r$ be the first round in which an agent moves to such a node. Before round $r$, the agents use only $\Home$ and its two neighbors. Hence, the first $r$ rounds are the same on every ring with at least five nodes, where $\Home$, its two neighbors and the two nodes at distance two are distinct. If some agent moves from $v$ to $w$ in round $r$, let $a$ be such an agent. Otherwise, an agent moves from $c_1$ to $c_2$, and we exchange the two directions in everything below. Hence we may assume that agent $a$ moves from $v$ to $w$ in round $r$. Let $P$ be the set of agents that use $v$ in round $r$. Note that $P$ contains $a$.

\emph{Case 1: $|P|\ge2$.} Let the gray hole be $v$, and let the adversary activate it in round $r$ and in no earlier round. The gray hole destroys every agent of $P$, so at most two agents are left, and none of them stands on $v$. Up to round $r$, the agents used only $v$, $\Home$, $c_1$, and $c_2$, so \PE{} fails by \cref{lem:two-left}.

\emph{Case 2: $P=\{a\}$.} Suppose that the gray hole is $v$ or $w$, and that it is active in round $r$. In both cases, agent $a$ is destroyed in round $r$ - at $v$ if the gray hole is $v$, and when it enters $w$ if the gray hole is $w$. No other agent uses $v$ or $w$ in round $r$, so the other agents observe the same events in both cases. We now describe what the surviving agents do next, without fixing the ring size yet. Consider an infinite oriented path whose nodes are, in this order, $\dots,w,v,\Home=c_0,c_1,c_2,\dots$. After agent $a$ is destroyed in round $r$, the surviving agents follow their algorithm on this path, and no agent is destroyed again. We call this execution the \emph{quiet continuation}. It serves only as a tool, and we always use finitely many of its rounds at a time. Suppose that, up to some round, no survivor has moved from $\Home$ to $v$, and that the agents have used only the nodes $c_0,\dots,c_m$. Then, in these rounds, the agents make the same observations on every ring with more than $m+2$ nodes, whether the gray hole is $v$ or $w$. We now consider two cases: either some survivor moves from $\Home$ to $v$ at some point in the quiet continuation, or no survivor ever does.
\begin{itemize}
    \item \emph{Case 2a: in the quiet continuation, some agent eventually moves from $\Home$ to $v$.} Let $t$ be the first round of the quiet continuation in which an agent moves from $\Home$ to $v$, and let $c_1,\dots,c_D$ be the nodes on the counterclockwise side that the agents have used up to round $t$. Let the gray hole be $v$, on a ring with more than $D+2$ nodes, and let the adversary activate it in rounds $r$ and $t$ and in no other round up to round $t$. Before round $t$, no agent stands on $v$, so the gray hole destroys $a$ and exactly the agents that move from $\Home$ to $v$ in round $t$. Hence at most two agents are left, none of them on $v$, and \PE{} fails by \cref{lem:two-left}.

    \item \emph{Case 2b: no agent ever moves from $\Home$ to $v$ in the quiet continuation.} Then, for every $i\ge1$, some agent eventually enters $c_i$ in the quiet continuation. Otherwise, they stay on $c_0,\dots,c_D$ for some $D$, and on a ring with more than $\max\{D+3,n_0\}$ nodes whose gray hole is $w$, active only in round $r$, the safe node $c_{D+1}$ is never visited. Fix $d\ge\max\{4,n_0\}$. Let $\tau$ be the first round of the quiet continuation in which an agent moves from $c_d$ to $c_{d+1}$, and let $Y$ be the set of agents that use $c_d$ in round $\tau$. Before round $\tau$, the agents stand only on $c_0,\dots,c_d$. The adversary now uses the three rings of \cref{fig:three-rings-f2f}:
        \begin{itemize}
            \item $R_1$ has $d+1$ nodes, and its gray hole is $c_d=v$;
            \item $R_2$ has $d+2$ nodes, and its gray hole is $c_d=w$, so that $v=c_{d+1}$ is safe;
            \item $R_3$ has $d+2$ nodes, and its gray hole is $c_{d+1}=v$, so that $c_d=w$ is safe.
        \end{itemize}
    In all three rings, the adversary activates the gray hole in rounds $r$ and $\tau$, and in no other round up to round $\tau$. Since $d\ge4$, all three rings have at least five nodes, so the first $r$ rounds are the same on them. In round $r$, the gray hole destroys only $a$, the only agent that uses $v$ or $w$: at $v$ in $R_1$ and $R_3$, and when it enters $w$ in $R_2$. By \cref{obs:same-rings} with $s=r$ and $m=d$, the three executions agree with each other up to round $\tau-1$, and in these rounds the surviving agents make the same moves as in the quiet continuation. In round $\tau$, the gray hole destroys every agent of $Y$ in $R_1$ and $R_2$, where it is $c_d$. In $R_3$, it destroys only the agents of $Y$ that move to $c_{d+1}$, among them the agent that defines $\tau$.

    \par\emph{Case 2b with $|Y|\ge2$.} In $R_1$ and $R_2$, at most one agent $S$ is left, and it has the same view in both rings. If no agent is left, \PE{} fails. Otherwise, from round $\tau+1$ on, the adversary activates the gray hole in both rings whenever $S$ enters $c_d$ (in $R_1$, where $c_d=v$, this includes every move from $\Home$ to $v$). By \cref{obs:same-rings} with $m=d-1$, agent $S$ makes the same moves in both rings until it enters $c_d$ or moves from $\Home$ to $v$. Each choice of $S$ fails:
        \begin{itemize}
            \item if $S$ enters $c_d$, it is destroyed in both rings;
            \item if $S$ moves from $\Home$ to $v$, it is destroyed in $R_1$;
            \item if $S$ does neither, it never visits $v$, a safe node of $R_2$ whose neighbors are $\Home$ and $c_d$.
        \end{itemize}
    Hence \PE{} fails in $R_1$ or in $R_2$.

    \par\emph{Case 2b with $Y=\{Z\}$.} Agent $Z$ is destroyed in all three rings, and at most two agents are left, with the same view in all three rings. We call a move of one of these agents a \emph{distinguishing move} if it enters $c_d$ or moves from $\Home$ to $v$. A move from $\Home$ to $v$ is safe only in $R_2$, and entering $c_d$ is safe only in $R_3$. From round $\tau+1$ on, in each ring, the adversary activates the gray hole of that ring in every round in which an agent enters this gray hole. As long as no agent makes a distinguishing move, the agents stand only on $c_0,\dots,c_{d-1}$, which are safe in all three rings, so the three executions agree by \cref{obs:same-rings}. Each choice of the agents fails. In the last three items, consider the first round in which an agent makes a distinguishing move:
        \begin{itemize}
            \item if no agent ever makes a distinguishing move, the safe node $c_d$ of $R_3$ is never visited again, and \PE{} fails in $R_3$;
            \item if both agents make a distinguishing move in the same round, both enter the gray hole of $R_1$, where $c_d=v$, and \PE{} fails in $R_1$;
            \item if a single agent moves from $\Home$ to $v$, it is destroyed in $R_1$ and $R_3$, and the other agent $S$, if any, keeps the same view in these two rings; then $S$ faces the same choice as the last agent in the proof of \cref{lem:two-left}, with $\delta=d$: it is destroyed in both rings if it moves from $\Home$ to $v$, it is destroyed in $R_1$ if it enters $c_d$, and otherwise the safe node $c_d$ of $R_3$ is never visited again;
            \item if a single agent enters $c_d$, it is destroyed in $R_1$ and $R_2$, and the other agent $S$, if any, keeps the same view in these two rings; then $S$ is destroyed in both rings if it enters $c_d$, it is destroyed in $R_1$ if it moves from $\Home$ to $v$, and otherwise it never visits the safe node $v$ of $R_2$.
\end{itemize}

\end{itemize}

In every case, \PE{} fails on a ring with at least $n_0$ nodes: in Case~2b, every ring has at least $d+1>n_0$ nodes, and in Cases~1 and~2a, \cref{lem:two-left} gives such a ring.
\end{proof}

\subsection{Five Honest Agents Suffice}
\label{sec:f2f-unknown-five}

The lower bound of \cref{thm:f2f-unknown-lower} leaves the last two agents unable to resolve the possible ring sizes and gray-hole locations. Our five-agent algorithm keeps two agents at $\Home$ while three agents walk clockwise as a chain. Every forward departure is observed by a partner at the departure node, so a missed meeting identifies the gray hole or leaves two adjacent candidates without a warning walk around the ring. After a destruction during the first crossing, a survivor beyond the gray hole walks on to $\Home$ through safe nodes and learns $n$ by counting its moves. A survivor before the gray hole identifies its location exactly. If the survivor still does not know $n$, it returns to $\Home$ and starts a counterclockwise chain with the two waiting agents. Identifying the gray hole at counterclockwise distance $h$ then gives $n=g+h$, where $g$ is its known clockwise distance. If the gray hole never destroys an agent of the first chain, the chain visits every safe node infinitely often. Otherwise, the gray hole destroys at most two agents in each run, and some honest agent eventually walks the omission tour of the gray hole. The agent on the correct omission tour need not distinguish the gray hole from another candidate.

We next specify the roles and schedule, explain how agents respond to missed meetings and decide at $\Home$, and describe the second run. An example illustrates the rules before we give the variables, pseudocode, and correctness proof.

Let $a_1,\dots,a_5$ be the agents in increasing order of identifier. In the \emph{first run}, $a_1$, $a_2$, and $a_3$ form a chain that walks clockwise from $\Home$, in the roles $A$, $B$, and $C$, while $a_4$ and $a_5$ wait at $\Home$. An agent of a chain does not know $n$, so it locates itself by counting. Its \emph{coordinate} is $0$ at $\Home$ when the run starts, grows by one with every move in the direction of the run, and decreases by one with every move against it. Coordinate $i$ thus denotes the node reached after $i$ moves in the direction of the run. When an agent first comes back to $\Home$ with a positive coordinate, this coordinate equals $n$, and the agent learns $n$. Agents of a chain tell each other $n$ when they meet, and they ignore agents outside their run.

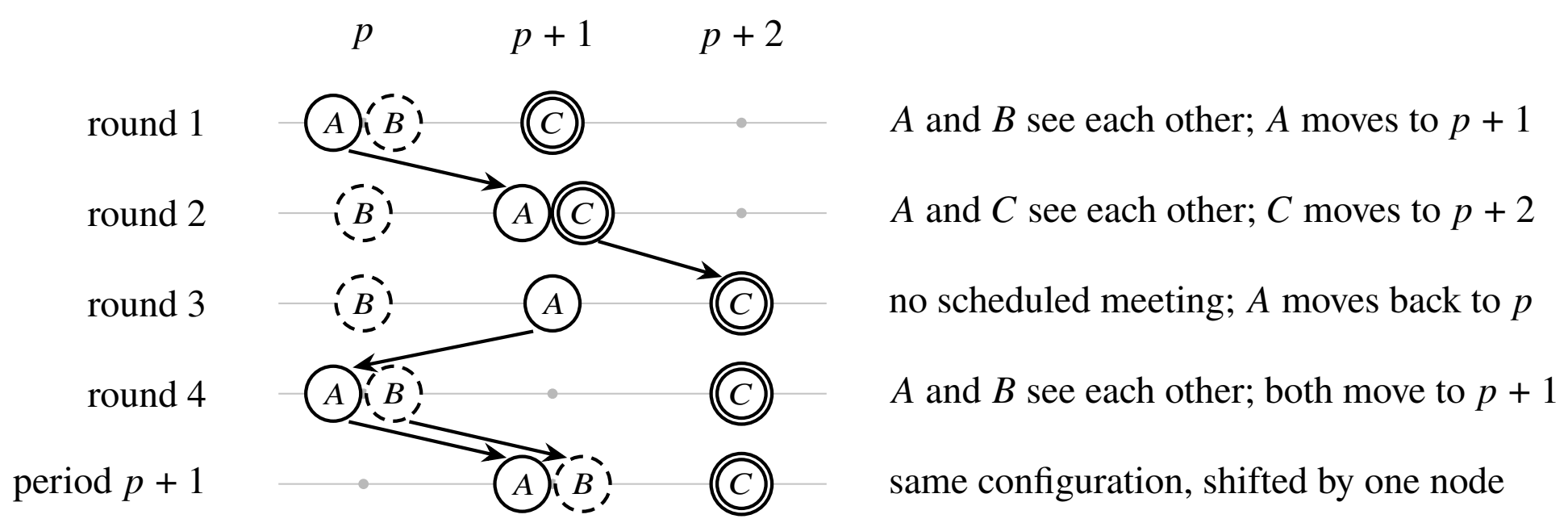
\begin{figure}[t]
\centering
\begin{tikzpicture}[x=1.55cm,y=0.74cm]
  \foreach \c/\l in {0/p,1/{p+1},2/{p+2}} \node at (\c,0.95) {$\l$};
  \foreach \j/\t in {0/{round 1},1/{round 2},2/{round 3},3/{round 4},4/{period $p+1$}}{
    \node[anchor=east] at (-0.75,-\j) {\t};
    \draw[tgrid] (-0.45,-\j)--(2.45,-\j);
    \foreach \c in {0,1,2} \node[dot] at (\c,-\j) {};}
  \node[agA] at (-0.16,0) {$A$}; \node[agB] at (0.16,0) {$B$}; \node[agC] at (1,0) {$C$};
  \node[agB] at (0,-1) {$B$}; \node[agA] at (0.84,-1) {$A$}; \node[agC] at (1.16,-1) {$C$};
  \node[agB] at (0,-2) {$B$}; \node[agA] at (1,-2) {$A$}; \node[agC] at (2,-2) {$C$};
  \node[agA] at (-0.16,-3) {$A$}; \node[agB] at (0.16,-3) {$B$}; \node[agC] at (2,-3) {$C$};
  \node[agA] at (0.84,-4) {$A$}; \node[agB] at (1.16,-4) {$B$}; \node[agC] at (2,-4) {$C$};
  \draw[move] (-0.1,-0.3)--(0.78,-0.72);
  \draw[move] (1.22,-1.3)--(1.92,-1.72);
  \draw[move] (0.92,-2.3)--(-0.08,-2.72);
  \draw[move] (-0.1,-3.3)--(0.78,-3.72);
  \draw[move] (0.22,-3.3)--(1.1,-3.72);
  \node[anchor=west,lab] at (2.7,0) {$A$ and $B$ see each other; $A$ moves to $p+1$};
  \node[anchor=west,lab] at (2.7,-1) {$A$ and $C$ see each other; $C$ moves to $p+2$};
  \node[anchor=west,lab] at (2.7,-2) {no scheduled meeting; $A$ moves back to $p$};
  \node[anchor=west,lab] at (2.7,-3) {$A$ and $B$ see each other; both move to $p+1$};
  \node[anchor=west,lab] at (2.7,-4) {same configuration, shifted by one node};
\end{tikzpicture}
\caption{Period $p$ of the tight chain. Each row shows where the agents stand at the start of a round, and the arrows show the moves of the round. Agent $A$ (solid outline) relays information between $B$ (dashed outline), which follows behind, and $C$ (double outline), which moves ahead. Every agent that moves forward is seen by a partner at the start of the same round: $B$ sees $A$ in round~1, $A$ sees $C$ in round~2, and $A$ and $B$ see each other in round~4. After four rounds, the chain has advanced by one node.}
\label{fig:tight-period}
\end{figure}

The movement schedule starts with one start-up round, in which $C$ moves from coordinate $0$ to coordinate $1$. The chain then repeats \emph{periods} of four rounds. As in \cref{sec:f2f-algorithms}, a period is one repetition of the chain schedule that advances the chain by one node; the schedule here uses four rounds instead of seven steps. We number the rounds of a period from 1 to 4 and write \emph{round $j$ of period $p$}; ``round~$j$'' alone refers to the current period. Period $p\ge0$ starts with $A$ and $B$ on $p$ and $C$ on $p+1$ (\cref{fig:tight-period}). In round~1, $A$ moves from $p$ to $p+1$. In round~2, $C$ moves from $p+1$ to $p+2$. In round~3, $A$ moves from $p+1$ back to $p$. In round~4, $A$ and $B$ move from $p$ to $p+1$ together. The chain has then advanced by one node, and period $p+1$ starts. A \emph{lap} consists of the $n$ periods in which the chain advances once around the ring. We call the resulting three-agent chain the \emph{tight chain}. Agent $C$ stays ahead, $B$ stays behind, and $A$ relays information between them. Every forward move begins in the presence of a partner: $B$ sees $A$ leave in round~1, $A$ sees $C$ leave in round~2, and $A$ and $B$ move together in round~4. The partner therefore knows exactly which nodes the leaving agent uses until their next meeting. Each period has two scheduled meetings. At the start of round~2, $A$ and $C$ expect each other on $p+1$, and at the start of round~4, $A$ and $B$ expect each other on $p$. Agents $A$ and $B$ also stand together at the start of round~1, but they arrived together, so the meeting at the start of round~1 needs no check.

An agent that misses its partner at a scheduled meeting looks at the nodes that the partner has used since the two agents last saw each other. \Cref{tab:tight-rules} lists the four possible missed meetings, what each of them proves, and what happens next. In rules~(i) to~(iii), an agent identifies the gray hole exactly. In rule~(iv), $C$ keeps two adjacent candidates as coordinates and removes $\Home$ whenever it can recognize $\Home$ among them. In rules~(i) and~(iii), $A$ \emph{relays} its identification to its surviving partner, $B$ in rule~(i) and $C$ in rule~(iii), within at most two moves; the partner then leaves the chain together with $A$. A \emph{crossing} is the chain's scheduled passage through one occurrence of the gray-hole coordinate. Relative to a crossing, an agent leaves on the \emph{home side} if its coordinate is smaller than the gray-hole coordinate at that crossing, and on the \emph{far side} if its coordinate is larger. These terms specify a direction relative to the current run, not two fixed sides of the ring. Each rule fixes the single move that the agent applying the rule makes in the round in which the rule applies, and an agent that receives a relayed identification may start its walk home or its omission tour as its move in the round when it receives the identification. Agent $A$ completes its relay before it leaves the chain. The relay ends at the start of round~3 in rule~(i), and at the start of round~2 of period $p+1$ in rule~(iii). \Cref{lem:tight-missed} proves the conclusions of the table.

\begin{table}[t]
\centering
\small
\renewcommand{\arraystretch}{1.15}
\begin{tabular}{@{}c p{3.5cm} p{3.0cm} p{7.6cm}@{}}
\toprule
\textbf{Rule} & \textbf{Missed meeting} & \textbf{Conclusion} & \textbf{What happens next} \\
\midrule
(i) & $A$ misses $C$ on $p+1$, round~2 & $C$ was destroyed at $p+1$ & $A$ moves back to $p$ in the same round and tells $B$ at the start of round~3. Both leave on the home side, at $p$. \\
(ii) & $B$ misses $A$ on $p$, round~4 & $A$ was destroyed at $p+1$ & $B$ does not move to $p+1$ and leaves on the home side, at $p$. \\
(iii) & $A$ misses $B$ on $p$, round~4 & $B$ was destroyed at $p$ & $A$ makes its next two scheduled moves, to $p+1$ and $p+2$, and tells $C$ at the start of round~2 of period $p+1$. Both leave on the far side, at $p+2$. \\
(iv) & $C$ misses $A$ on $p+1$, round~2 & $A$ was destroyed at $p-1$ or at $p$ & $C$ keeps the candidates $p-1$ and $p$, without $\Home$, and leaves on the far side, at $p+1$. \\
\bottomrule
\end{tabular}
\caption{The four missed meetings of the tight chain in period $p$, in coordinates along the run. Each meeting is checked at the start of the round named in the second column. \Cref{lem:tight-missed} proves the third column.}
\label{tab:tight-rules}
\end{table}

An agent that leaves the chain and knows both $n$ and the exact gray hole starts the omission tour of the gray hole at once. Any other agent walks to $\Home$. On the home side, it walks back against the direction of the run. On the far side, it walks on in the direction of the run and learns $n$ when it reaches $\Home$. Both walks use only safe nodes, since the home side lies before the gray hole and the far side lies after it. An agent records its candidates as coordinates. Once it knows $n$, it reduces them modulo $n$, removes $\Home$, and converts them to clockwise distances, so that all evidence refers to the same node labels. When agents exchange candidate sets, they intersect these sets after conversion; an exact identification is a candidate set containing one node.

The waiting agents and the agents that have come back from a chain decide together at $\Home$. No other agent takes part, and an agent that arrives at the end of a round takes part from the next round on. In every round, the agents that take part exchange what they know and apply the first of the following rules that fits.
\begin{enumerate}[label=(H\arabic*)]
  \item If they know $n$ and the gray hole, they all walk the omission tour of the gray hole.
  \item If they know $n$ and two candidates, and at least two agents take part, they sort the candidates by clockwise distance and assign their omission tours alternately, in increasing order of identifier, so that both tours are walked.
  \item If they know the clockwise distance $g$ of the gray hole but not $n$, and at least three agents take part, the three agents with the smallest identifiers start the \emph{second run}, in the roles $A$, $B$, and $C$, in increasing order of identifier. The other agents keep waiting.
\end{enumerate}
If no rule fits, the agents wait at $\Home$.

The second run is a tight chain that walks counterclockwise. Its agents count new coordinates from $0$ and react only to missed meetings of the second run, but they keep $g$. A survivor that identifies the gray hole at counterclockwise distance $h$ learns $n=g+h$, and a survivor on the far side walks on to $\Home$ and learns $n$ there. If the gray hole destroys no agent of the second run, the run goes around the ring: $C$ reaches $\Home$ in round~2 of period $n-2$, $A$ reaches $\Home$ in round~1 of period $n-1$, and $B$ hears $n$ from $A$ at the start of round~4 of period $n-1$. The second run therefore has a \emph{final exit} scheduled after its first lap: at the start of round~4 of period $n-1$, after the agents have exchanged their messages, every agent of the second run that knows $n$ starts the omission tour of $g$. The final exit takes priority over any scheduled move or relay. An agent that survives on the gray hole until the message exchange at the final exit leaves it in the same round. The departure is safe because survival until the message exchange implies that the gray hole is inactive for the round.

\begin{figure}[t]
\centering
\begin{tikzpicture}
\foreach \xx in {0,8}{
  \begin{scope}[xshift=\xx cm]
    \draw[edge] (0,0) circle (1.8);
    \foreach \v in {1,...,9} \node[vtx] at ({90-36*\v}:1.8) {$u_{\v}$};
    \node[home] at (90:1.8) {$\Home$};
    \node[gh] at (-54:1.8) {$u_{4}$};
    \draw[walk] (-18:1.05) arc (-18:72:1.05);
  \end{scope}}
\begin{scope}[xshift=0cm]
  \node[lab] at (0,3.05) {(a) First run, clockwise};
  \draw[dir] (78:2.35) arc (78:40:2.35);
  \node[ag] at (-0.5,2.55) {$a_4$}; \node[ag] at (-1.08,2.42) {$a_5$};
  \node[agB] at (-18:2.45) {$a_2$};
  \node[note,anchor=west] at (1.45,-1.72) {$a_3$ and $a_1$\\destroyed at $u_4$};
  \node[note] at (0.05,0.1) {$a_2$ returns\\to $\Home$};
  \node[note,anchor=north] at (0,-2.6) {$a_2$ knows $g=4$ but not $n$};
\end{scope}
\begin{scope}[xshift=8cm]
  \node[cand] at (-90:1.8) {$u_{5}$};
  \node[lab] at (0,3.05) {(b) Second run, counterclockwise};
  \draw[dir] (102:2.35) arc (102:140:2.35);
  \node[agC] at (-18:2.45) {$a_5$};
  \node[note,anchor=west] at (1.45,-1.72) {$a_2$ and $a_4$\\destroyed at $u_4$};
  \node[note] at (0.05,0.1) {$a_5$ continues\\to $\Home$};
  \node[note,anchor=north] at (0,-2.6) {$a_5$ counts $7+3=10$ moves: $n=10$};
\end{scope}
\end{tikzpicture}
\caption{The five-agent algorithm on a ring of ten nodes whose gray hole is $u_4$ (dark gray), in the worst case, in which four agents are destroyed. The arrow near $\Home$ shows the direction of the run. (a)~In the first run ($A=a_1$, $B=a_2$, $C=a_3$), the gray hole destroys $C$ while it waits alone on $u_4$, and then $A$ when $A$ enters $u_4$ (outcome~(b) of \cref{lem:tight-outcomes}). Agent $a_2$ identifies $u_4$ and walks back to $\Home$, where $a_4$ and $a_5$ wait. (b)~In the second run ($A=a_2$, $B=a_4$, $C=a_5$), the gray hole destroys $a_2$ and $a_4$ together at $u_4$ (outcome~(d) of \cref{lem:tight-outcomes}). Agent $a_5$ keeps the candidates $u_5$ (light gray) and $u_4$, walks on from $u_3$ to $\Home$, and learns $n=10$. With $g=4$ from the first run, it walks the omission tour of $u_4$. In both panels, the dashed arrow is the same walk from $u_3$ to $\Home$: a walk on the home side in the first run and on the far side in the second.}
\label{fig:five-example}
\end{figure}
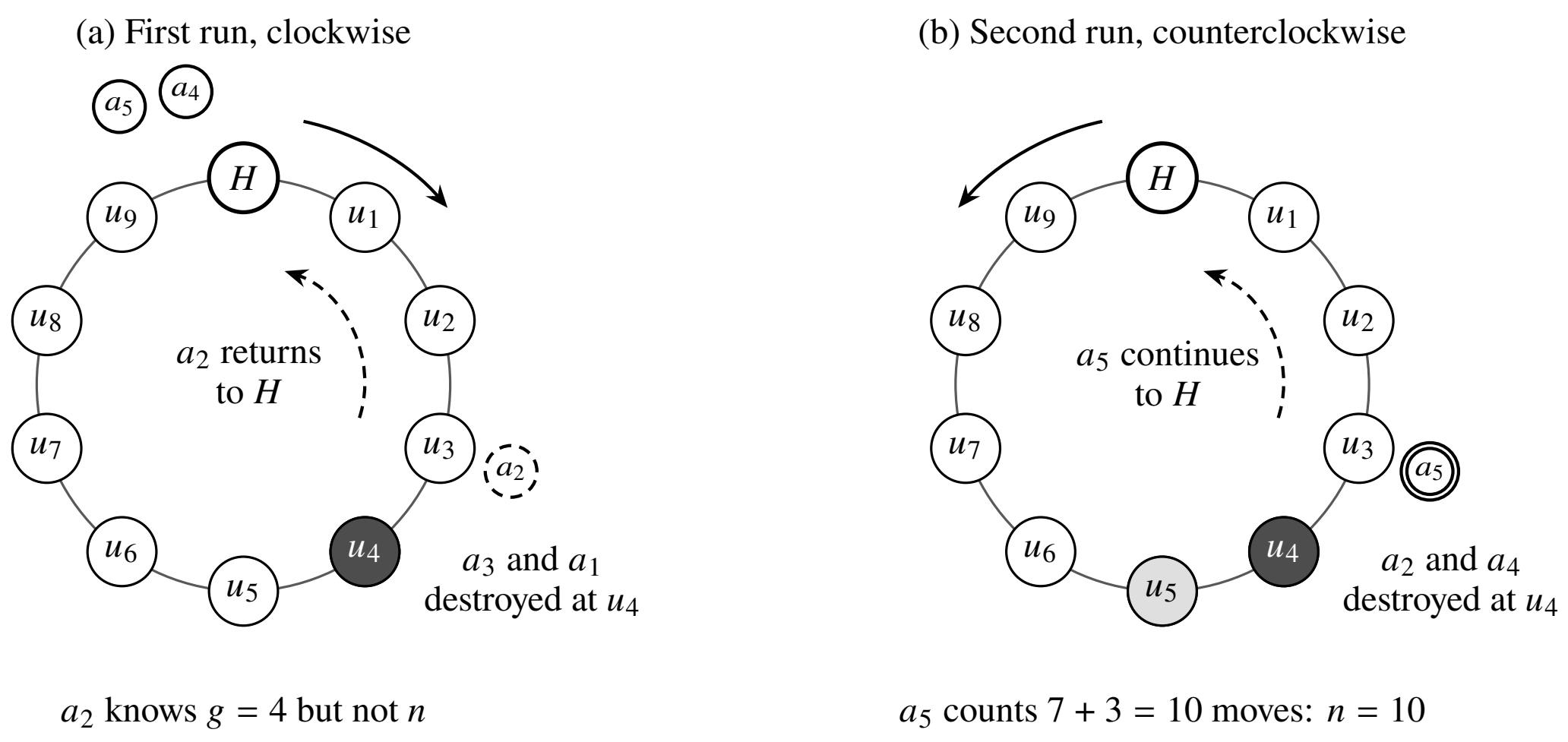

\Cref{fig:five-example} follows the algorithm on a ring of ten nodes whose gray hole is $u_4$, in the worst case, in which the gray hole destroys four agents. In the first run, $C=a_3$ reaches $u_4$ and waits there alone, and the gray hole destroys it. When $A=a_1$ enters $u_4$ to meet $C$, the gray hole is active again and destroys $A$ as well. Agent $B=a_2$ saw $A$ move from $u_3$ to $u_4$, misses $A$ at their next meeting, and identifies $u_4$ as the gray hole by rule~(ii). It knows $g=4$ but not $n$. It walks back to $\Home$, where rule~(H3) starts the second run with $a_2$, $a_4$, and $a_5$ in the roles $A$, $B$, and $C$. The second run walks counterclockwise through $u_9,u_8,\dots,u_5$, so it reaches $u_4$ at counterclockwise distance $6$. In the second run, the gray hole destroys $A=a_2$ and $B=a_4$ together when they enter $u_4$. Agent $C=a_5$ stands on $u_3$, at counterclockwise distance $7$. It misses $A$ and keeps the candidates $u_5$ and $u_4$ by rule~(iv). It walks on through $u_2$ and $u_1$ to $\Home$ and counts $7+3=10$ moves, so $n=10$. It learned $g=4$ from $a_2$ at $\Home$ before the second run, so rule~(H1) applies, and $a_5$ walks the omission tour of $u_4$, back and forth between $u_3$ and $u_5$ through $\Home$. The walk from $u_3$ to $\Home$ is the same in both runs, but it plays two roles: in the first run, $u_3$ lies on the home side, and in the second run, it lies on the far side.

To implement the rules, each agent stores the ring size $n$, unknown at first. It learns $n$ when it first comes back to $\Home$ with a positive coordinate, from another agent, or by computing $n=g+h$ after identifying the gray hole at counterclockwise distance $h$ in the second run. It stores its candidates of each run and the clockwise distance $g$ of the gray hole, once it knows $g$. Before each call to \textsc{Chain}, the selected agents assign roles $A,B,C$ in identifier order and agree on the current round as the common start-up round. Each stores its role, coordinate, and start-up round, from which it computes the current period $p$ and round within that period. In rule~(H3), only the selected three agents call \textsc{Chain}; the others keep waiting. \Cref{alg:five} summarizes the algorithm.

\begin{algorithm}[tbp]
\caption{Perpetual exploration with five honest agents and unknown ring size, for an agent $X$.}
\label{alg:five}
\begin{algorithmic}[1]
\If{$X\in\{a_1,a_2,a_3\}$} \Call{Chain}{clockwise} \Else{} \Call{AtHome}{} \EndIf
\Statex
\Procedure{Chain}{direction}
  \State start a new coordinate for this run, keeping earlier evidence
  \Repeat \Comment{one round, including the start-up round}
    \State exchange $n$ with the partners on the node, and deliver any relayed identification that is due
    \If{counterclockwise, $n$ known, $p=n-1$, round~4} \Call{Tour}{$u_g$} \Comment{final exit has priority} \EndIf
    \State process a relayed identification or a missed scheduled meeting by rules~(i)--(iv)
    \State choose and make at most one move: the required relay move, the first move after leaving the chain, or the scheduled move
  \Until{$X$ leaves the chain by rules~(i)--(iv)}
  \State continue the selected action from the next round if a move was already made in the current round
  \If{$n$ and the exact gray hole $x$ are known} \Call{Tour}{$x$}
  \Else{} walk to $\Home$ (against the run on the home side, in the run direction on the far side); \Call{AtHome}{}
  \EndIf
\EndProcedure
\Statex
\Procedure{AtHome}{}
  \Loop
    \State exchange $n$ and the candidates with all agents that take part at $\Home$
    \If{$n$ and the exact gray hole $x$ are known} \Call{Tour}{$x$} \Comment{(H1)}
    \ElsIf{$n$ and two candidates are known, and at least two agents take part}
      \State \Call{Tour}{the candidate assigned to $X$ by identifier} \Comment{(H2)}
    \ElsIf{$g$ is known, $n$ is not, and at least three agents take part}
      \State the three smallest identifiers, as $A$, $B$, $C$: \Call{Chain}{counterclockwise} \Comment{(H3)}
    \EndIf
    \State wait for the next round if $X$ is still at $\Home$
  \EndLoop
\EndProcedure
\Statex
\Procedure{Tour}{$x$}
  \If{$X$ stands on $x$} move to a neighbor in this round; continue the tour from the next round \EndIf
  \State walk the omission tour $O(x)$ forever
\EndProcedure
\end{algorithmic}
\end{algorithm}

The correctness proof has three parts: the information obtained from each missed meeting, the possible outcomes of crossing the gray hole, and the progress of the two runs. Throughout, we use the following consequence of the activation rule. An active gray hole destroys every agent that uses it, so if an agent used a node $u$ in a round and survived that round, the gray hole did not destroy its partner at $u$ in the same round. The partner could still have been destroyed at another node that it used in the same round.

\begin{lemma}\label{lem:tight-missed}
Consider a scheduled meeting in period $p$ of the tight chain of \cref{alg:five}, with all three agents honest. Suppose that both agents of this meeting have followed the schedule of the tight chain since they last met, unless they were destroyed. If one of them misses the other, then the missing agent was destroyed at the node, or at one of the two nodes, given below:
\begin{itemize}
  \item if $A$ misses $C$ in round~2, then $C$ was destroyed at node $p+1$;
  \item if $B$ misses $A$ in round~4, then $A$ was destroyed at node $p+1$;
  \item if $A$ misses $B$ in round~4, then $B$ was destroyed at node $p$;
  \item if $C$ misses $A$ in round~2, then $p\ge1$, and $A$ was destroyed at node $p-1$ or at node $p$.
\end{itemize}
\end{lemma}

\begin{proof}
Suppose that $A$ misses $C$ in round~2. Agent $A$ saw $C$ on node $p$ at the start of round~2 of period $p-1$, or at $\Home$ before the start-up round if $p=0$, and the gray hole did not destroy $C$ at node $p$ in that round, since $A$ survived on the same node. From its entry into $p+1$ until the missed meeting, $C$ used only $p+1$. Suppose next that $B$ misses $A$ in round~4. Agent $B$ saw $A$ on $p$ at the start of round~1. Since then, $A$ used $p+1$ in rounds~1 to~3 and entered $p$ in round~3. Agent $B$ stood on $p$ in all these rounds and survived, so the gray hole did not destroy $A$ at $p$, and hence $A$ was destroyed at $p+1$. Suppose now that $A$ misses $B$ in round~4. Agent $A$ saw $B$ on $p$ at the start of round~1, and $B$ has used only $p$ since. Suppose finally that $C$ misses $A$ in round~2. For $p=0$, agent $A$ used only $\Home$ before it entered node~$1$, where $C$ stood and survived, so $A$ cannot be missing. For $p\ge1$, agent $C$ saw $A$ on node $p$ at the start of round~2 of period $p-1$. Since then, $A$ used $p$ and $p-1$, and it entered $p+1$ in round~1 of period $p$. Agent $C$ stood on $p+1$ in round~1 of period $p$ and survived, so $A$ was destroyed at $p-1$ or at $p$.
\end{proof}

We write $x$ for the coordinate of the gray hole at the crossing under consideration. In the first run, $x=g+kn$ for some lap index $k\ge0$; in particular, $x=g$ on the first crossing. In the second run, the first crossing has coordinate $x=h=n-g$. For a crossing with $x\ge2$, if no agent has yet left the tight chain or been destroyed, the scheduled visits to the gray hole occupy eleven consecutive rounds. The start-up case $x=1$ is described below. Grouping consecutive rounds with the same users gives six blocks:
\[
\underbrace{C}_{3\text{ rounds}}\;\longrightarrow\;
\underbrace{A,C}_{2\text{ rounds}}\;\longrightarrow\;
\underbrace{A}_{1\text{ round}}\;\longrightarrow\;
\underbrace{A,B}_{2\text{ rounds}}\;\longrightarrow\;
\underbrace{B}_{1\text{ round}}\;\longrightarrow\;
\underbrace{A,B}_{2\text{ rounds}}
\]
First, $C$ enters $x$ in round~2 of period $x-2$ and waits there alone. In period $x-1$, agent $A$ joins $C$ in round~1, both stand on $x$ in round~2, and $A$ stands there alone in round~3. Agents $A$ and $B$ enter $x$ together in round~4 and stand on it in round~1 of period $x$. Agent $B$ is alone on $x$ in round~2, $A$ comes back in round~3, and both leave in round~4. For $x=1$, the start-up round replaces the entry in period $x-2$, and the first block has one round, giving nine scheduled usage rounds in total. We call an activation of the gray hole that destroys at least one agent a \emph{destructive activation}. Before the first destructive activation, all three agents still follow the schedule. The first destructive activation therefore destroys all agents listed in its block. When a survivor leaves the chain, it stands at coordinate $x-1$ on the home side, or at $x+1$ or $x+2$ on the far side. We call the destruction of $A$ and $B$ in the fourth block an \emph{early loss}. A \emph{late loss} destroys $B$ in the fifth or sixth block and $A$ in the sixth block. We call $\{x-1,x\}$ the \emph{early pair} and $\{x,x+1\}$ the \emph{late pair}; the following lemma shows when these pairs are the candidates of $C$.

The same tight-chain rules apply when the actual ring has $n=3$; the agents do not switch to the known-size triangle algorithm. The exit coordinates $x-1$, $x+1$, and $x+2$ all denote safe nodes because none equals $x$ modulo three. Although $x-1$ and $x+2$ denote the same physical node, the exit rule records whether the agent must move against or in the direction of its run. The agent learns $n$ by the same return-to-$\Home$ rule used on larger rings.

\begin{figure}[t]
\centering
\begin{tikzpicture}[x=0.86cm,y=0.86cm]
\foreach \sx/\sy in {0/0,5.9/0,11.8/0,2.95/-3.5,8.85/-3.5}{
  \begin{scope}[shift={(\sx,\sy)}]
    \draw[edge] (-0.6,0)--(3.6,0);
    \foreach \i/\l in {0/{x-1},1/x,2/{x+1},3/{x+2}} \node[lab] at (\i,-0.62) {$\l$};
    \node[vtx] at (0,0) {}; \node[gh] at (1,0) {}; \node[vtx] at (2,0) {}; \node[vtx] at (3,0) {};
  \end{scope}}
\begin{scope}[shift={(0,0)}]
  \node[ttl] at (-0.9,1.5) {(a) $C$ is destroyed};
  \node[agA] at (-0.28,0.75) {$A$}; \node[agB] at (0.28,0.75) {$B$};
  \draw[walk] (-0.62,0.75)--(-1.25,0.75);
  \node[lab] at (1.5,-1.15) {$A$ and $B$ identify $x$};
\end{scope}
\begin{scope}[shift={(5.9,0)}]
  \node[ttl] at (-0.9,1.5) {(b) $C$ and $A$ are destroyed};
  \node[agB] at (0,0.75) {$B$};
  \draw[walk] (-0.35,0.75)--(-1.0,0.75);
  \node[lab] at (1.5,-1.15) {$B$ identifies $x$};
\end{scope}
\begin{scope}[shift={(11.8,0)}]
  \node[ttl] at (-0.9,1.5) {(c) $A$ is destroyed};
  \node[cand] at (0,0) {};
  \node[agB] at (0,0.75) {$B$}; \node[agC] at (2,0.75) {$C$};
  \draw[walk] (-0.35,0.75)--(-1.0,0.75);
  \draw[walk] (2.35,0.75)--(3.0,0.75);
  \node[lab,align=center] at (1.5,-1.3) {$B$ identifies $x$;\\$C$: candidates $x-1$, $x$};
\end{scope}
\begin{scope}[shift={(2.95,-3.5)}]
  \node[ttl] at (-0.9,1.5) {(d) $A$ and $B$ are destroyed};
  \node[cand] at (0,0) {};
  \node[agC] at (2,0.75) {$C$};
  \draw[walk] (2.35,0.75)--(3.0,0.75);
  \node[lab,align=center] at (1.5,-1.3) {$C$: early loss, $\{x-1,x\}$;\\late loss, $\{x,x+1\}$};
\end{scope}
\begin{scope}[shift={(8.85,-3.5)}]
  \node[ttl] at (-0.9,1.5) {(e) $B$ is destroyed};
  \node[agA] at (2.72,0.75) {$A$}; \node[agC] at (3.28,0.75) {$C$};
  \draw[walk] (3.62,0.75)--(4.25,0.75);
  \node[lab] at (1.5,-1.15) {$A$ and $C$ identify $x$};
\end{scope}
\end{tikzpicture}
\caption{The five outcomes of \cref{lem:tight-outcomes}, in coordinates along the run. The dark-gray node is the gray hole, at coordinate $x$, a light-gray node is a candidate, and the direction of the run is to the right. Each panel shows the survivors where they leave the chain, and the text below states what they know. A dashed arrow shows the direction of a survivor's walk to $\Home$: backward on the home side, forward on the far side. Panel~(d) shows an early loss of $A$ and $B$. After a late loss, $C$ leaves at $x+2$ with the candidates $\{x,x+1\}$. In panels~(c) and~(d), $C$ removes $\Home$ from its candidates whenever it recognizes it.}
\label{fig:tight-outcomes}
\end{figure}
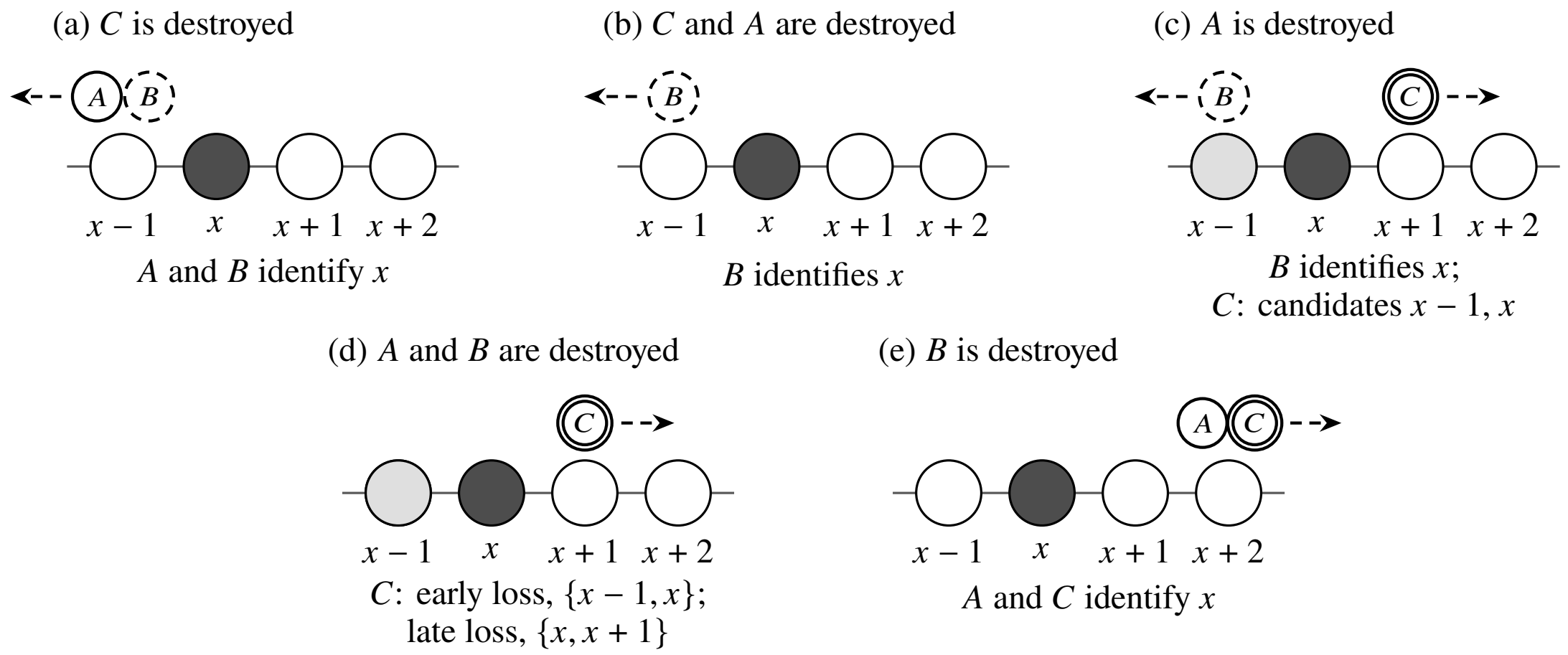

\begin{lemma}\label{lem:tight-outcomes}
Consider the tight chain of \cref{alg:five}, with all three agents honest and following rules~(i) to~(iv). Suppose that the gray hole destroys an agent of the chain while the chain crosses it at coordinate $x$. Then the gray hole destroys at most two agents of the chain, every survivor leaves the chain on a safe node, and one of the following five outcomes holds (\cref{fig:tight-outcomes}).
\begin{enumerate}[label=(\alph*)]
  \item $C$ is destroyed. $A$ and $B$ are on the home side and identify $x$ as the gray hole.
  \item $C$ and $A$ are destroyed. $B$ is on the home side and identifies $x$ as the gray hole.
  \item $A$ is destroyed. $B$ is on the home side and identifies $x$ as the gray hole. $C$ is on the far side, and its candidates are $x-1$ and $x$, except $\Home$.
  \item $A$ and $B$ are destroyed. $C$ is on the far side. Its candidates form the early pair $\{x-1,x\}$ after an early loss, or the late pair $\{x,x+1\}$ after a late loss. In both cases, $\Home$ is removed from the candidates.
  \item $B$ is destroyed. $A$ and $C$ are on the far side and identify $x$ as the gray hole.
\end{enumerate}
\end{lemma}

\begin{proof}
We consider the block of the first destructive activation. Suppose first that it destroys $C$ alone. Agent $A$ then enters $x$ in round~1 of period $x-1$ and stands on $x$ in round~2. If the gray hole is inactive in both rounds, $A$ misses $C$, identifies $x$ by \cref{lem:tight-missed}, and moves back to $x-1$ in round~2, in which $x$ is inactive. Agent $A$ tells $B$ at the start of round~3, before $B$ is scheduled to move, which gives outcome~(a). Otherwise, the gray hole destroys $A$ as well. Agent $B$ then misses $A$ in round~4, identifies $x$, and stays on $x-1$, which gives outcome~(b). The same holds if the first destructive activation destroys $A$ and $C$ together. Suppose next that it destroys $A$ alone, in round~3 of period $x-1$. Agent $C$ has moved on to $x+1$. Agent $B$ misses $A$ in round~4 and identifies $x$, and $C$ misses $A$ in round~2 of period $x$ and keeps the candidates $\{x-1,x\}$, which gives outcome~(c). If the first destructive activation destroys $A$ and $B$ when they enter $x$ or one round later, $C$ misses $A$ in the same way, which gives outcome~(d) after an early loss. Suppose now that it destroys $B$ alone, in round~2 of period $x$. Agent $A$ enters $x$ in round~3 and stands on $x$ in round~4. If the gray hole is inactive in both rounds, $A$ misses $B$ and identifies $x$ by \cref{lem:tight-missed}. Its next two scheduled moves lead to $x+1$ and $x+2$, where $A$ tells $C$, which gives outcome~(e). Otherwise, the gray hole destroys $A$ as well, and $C$ misses $A$ in round~2 of period $x+1$ and keeps the candidates $\{x,x+1\}$, which gives outcome~(d) after a late loss. The same holds if the first destructive activation destroys $A$ and $B$ in the last block. Agent $A$ can be destroyed later only if the first destructive activation destroys $C$ alone or $B$ alone, so at most two agents are destroyed. Once the checks and relays end, every survivor stands at $x-1$, $x+1$, or $x+2$, and these nodes are safe for every $n\ge3$.
\end{proof}

Thus every survivor on the home side has identified the gray hole exactly, and a survivor with two candidates is always on the far side. If the gray hole is the first node in the direction of the run, the early pair $\{0,1\}$ reduces to $\{1\}$ because coordinate $0$ is $\Home$. We now show how the survivors use their information to complete perpetual exploration, either after the first run or after a second run.

\begin{lemma}\label{lem:first-run}
Suppose that five honest agents execute \cref{alg:five} without knowing $n$. In the first run, one of the following three cases occurs.
\begin{enumerate}[label=(\arabic*)]
  \item The gray hole never destroys an agent of the chain, so the honest agents of the chain visit every safe node infinitely often.
  \item Some agent eventually walks the omission tour of the gray hole.
  \item At most two agents are destroyed, and three agents that know the clockwise distance $g$ of the gray hole start the second run.
\end{enumerate}
\end{lemma}

\begin{proof}
If the gray hole never destroys an agent of the chain, the chain advances by one node in every period and visits every node infinitely often. Otherwise, \cref{lem:tight-outcomes} applies, and at most two agents of the chain are destroyed. The waiting agents stand on the safe node $\Home$ and are never destroyed. On the first crossing of the gray hole, a survivor on the home side stands at $g-1$ and walks back to $\Home$ along safe nodes, and a survivor on the far side that still needs $n$ walks on to $\Home$ along safe nodes. On the first crossing, a survivor on the far side has passed $\Home$ when it leaves the chain only if $g=n-1$ and it leaves at coordinate $g+2=n+1$. Such a survivor already knows $n$, and it has either an exact identification or the late pair $\{n-1,n\}$, which reduces to $\{n-1\}$ once $\Home$ is removed, so it starts the correct omission tour. The analysis of the first crossing applies to every later crossing, with coordinates taken modulo $n$. If no agent is destroyed in the first lap, then by its end every agent of the chain knows $n$, and a period shifted by $n$ has the same meetings and candidates, and its agents leave the chain in the same directions.

Now consider what the survivors know when they reach $\Home$. A survivor that knows $n$ and the exact gray hole starts its omission tour after its relay, if any, and (2) holds. In outcomes~(a) and~(b), if no survivor already knows $n$, the survivors bring the exact $g$ to $\Home$, where the two waiting agents still stand, and (H3) starts the second run, so (3) holds. In outcome~(e), $A$ and $C$ reach $\Home$ with the exact $g$ and with $n$, and (H1) applies. In outcome~(d), $C$ reaches $\Home$ with $n$ and its candidates. If one candidate is left, (H1) applies. If two are left, (H2) applies with the two waiting agents, and the agent assigned to the gray hole walks its omission tour. In outcome~(c), $B$ comes back with the exact $g$, and $C$ walks on to $\Home$ with $n$ and at most two candidates. If $B$ takes part first, (H3) starts the second run. If $C$ takes part first, (H1) or (H2) applies. If both take part in the same round, they exchange what they know, and (H1) applies. In every case, (2) or (3) holds.
\end{proof}

\begin{lemma}\label{lem:second-run}
Suppose that three honest agents know the clockwise distance $g$ of the gray hole and start the second run of \cref{alg:five}. Then some agent eventually walks the omission tour of the gray hole.
\end{lemma}

\begin{proof}
Every agent of the second run knows $g$. If an agent takes the final exit, it knows $g$ and $n$ and starts the correct omission tour, leaving the gray hole in the same round if it stands on it. The final-exit argument covers a lap without destruction and every destruction or relay that overlaps the final exit. For instance, if $g=1$, the second run reaches the gray hole at counterclockwise distance $n-1$, in the last periods of its lap. If its agent $B$ is destroyed alone in round~2 of period $n-1$, then $A$ misses $B$ in round~4 in the same round in which $C$ takes the final exit, and $A$ and $C$, which both know $n$, start their tours in that round. Otherwise, \cref{lem:tight-outcomes} applies to the second run with the directions exchanged and with $x=h$, where $h$ is the counterclockwise distance of the gray hole, and at least one agent of the chain survives. A survivor that identifies the gray hole at counterclockwise distance $h$ learns $n=g+h$ if it does not know $n$ yet. A survivor with two candidates is on the far side, so it walks on to $\Home$ along safe nodes and learns $n$ there. In both cases, the survivor knows $g$ and $n$ and walks the omission tour of the gray hole. Agents of the first run that reach $\Home$ after the second run has started are not needed.
\end{proof}

\begin{theorem}\label{thm:f2f-unknown-five}
Suppose that the agents do not know $n$ and that all agents are honest. Then \cref{alg:five} solves \PE{} in the face-to-face model with five agents.
\end{theorem}

\begin{proof}
By \cref{lem:first-run}, either the agents of the first run visit every safe node infinitely often, or some agent walks the omission tour of the gray hole, or three agents that know $g$ start the second run. In the last case, some agent walks the omission tour of the gray hole by \cref{lem:second-run}. An agent on the omission tour of the gray hole never enters the gray hole, so it is never destroyed. It walks back and forth along the safe path, so it visits every safe node infinitely often.
\end{proof}

\begin{corollary}\label{cor:f2f-unknown-honest}
Suppose that the agents do not know $n$ and that all agents are honest. Then, in the face-to-face model, five agents are necessary and sufficient for \PE{}.
\end{corollary}

\begin{proof}
Four or fewer honest agents cannot solve \PE{} when $n$ is unknown, by \cref{thm:f2f-unknown-lower}. The five-agent algorithm solves \PE{} in the same setting, by \cref{thm:f2f-unknown-five}.
\end{proof}

\subsection{Tolerating \texorpdfstring{$f$}{f} Byzantine Agents}
\label{sec:f2f-unknown-byz}

We now turn to Byzantine agents. With whiteboards, $3f+3$ agents are not enough when $n$ is unknown (\cref{thm:3f+3-unknown}). The same bound holds in the face-to-face model, and its proof uses the same three rings, illustrated in \cref{fig:three-rings}. In the construction, $z=c_d$ is a node far from $\Home$ on the counterclockwise side. Every agent that could tell the rings apart moves into $v$, $w$, or $z$, where it is destroyed or, if it is Byzantine, stops acting, so the agents on the common safe path never meet it again.

\begin{theorem}\label{thm:f2f-unknown-byz-lower}
Suppose that the agents do not know $n$ and that at most $f$ agents are Byzantine. Then no algorithm solves \PE{} in the face-to-face model with $3f+3$ or fewer agents. This remains true even if the agents know that $n$ is at least some fixed number $n_0$.
\end{theorem}

\begin{proof}
Fix a deterministic face-to-face algorithm for at most $3f+3$ agents and an integer $d\ge\max\{3,n_0\}$. Let $E_0$ be the execution in which all agents are honest and the gray hole is never active, and let $r$ be the first round in which an agent moves from a node other than $\Home$ to an unvisited node. Before round $r$, the agents have visited only $\Home$ and its neighbors. By exchanging the two directions if necessary, let the move be from the clockwise neighbor $v$ to the next node $w$, and let $M$ be the nonempty set of agents scheduled to make it.

The three ring instances illustrated in \cref{fig:three-rings} also support an indistinguishability argument in the face-to-face model. In $I_1$, the ring has $d+1$ nodes and the gray hole is $v=z=c_d$. In $I_2$, it has $d+2$ nodes and the gray hole is $w=z=c_d$. In $I_3$, it has $d+3$ nodes, the gray hole is $v$, and $z=c_d$ is safe. Their common safe path is $\Home,c_1,\dots,c_{d-1}$. The gray hole is inactive before round $r$ and active from round $r$ onward. Let $P$ contain the agents outside $M$ that stand on $v$ in round $r$ or subsequently enter $v$ from $\Home$, including arrivals in round $r$. Let $F$ contain the agents that enter $z$ from $c_{d-1}$. Every agent of $P$ is destroyed in $I_1$ and $I_3$; in the construction of $I_2$, it is Byzantine and stops acting at $v$. Every agent of $F$ is destroyed in $I_1$ and $I_2$; in the construction of $I_3$, it is Byzantine and stops acting at $z$. The agents of $M$ are destroyed in all three instances. The counting below selects a construction using at most $f$ Byzantine agents.

The first difference arises in round $r$. In $I_1$ and $I_3$, the gray hole destroys the agents on $v$ at the start of round $r$, before they act. In $I_2$, the node $v$ is safe, so the agents on $v$ meet there in round $r$. We let every agent of $P$ on $v$ send in round $r$ the same messages as in $E_0$. The agents of $P$ on $v$ stop acting only after sending these messages. Hence the agents of $M$ hear the same messages as in $E_0$, move to $w$, and are destroyed there. Only agents on $v$ hear these messages, and all of them belong to $M\cup P$.

After round $r$, every agent outside $M\cup P\cup F$ stays on the $d$ safe nodes $\Home,c_1,\dots,c_{d-1}$. An agent leaves these nodes only when it joins $P$ or $F$, and then it never returns: it is destroyed at $v$, $w$ or $z$, or it stops acting at node $v$ or node $z$ and sends no more messages. Hence, by induction on the rounds, every agent outside $M\cup P\cup F$ meets the same agents and hears the same messages in $I_1$, $I_2$ and $I_3$, and it makes the same moves in all three executions. The identical observations and moves on the common safe path therefore determine the same disjoint sets $M$, $P$, and $F$ in all three constructions. Since the algorithm is deterministic, the sets can be specified before the failing execution begins. No Byzantine identity is changed during the execution. A stopped Byzantine agent remains visible at its node, but an agent outside $M\cup P\cup F$ never visits the node of a stopped agent.

The three possible counts give a legal failing execution as follows:
\begin{itemize}
  \item if $|F|\le f$, the safe node $z$ of $I_3$ is entered only by Byzantine agents, and \PE{} fails in $I_3$;
  \item if $|P|\le f$, honest agents visit the safe node $v$ of $I_2$ only finitely often, and \PE{} fails in $I_2$;
  \item otherwise, at least $|M|+2(f+1)\ge 2f+3$ agents are destroyed in $I_1$, at most $f$ agents are left, and the adversary chooses them to be Byzantine, so \PE{} fails in $I_1$.
\end{itemize}
In each case, the failing execution has at most $f$ Byzantine agents. If it has fewer, the adversary chooses further agents to be Byzantine and lets them follow the algorithm, which changes nothing.
\end{proof}

We extend the five-agent algorithm to Byzantine agents by splitting the team into independent groups, ensuring that at least one group is entirely honest.

\begin{corollary}\label{cor:f2f-unknown-five}
Suppose that the agents do not know $n$ and that at most $f$ agents are Byzantine. Then, in the face-to-face model, $5f+5$ agents suffice for \PE{}, and at least $\max\{5,3f+4\}$ agents are necessary.
\end{corollary}

\begin{proof}
At $\Home$, the agents sort their identifiers, and the team splits into $f+1$ independent groups of five consecutive agents, at least one of which is entirely honest. Each group runs the algorithm of \cref{thm:f2f-unknown-five} and ignores the other agents. No agent can forge the identifier of another agent, so the other groups cannot disturb the entirely honest group, and \cref{thm:f2f-unknown-five} applies to it. For $f=0$, the lower bound is \cref{thm:f2f-unknown-lower}. For $f\ge1$, it is \cref{thm:f2f-unknown-byz-lower}.
\end{proof}

The bound of $5f+5$ agents improves the baseline of $8f+8$ agents and is tight for $f=0$. For $f\ge1$, a gap remains between $3f+4$ and $5f+5$ agents. The proof of \cref{thm:f2f-unknown-lower} uses only honest agents, and each distinguishing move in that construction can cause the destruction of a single agent. With Byzantine agents, one disappearance does not establish that an honest agent was destroyed, because a Byzantine agent may deliberately stop acting. A direct adaptation of the four-agent counting argument would therefore need to account for agents that deliberately stop acting at each distinguishing move. The proof of \cref{thm:f2f-unknown-byz-lower} does so for its two distinguishing moves, the move into $v$ and the move into $z$: in its final counting case, each of the sets $P$ and $F$ has at least $f+1$ agents.

\section{Conclusion}
\label{sec:conclusion}

We studied perpetual exploration of an oriented ring that contains one gray hole and up to $f$ Byzantine agents, where the same adversary controls the gray hole and the Byzantine agents.
Our results leave several questions open. First, in the whiteboard model with unknown $n$, we do not know whether $3f+4$ agents suffice. Second, in the face-to-face model, the gaps are larger. For known $n\ge4$, the minimum number of agents lies between $2f+3$ and $3f+3$; for the known triangle, it is $2f+2$. For unknown $n$, five honest agents are necessary and sufficient in the worst case over all ring sizes. The requirement of five honest agents is two more than the three needed for known $n\ge4$. With $f\ge1$ Byzantine agents, the minimum lies between $3f+4$ and $5f+5$. Our face-to-face algorithms for Byzantine agents use independent groups; allowing agents to cooperate across groups is one possible approach to improving the upper bounds. Third, we do not know whether our proof that four honest agents are not enough also holds in the whiteboard model. For $f=0$, a whiteboard version of the four-agent lower bound would close the gap between four and five agents in the whiteboard model. Several further directions remain open. Our whiteboard algorithms rely on every activation of the gray hole erasing its whiteboard. Hence the Byzantine black hole of Goswami et al.~\cite{DBLP:conf/opodis/GoswamiBD024}, which may keep its whiteboard, is a natural next model. Other directions include agents that start at different nodes, networks other than rings, and the time and memory that the agents need.

\par\medskip\noindent\begin{minipage}{\linewidth}
\section*{Acknowledgment of AI Use}
We contributed most of the ideas in this paper. AI assistants (Claude and ChatGPT) helped us refine the ideas and write the paper, including the polishing of the text, the figures, and the checking of references. We take full responsibility for the write-up.
\end{minipage}

\bibliographystyle{unsrtnat}
\bibliography{references}

\end{document}